\documentclass[journal]{IEEEtran}
\usepackage{array}
\usepackage{color}
\usepackage{epsf}
\usepackage{times}
\usepackage{epsfig}
\usepackage{graphicx}
\usepackage{epstopdf}
\usepackage{hyperref}
\usepackage{cite}
\usepackage{amsmath}
\usepackage{amssymb}
\usepackage{amsxtra}
\usepackage{amsthm}
\usepackage{bbm}
\usepackage{algorithmic}
\usepackage{algorithm}
\usepackage{subfigure}
\usepackage{authblk}
\usepackage{booktabs}
\usepackage{longtable}
\usepackage[table,xcdraw]{xcolor}
\usepackage{tikz}
\usepackage{adjustbox}
\newcommand{\tabitem}{~~\llap{\textbullet}~~}
\usepackage[normalem]{ulem}
\usepackage{tabularray}
\useunder{\uline}{\ul}{}
\usepackage[colorinlistoftodos,bordercolor=orange,backgroundcolor=orange!20,linecolor=orange,textsize=scriptsize]{todonotes}
\usepackage{afterpage}

\usepackage{amssymb}
\usepackage{pifont}
\newcommand{\cmark}{\ding{51}}%
\newcommand{\xmark}{\ding{55}}%

\ifCLASSINFOpdf
  \usepackage{graphicx}
\else
\fi

\begin{document}

\title{On the Interplay of Artificial Intelligence and Space-Air-Ground Integrated Networks: A Survey}

\author{Adilya~Bakambekova,~\IEEEmembership{Student Member,~IEEE,}
        Nour~Kouzayha,~\IEEEmembership{Member,~IEEE,}
        and~Tareq~Al-Naffouri,~\IEEEmembership{Senior Member,~IEEE}
        \thanks{The authors are with the Electrical and Computer Engineering program, CEMSE division, King Abdullah University of Science and Technology (KAUST), Thuwal, 23955-6900, Kingdom of Saudi Arabia.}
}


\maketitle

\begin{abstract}
Space-Air-Ground Integrated Networks (SAGINs), which incorporate space and aerial networks with terrestrial wireless systems, are vital enablers of the emerging sixth-generation (6G) wireless networks. Besides bringing significant benefits to various applications and services, SAGINs are envisioned to extend high-speed broadband coverage to remote areas, such as small towns or mining sites, or areas where terrestrial infrastructure cannot reach, such as airplanes or maritime use cases. However, due to the limited power and storage resources, as well as other constraints introduced by the design of terrestrial networks, SAGINs must be intelligently configured and controlled to satisfy the envisioned requirements.
Meanwhile, Artificial Intelligence (AI) is another critical enabler of 6G. Due to massive amounts of available data, AI has been leveraged to address pressing challenges of current and future wireless networks. By adding AI and facilitating the decision-making and prediction procedures, SAGINs can effectively adapt to their surrounding environment, thus enhancing the performance of various metrics. In this work, we aim to investigate the interplay of AI and SAGINs by providing a holistic overview of state-of-the-art research in AI-enabled SAGINs. Specifically, we present a comprehensive overview of some potential applications of AI in SAGINs. We also cover open issues in employing AI and detail the contributions of SAGINs in the development of AI. Finally, we highlight some limitations of the existing research works and outline potential future research directions.
\end{abstract}


\IEEEpeerreviewmaketitle

\section{Introduction}
\IEEEPARstart{I}{n} recent years, substantial scientific and industrial interest has been devoted to the sixth generation (6G) of wireless networks. 6G systems will foster next-generation vertical services and facilitate the establishment of an intelligent and fully connected society by seamlessly connecting people, things, data, applications, smart cities, and transportation systems. Thanks to its enhanced capabilities, 6G is poised to accommodate a multitude of devices and manage extensive data, prioritizing the delivery of services that augment the user experience and cater to emerging applications, including the Internet of Things (IoT), Augmented Reality (AR), Virtual Reality (VR), holographic communications, immersive media streaming, and smart societies~\cite{dang2020should}. To fulfill the various quality-of-service (QoS) requirements of the newly added services and applications, 6G networks are anticipated to offer ubiquitous connectivity, low latency, high capacity, and high reliability. Existing terrestrial wireless networks, constrained by limited coverage and capacity, face challenges in universally delivering reliable and cost-effective wireless services, particularly in challenging locations like disaster-stricken areas, mountainous regions, and oceans~\cite{geraci2022integrating}.

Space-Air-Ground Integrated Networks (SAGINs) will play an increasingly vital role in meeting the demand for seamless connectivity in 6G systems~\cite{sharrab2023}. SAGINs consist of three network segments, i.e., the communication satellites form the space sub-network; aerial communication devices such as balloons, airships, high-altitude platforms (HAPS), and unmanned aerial vehicles (UAVs) form the aerial sub-network; and terrestrial communication networks form the ground sub-network. A key advantage of SAGINs is their ability to extend the coverage of terrestrial networks and alleviate congestion caused by a growing number of connected devices. This capability is crucial as deploying emerging technologies, such as autonomous aerial fleets, cargo drones, and flying cars, becomes more prevalent~\cite{mozaffari2019}.
\begin{figure*}
    \centering
    \includegraphics[width=0.75\textwidth]{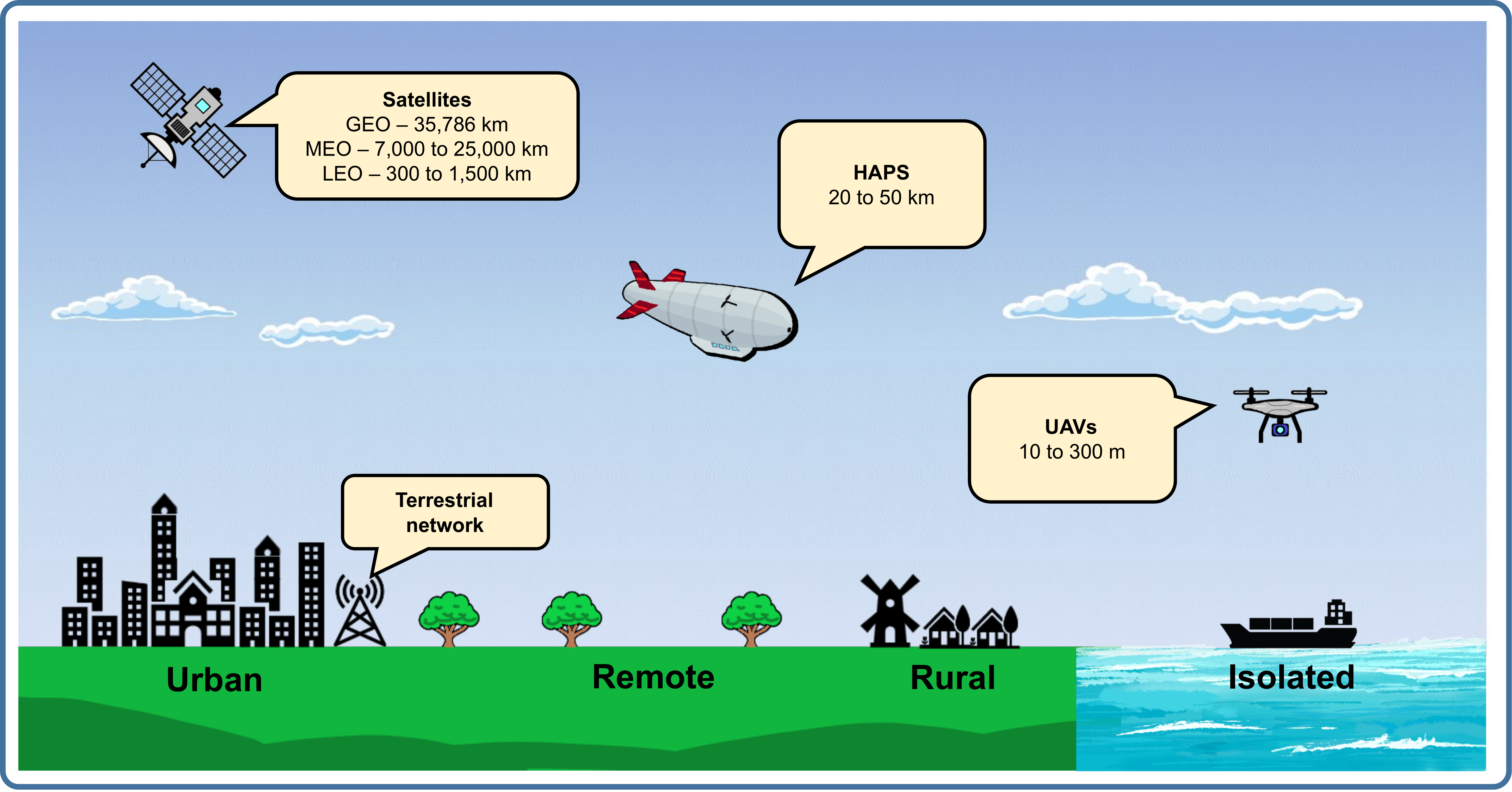}
    \caption{Various components of SAGIN (satellites, HAPS, UAVs, and the terrestrial network) serving users in urban, remote, rural, and isolated areas.}
    \label{fig:img1}
    \vspace{-0.2cm}
\end{figure*}

Fig.~\ref{fig:img1} provides an overview of how different elements of SAGINs are distributed on Earth and in the atmosphere. The ground sub-network consists mainly of base stations (BSs), with coverage limited to nearby users. Consequently, users located further away or in regions lacking terrestrial network infrastructure may experience performance issues or a complete lack of service. A viable option to enhance ground networks, satellite communication systems became essential for enabling worldwide connectivity and closing the digital divide~\cite{kodheli2021satellite}. Next-generation SAGINs, especially those involving UAVs, are expected to be supported by satellite networks for terrestrial and aerial communications~\cite{yun2023dynamic}. Low-Earth orbit (LEO) constellations, which are usually deployed at altitudes of $300$ to $1500$~km, are the key elements in the future vision of global communication and coverage and are taking the leading role towards the integration of satellites in the SAGIN architecture~\cite{salem2023exploiting}. Geostationary (GEO) and medium-earth orbit (MEO) are yet other types of satellites that usually orbit the Earth at altitudes of $35786$~km and $7000$ to $25000$~km, respectively~\cite{hraishawi2023survey}.

Recent years have witnessed significant strides in satellite technology, notably with private initiatives like SpaceX, OneWeb, and Amazon leading the charge in satellite constellation design, operation, and research~\cite{giordani_2021}. While efforts are underway to standardize the next generation of satellite networks, several fundamental challenges persist. These challenges include resource management, network control, network security, spectrum management, and energy usage~\cite{ahmad2022security} and are mainly due to the distinctive attributes of satellites, such as rapid orbital motion, varying coverage range, and evolving topology. Meeting the demand for adaptability in a dynamically changing radio environment inherent in satellite networks necessitates the development of agile methodologies capable of rapid and near-instantaneous adaptation~\cite{fourati_artificial_2021}.

Apart from the space sub-network, the two main components of the aerial sub-network are the HAPSs and UAVs. HAPSs, usually deployed in the stratosphere at cloud-free atmospheric altitudes between $20$ and $50$~km, can vary in design features, operability (manned or unmanned), stability (aerostatic or aerodynamic), mobility (free-flying or controlled), autonomy (autonomous or tethered), and energy source (motorized or non-motorized)~\cite{mershad2021cloud}. The high deployment altitude of HAPSs makes them more practical in providing stable access to large numbers of users in remote areas and lower transmission energy costs compared to traditional fixed BSs. Furthermore, HAPSs are promoted to provide mobile edge computing (MEC) services to swiftly manage computational tasks and effectively handle the surge in traffic spurred by the expanded adoption of the IoT. HAPSs can also act as relay nodes between the satellites and the terrestrial network~\cite{Jia2022Toward}. Because of their unique benefits, such as adaptable deployment, cost efficiency, simplified maintenance, and broader service coverage, HAPSs have the potential to enhance system throughput, thereby boosting the scalability of SAGINs.

UAVs can operate at various altitudes and exhibit different design features, such as fixed-wing or multi-rotor configurations. They can be operated remotely or autonomously, and their energy source can be battery-powered or fueled~\cite{delavarpour2021}. UAVs can also be equipped with lightweight BS equipment and act as airborne BSs to serve users in challenging locations and when the terrestrial network is down~\cite{kouzayha2021analysis}. This solution extends further to scenarios that require high capacity and enhanced service quality, including highly populated areas and hotspots~\cite{mozaffari2019}. UAVs can, therefore, provide on-demand support to the terrestrial network, effectively managing unexpected traffic spikes and emergency cases~\cite{Kouzayha2020Stochastic}. On the other side, UAVs can be leveraged as aerial users that use existing infrastructure (connect to either the terrestrial or the space sub-networks) to serve diverse tasks such as parcel delivery, search and rescue missions, inspections, surveillance, and event streaming~\cite{Geraci2022What}. In both scenarios, UAVs face several challenges that must be addressed for efficient integration in SAGIN systems. These objectives include but are not limited to optimizing service quality, reducing energy usage, guaranteeing user and core network connectivity, and mitigating interference. In this context, UAVs, for instance, must optimize the flying trajectory to meet performance requirements and provide the expected service while taking system constraints into consideration~\cite{Zeng2019Accessing}. UAVs also suffer from limited flight endurance and payload capacity, which can impact the range and breadth of the communication systems they can support~\cite{delavarpour2021}. Despite these challenges, UAVs offer the advantage of rapid deployment and mobility, making them an indispensable component of the SAGIN architecture. 

Leveraging SAGINs, 6G systems can benefit from their ability to maintain reliable, uninterrupted, and low-latency communication, mainly in difficult-to-reach locations and densely populated cities. These diverse applications also highlight the need for intelligent algorithms and improved orchestration techniques in SAGINs to address the challenges in task allocation, mobility handling, and radio resource management (RRM)~\cite{zhou2023}. Optimization efforts are also mandatory to balance the different targeted objectives, including energy efficiency, coverage region, and user satisfaction. Furthermore, the intrinsic dynamics of SAGIN environments and fluctuating user demands require high adaptability and intelligence within SAGINs. Traditional optimization techniques and heuristic algorithms fail to meet these requirements due to their constraints in flexibility and adaptability, especially given the evolving and diverse nature of SAGINs~\cite{Homssi2023Artificial}.

In another context, Artificial Intelligence (AI) has emerged in the last few years as a collection of algorithms that mimic a human's behavior, using predictions based on available data. Moreover, AI has become prevalent in scientific research for decision-making and optimization problems thanks to the huge amount of available data nowadays and high-performance computing systems. AI is expected to be a key pillar for 6G networks, providing numerous benefits and driving the evolution of future wireless communications~\cite{letaief2019}. Unlike conventional, reactive methods, AI techniques empower wireless networks to learn and adapt in an autonomous fashion. This self-driven intelligence allows them to analyze non-linear network behaviors and automatically adjust configurations for optimal efficiency. Recently, AI algorithms have found their way into various applications in SAGINs. Specifically, AI algorithms can address the limitations of conventional optimization approaches in SAGINs by efficiently managing resources, intelligently allocating tasks, and autonomously making decisions while accounting for the dynamic, heterogeneous, and distributed characteristics of the SAGIN environment~\cite{michailidis_ai-inspired_2020}.

Among the different existing AI algorithms, Reinforcement Learning (RL) has demonstrated remarkable success in optimizing various aspects of SAGINs~\cite{naous2023}. RL-based algorithms, through their ability to discern intricate interdependencies between network parameters and environmental cues, offer an optimal and adaptive solution for dynamic SAGIN environments. RL is mainly used to enhance the orchestration and control of SAGINs by optimizing factors like trajectory planning and resource allocation. Furthermore, recent advancements in Machine Learning (ML), Deep Learning (DL), and Neural Networks (NNs) can effectively address scalability issues in large-scale SAGINs, facilitating. As a result, the development of adaptable, efficient, and intelligent SAGINs~\cite{Bai2023Towards}. Federated learning (FL) is yet another efficient AI algorithm that can adapt to the distributed architecture of SAGINs by involving different SAGIN layers in the learning process, further enhancing the scalability and adaptability of SAGIN systems~\cite{Tang2023Federated}.

While the notion of AI driving SAGINs prevails, the possibility of a reverse influence requires further investigation. Certainly, AI algorithms are crucial for SAGINs, extracting insights from vast data and optimizing performance. However, we propose that SAGINs can potentially enhance several AI domains. For instance, they offer a rich source of diverse, high-quality data vital for improving model accuracy. In addition, their edge computing capabilities can significantly accelerate AI training. Given the increasing importance of AI in SAGINs, there is a need to provide a comprehensive survey of the state-of-the-art research on this topic. In light of this, this paper aims to serve this need by providing a holistic overview of the role and applications of AI in future SAGINs, summarizing and identifying the key challenges for different SAGIN layers, highlighting the interplay between AI and SAGIN, and discussing potential future directions. 

\begin{table*}[t!]
\caption{Summary of existing relevant surveys}
\resizebox{\textwidth}{!}{%
\begin{tabular}{clllccccc}
\hline
Publication & \multicolumn{1}{c}{Title} & \multicolumn{1}{c}{Year} & \multicolumn{1}{c}{Highlight} & \begin{tabular}[c]{@{}c@{}}AI in Comm. \\ Systems\end{tabular} & Satellites & HAPS & UAV & Integrating SAGINs \\ \hline
\cite{rinaldi_non-terrestrial_2020} & \begin{tabular}[c]{@{}l@{}}Non-Terrestrial Networks in \\ 5G \& Beyond: A Survey\end{tabular} & 2020 & \begin{tabular}[c]{@{}l@{}} \tabitem SAGIN evolution from 1G to 5G;\\ \tabitem Radio resource allocation and \\ mobility management\\ \tabitem Role of SAGIN in cellular \\ communications\end{tabular} & \xmark & \cmark & $\partial$ & $\partial$ & $\partial$ \\ \hline
\cite{michailidis_ai-inspired_2020} & \begin{tabular}[c]{@{}l@{}}AI-Inspired Non-Terrestrial \\ Networks for IIoT: Review on \\ Enabling Technologies and \\ Applications\end{tabular} & 2020 & \begin{tabular}[c]{@{}l@{}} \tabitem UAV-enabled and AI-enhanced \\ industrial IoT solutions\\ \tabitem Challenges of SAGIN-enabled \\ Industrial IoT\end{tabular} & \cmark & $\partial$ & $\partial$ & \cmark & \xmark \\ \hline
\cite{jiang_road_2021} & \begin{tabular}[c]{@{}l@{}}The Road Towards 6G: A \\ Comprehensive Survey\end{tabular} & 2021 & \begin{tabular}[c]{@{}l@{}} \tabitem Drivers and challenges of 6G\end{tabular} & $\partial$ & \xmark & \xmark & \xmark & $\partial$ \\ \hline
\cite{fourati_artificial_2021} & \begin{tabular}[c]{@{}l@{}}Artificial intelligence for \\ satellite communication: A review\end{tabular} & 2021 & \begin{tabular}[c]{@{}l@{}} \tabitem Satellite communications \\ challenges and solutions\end{tabular} & \cmark & \cmark & \xmark & \xmark & \xmark \\ \hline
\cite{lahmeri_artificial_2021} & \begin{tabular}[c]{@{}l@{}}Artificial Intelligence for \\ UAV-Enabled Wireless Networks: \\ A Survey\end{tabular} & 2021 & \begin{tabular}[c]{@{}l@{}} \tabitem UAV communications \\ challenges and solutions\end{tabular} & \cmark & \xmark & \xmark & \cmark & \xmark \\ \hline
\cite{vaezi_cellular_2022} & \begin{tabular}[c]{@{}l@{}}Cellular, Wide-Area, and \\ Non-Terrestrial IoT: A Survey \\ on 5G Advances and the Road \\ Toward 6G\end{tabular} & 2022 & \begin{tabular}[c]{@{}l@{}} \tabitem Different wireless technologies \\ (including SAGIN) used for IoT \\ \tabitem Use of DL for IoT\end{tabular} & \cmark & $\partial$ & \xmark & $\partial$ & \xmark \\ \hline
\cite{xiao2022guest} & \begin{tabular}[c]{@{}l@{}}Antenna Array Enabled \\ Space/Air/Ground Communications \\ and Networking for 6G\end{tabular} & 2022 & \begin{tabular}[c]{@{}l@{}} \tabitem Antenna array enabled SAGINs\\ \tabitem Respective designs, challenges \\ opportunities, and future research directions\end{tabular} & \xmark & $\partial$ & $\partial$ & $\partial$ & $\partial$ \\ \hline
\cite{zhu2022survey} & \begin{tabular}[c]{@{}l@{}}Integrated Satellite-Terrestrial Networks \\ Toward 6G: Architectures, Applications, \\ and Challenges\end{tabular} & 2022 & \begin{tabular}[c]{@{}l@{}} \tabitem Categorization and overview of \\ the satellite-terrestrial integration architecture\\ \tabitem Typical applications of the integrated \\ architecture\\ \tabitem Main challenges such as the long \\ transmission delay, complicated link conditions, \\ and high dynamics of the network structure\end{tabular} & \xmark & $\partial$ & \xmark & \xmark & $\partial$ \\ \hline
\cite{naous2023} & \begin{tabular}[c]{@{}l@{}}Reinforcement Learning in the Sky: \\ A Survey on Enabling Intelligence \\ in NTN-Based Communications\end{tabular} & 2023 & \begin{tabular}[c]{@{}l@{}} \tabitem Reinforcement learning-based solutions \\ for SAGINs-related wireless aspects\\ \tabitem Markov decision process formulations \\ of the control-related problems\end{tabular} & $\partial$ & \cmark & \cmark & \cmark & \xmark \\ \hline
\cite{zhou2023} & \begin{tabular}[c]{@{}l@{}}Aerospace Integrated Networks \\ Innovation for Empowering 6G: A Survey \\ and Future Challenges\end{tabular} & 2023 & \begin{tabular}[c]{@{}l@{}} \tabitem System architecture and key components \\ of SAGINs\\ \tabitem Modeling, performance, and system \\ optimization for resource adaption \\ \tabitem Challenges and future directions, \\ with emphasis on 6G SAGINs with \\ ultra-dense satellite constellations \end{tabular} & $\partial$ & \cmark & \cmark & \cmark & $\partial$ \\ \hline
\cite{Bai2023Towards} & \begin{tabular}[c]{@{}l@{}}Towards Autonomous Multi-UAV \\ Wireless Network: A Survey of Reinforcement \\ Learning-Based Approaches\end{tabular} & 2023 & \begin{tabular}[c]{@{}l@{}} \tabitem Challenges of Multi-UAV wireless networks \\ \tabitem Recent advances of RL for Multi-UAV \\ wireless networks\end{tabular} & \cmark & \xmark & \xmark & \cmark & \xmark \\ \hline
\cite{Kurunathan2023Machine} & \begin{tabular}[c]{@{}l@{}}Machine Learning-Aided Operations \\and Communications of Unmanned Aerial \\Vehicles: A Contemporary Survey\end{tabular} & 2023 & \begin{tabular}[c]{@{}l@{}} \tabitem Growth areas, challenges and \\ research gaps of ML-aided UAV \\ operations and communications \\ \tabitem Classification of latest ML tools \end{tabular} & \cmark & \xmark & \xmark & \cmark & \xmark \\ \hline
\cite{mahboob2023revolutionizing} & \begin{tabular}[c]{@{}l@{}}Revolutionizing Future Connectivity: \\ A Contemporary Survey on AI-empowered \\ Satellite-based Non-Terrestrial Networks \\ in 6G\end{tabular} & 2023 & \begin{tabular}[c]{@{}l@{}} \tabitem Challenges and potential of \\ non-Terrestrial networks in the \\ context of 6G \\ \tabitem ML and DL techniques for \\ satellite-based non-terrestrial\\ networks  \end{tabular} & \cmark & \cmark & \xmark & \xmark & $\partial$ \\ \hline
\multicolumn{3}{c}{\textbf{This paper}} & \begin{tabular}[c]{@{}l@{}}Our contributions:\\ \tabitem Systematic review of AI-enabled \\ solutions for different SAGIN components\\ \tabitem Summary of AI-aided SAGIN \\ integration optimization  techniques\\ \tabitem Main future drivers and open \\ challenges\end{tabular} & \textbf{\cmark} & \textbf{\cmark} & \textbf{\cmark} & \textbf{\cmark} & \textbf{\cmark} \\ \hline
\end{tabular}
}
\label{table:surveys}
\end{table*}

\begin{figure*}[t!]
    \centering
    \includegraphics[width=\textwidth]{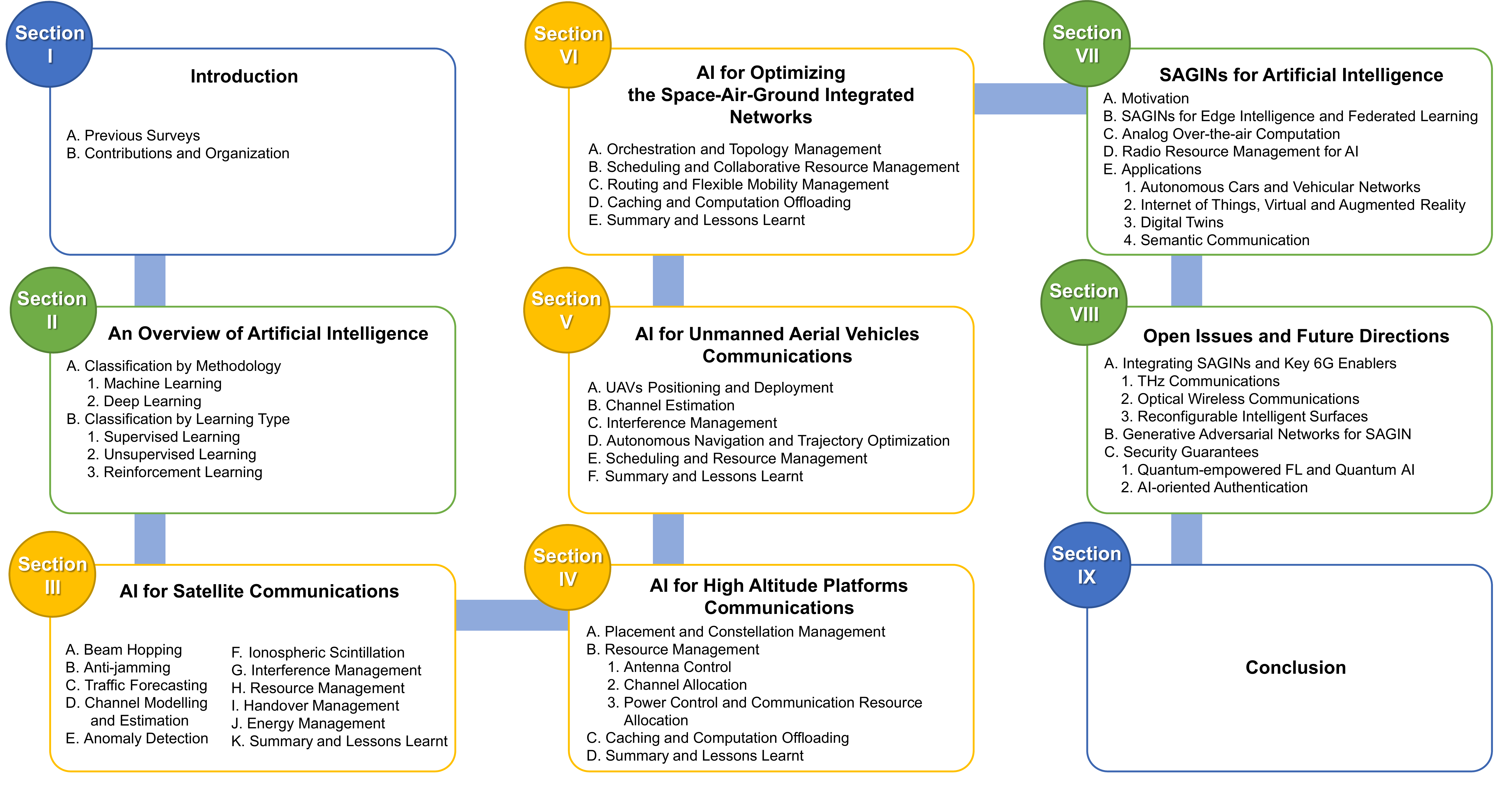}
    \caption{Survey organization and main sections.}
    \label{fig:taxonomy}
\end{figure*}
\begin{table}[htbp]
\caption{Acronyms used in the survey}
\centering
\begin{tabular}{|l|l|}
\hline
\textbf{Acronym}           & \textbf{Meaning}                                                                                               \\ \hline
\textbf{AI}       & Artificial Intelligence                                                                                        \\
\textbf{ANN}      & Artificial Neural Network                                           																					   \\
\textbf{AR}       & Augmented Reality                                                                                              \\
\textbf{BS}       & Bit Error Rate                                                                                                 \\
\textbf{BS}       & Base Station                                                                                                   \\
\textbf{CNN}      & Convolutional Neural Network                                                                                   \\
\textbf{CSI}      & Channel State Information                                                                                      \\
\textbf{DDPG}     & Deep Deterministic Policy Gradient                                                                             \\
\textbf{DL}       & Deep Learning                                                                                                  \\
\textbf{DNN}      & Deep Neural Network                                                                                               \\
\textbf{DRL}      & Deep Reinforcement Learning                                                                                    \\
\textbf{DQN}      & Deep Q-Network                                                                                               \\
\textbf{FL}       & Federated Learning                                                                                        \\
\textbf{GA}       & Genetic Algorithm                                                                                              \\
\textbf{GAN}      & Generative Adversarial Network                                                                              \\
\textbf{GEO}      & Geostationary Orbit                                                                                            \\
\textbf{GPS}      & Global Positioning System                                                                                           \\
\textbf{GRU}      & Gated Recurrent Unit                                                                                           \\
\textbf{HAPS}     & High Altitude Platforms                                                                                 \\
\textbf{IoT}      & Internet of Things                                                                                             \\
\textbf{KNN}      & k-Nearest Neighbors                                                                                            \\
\textbf{LEO}      & Low-Earth Orbit                                                                                                \\
\textbf{LoS}      & Line-of-Sight                                                                                                \\
\textbf{LSTM}     & Long Short-Term Memory                                                                                         \\
\textbf{MAE}     & Mean Absolute Error                                                                                \\
\textbf{MDP}      & Markov Decision Process                                                                                        \\
\textbf{MEC}      & Mobile Edge Computing                                                                                             \\
\textbf{MEO}      & Medium-Earth Orbit                                                                                             \\
\textbf{MIMO}       & Multiple-Input Multiple-Output                                                                                               \\
\textbf{ML}       & Machine Learning                                                                                               \\
\textbf{NN}       & Neural Network                                                                                               \\
\textbf{OWC}      & Optical Wireless Communication                                                                                 \\
\textbf{PPO}      & Proximal Policy Optimization                                                                                   \\
\textbf{PSO}      & Particle Swarm Optimization                                                                                   \\
\textbf{QoS}      & Quality of Service                                                                                             \\
\textbf{RIS}      & Reconfigurable Intelligent Surface                                                                                               \\
\textbf{RMSE}       & Root Mean Square Error                                                                                               \\
\textbf{RNN}      & Recurrent Neural Network                                                                                       \\
\textbf{RL}      & Reinforcement Learning                                                                                       \\
\textbf{RRM}      & Radio Resource Management                                                                                          \\
\textbf{SAC}      & Soft Actor-Critic                                                                                          \\
\textbf{SAGIN}    & Space-Air-Ground Integrated Networks	                                                                              \\
\textbf{SI}       & Swarm Intelligence                                                                                             \\
\textbf{SINR}     & Signal-to-Interference Noise Ratio                                                                                   \\
\textbf{SVM}      & Support Vector Machine                                                                                         \\
\textbf{UAV}      & Unmanned Aerial Vehicle                                                                                        \\
\textbf{VR}       & Virtual Reality                                                                                                \\ \hline
\end{tabular}
\label{table:notations}
\vspace{-0.5cm}
\end{table}

\subsection{Previous Surveys}
With the large number of published papers considering the integration of AI technologies in SAGINs, several surveys have tried to summarize the existing literature to provide a holistic overview of state-of-the-art research in AI-enabled SAGINs~\cite{fourati_artificial_2021,lahmeri_artificial_2021,michailidis_ai-inspired_2020,naous2023,Bai2023Towards,Kurunathan2023Machine,mahboob2023revolutionizing}. Furthermore, several survey papers discuss SAGINs and highlight their main potential applications,  challenges, and their role in 5G and 6G, envisioning diversified architecture where non-terrestrial stations seamlessly support the terrestrial infrastructure, paving the way for unprecedented connectivity~\cite{rinaldi_non-terrestrial_2020,vaezi_cellular_2022,xiao2022guest,zhu2022survey}. Moreover, several surveys covering 6G applications and challenges have promoted SAGINs and AI as key enabling technologies to maximize the coverage of terrestrial networks and introduce scalable and flexible management and control of the wireless network~\cite{jiang_road_2021}. For instance, the authors in~\cite{jiang_road_2021} provide a comprehensive survey about 6G and separately discuss SAGINs as an emerging architecture and AI as a tool for networking. 

The work done in~\cite{rinaldi_non-terrestrial_2020} outlines the advantages and challenges associated with SAGINs and mentions AI only as a tool that will be used for 6G technologies for real-time control and optimization. Another survey~\cite{vaezi_cellular_2022} offers a separate chapter on DL for SAGIN integration into IoT networks. Authors in~\cite{xiao2022guest} focus on the antenna design, challenges, and opportunities for the SAGINs. Work done in~\cite{zhu2022survey} presents a comprehensive overview of the challenges and opportunities the integrated satellite-terrestrial networks created in the 6G era. Finally, the work in~\cite{zhou2023} provides a comprehensive overview of the latest technical advances, challenges, and future directions in aerospace integrated networks (AINs) for empowering 6G, covering system architecture, enabling technologies, modeling, performance analysis, and optimization of AINs. 

In addition to the works mentioned above, several surveys are oriented toward the applications of AI in specific layers of the SAGIN architecture. For instance, the work in~\cite{fourati_artificial_2021} presents an extensive survey of the AI-based solutions proposed to address the challenges of satellite communications. Another work done in~\cite{mahboob2023revolutionizing} expands the discussion by addressing the AI-enabled satellite-based Non-Terrestrial Networks, its associated challenges, and future directions. Furthermore, the work in~\cite{lahmeri_artificial_2021} describes the main applications of AI in UAV networks and highlights the limitations of the existing works in this era. At the same time, another survey~\cite{Bai2023Towards} summarizes the main RL-driven applications for efficient multi-UAV wireless networks. The work done in~\cite{Kurunathan2023Machine} provides an overview of the transformative potantial of ML techniques in UAV operations and communications, highlighting crucial domains like UAV perception, feature collection and processing, trajectory planning, aerodynamic control, and operational management. On the other side, some existing surveys focus on specific AI algorithms and their applications in SAGINs. For instance, the authors in~\cite{naous2023} explore RL techniques to seamlessly integrate aerial platforms, including satellites, HAPS, and UAVs, with the existing terrestrial networks. Another paper~\cite{michailidis_ai-inspired_2020} sheds light on the potential role of AI in deploying and optimizing SAGIN-based industrial IoT solutions and provides a comprehensive review of the relevant research works.  

Table~\ref{table:surveys} summarizes previous surveys relevant to this paper's scope, highlights their main contributions, and compares them with our work. While many works have been published in the cross-section of the AI and SAGIN domains, there is no comprehensive survey discussing advances made by the research community. In this paper, we aim to fill this gap by providing a comprehensive survey of the state-of-the-art research in AI-enabled SAGINs, covering the main potential applications and challenges that AI can address and highlighting the research problems for seamless integration of AI in SAGINs.

\subsection{Contributions and Organization}
In this work, we aim to provide a holistic overview of state-of-the-art research on the role of AI in enhancing the performance of SAGINs. We specifically discuss the main challenges of satellites, HAPS, and UAV networks and highlight AI's main advantages to overcome these limitations. Furthermore, we provide a comprehensive overview of the work done on using AI to optimize integrated terrestrial and non-terrestrial network architectures. We finally highlight the main open issues and future trends related to the synergy between AI and SAGINs. To the best of the authors' knowledge, this is the first comprehensive survey that provides a systematic review of AI-enabled solutions for different components of SAGIN architecture and highlights the main AI-aided optimization techniques to fully realize the potential of AI in the context of SAGINs, thereby guiding future research efforts.

The survey is organized as shown in Fig.~\ref{fig:taxonomy} and includes the following sections:
\begin{itemize}
    \item In Section~\ref{sec:artificialintelligence}, we present a brief introduction to AI and its main algorithms for the reader's convenience.
    \item Section~\ref{sec:satellite}  reports the main works that aim at using AI in solving challenges present in satellite networks.
    \item In Section~\ref{sec:haps}, we highlight the main research efforts that consider AI for addressing challenges of efficient HAPS deployment and functionality.
    \item Section~\ref{sec:uavs} presents the main works that use AI to address challenges in UAV networks.
    \item Section~\ref{sec:optimizing} discusses how AI can optimize the integration of the space, aerial, and terrestrial subnetworks together toward the full operation of SAGINs.
    \item Section~\ref{sec:overlay} highlights how the recent advancements in SAGINs influence AI methods and applications.
    \item In Section~\ref{sec:openissues}, we outline the remaining issues and potential research directions to harness the full benefits of AI in efficiently implementing SAGINs. Finally, Section~\ref{sec:conclusion} concludes the paper.
\end{itemize}
The list of notations used throughout the paper is presented in Table~\ref{table:notations}.

\section{An Overview of Artificial Intelligence}
\label{sec:artificialintelligence}
AI has been extensively used to address various issues for many decades. At the same time, it has proved to be a highly complex and difficult-to-define subject. While the fundamentals of AI would likely be known to many readers, this section presents a comprehensive overview for completeness. Ever since John McCarthy et al.~\cite{mccarthy1955} coined the term "Artificial Intelligence" in 1955, researchers have actively explored and expanded this field. In a very broad sense, AI is a set of techniques and algorithms designed to teach a computer how to learn - imitating human abilities such as vision, speech, decision-making, problem-solving, analytical thinking, perception, and others through various AI methods to simulate intelligent behavior~\cite{russell2020artificial}.

AI algorithms can be classified into different categories based on their learning approach. One way to categorize AI algorithms is by their learning type, which includes supervised, unsupervised, and RL algorithms. Another way to classify AI algorithms is based on their methodology, providing for ML and DL algorithms. This section briefly discusses these different subsets of AI for the reader's convenience. Their key differences and examples of algorithms are listed in Table~\ref{table:algorithms} and Fig.~\ref{table:MLDL}, respectively.
\begin{table*}[t!]
\caption{Classification of Artificial Intelligence algorithms}
\resizebox{\textwidth}{!}{%
\begin{tabular}{llll}
\cline{2-4}
 &
  \multicolumn{1}{c}{\textbf{Supervised Learning}} &
  \multicolumn{1}{c}{\textbf{Unsupervised Learning}} &
  \multicolumn{1}{c}{\textbf{Reinforcement Learning}} \\ \hline
\textbf{Main problem types} &
  \begin{tabular}[c]{@{}l@{}}\tabitem Classification (object recognition, \\ sentiment analysis, anomaly detection) \\  \tabitem Regression (predictive analytics, modeling)\end{tabular} &
  \begin{tabular}[c]{@{}l@{}}\tabitem Clustering (customer segmentation,\\ genes clustering, document mining, \\ blind signal separation) \\ \tabitem Association (market basket analysis, \\ recommender systems, predictive \\ maintenance, generative modeling)\end{tabular} &
  \begin{tabular}[c]{@{}l@{}}\tabitem Exploitation Exploration (autonomous cars, \\ gaming, finance, robotics, national\\ language processing, business management, \\ adaptive control)\end{tabular} \\ \hline
\textbf{Key difference} &
  \begin{tabular}[c]{@{}l@{}}\tabitem Uses labeled data to learn an output \\ corresponding to certain input\end{tabular} &
  \begin{tabular}[c]{@{}l@{}} \tabitem Uses unlabeled data to learn underlying \\ patterns\end{tabular} &
  \begin{tabular}[c]{@{}l@{}}\tabitem Uses interactions with the environment and \\ corresponding rewards/punishments to learn \\ how to act\end{tabular} \\ \hline
\end{tabular}
}
\label{table:algorithms}
\end{table*}

\subsection{Classification by Methodology}
\subsubsection{Machine Learning}
\label{sec:ml}
ML is a branch of AI associated with developing and constructing analytical models that can automatically detect hidden and previously unknown patterns in data and independently acquire the properties necessary to recognize these patterns~\cite{bishop2006pattern}. The critical enabling factor of ML is the presence of a data set that describes the objects or processes under study and reflects their inherent properties and patterns. Such a dataset is called a training set, which can be obtained as a set of observations created by an expert or analyst based on hypotheses, analogies, personal experience, and possibly intuition.

Observations from the training set, called training examples, are sequentially presented to the trained model, and in the process, it acquires the necessary properties. This process is called model learning; it is an iterative procedure where a data object from the training sample is presented at each step, and the model parameters are adjusted by a rule called the learning algorithm. Training continues until the model can perform the required transformation with sufficient accuracy. Once the model is trained and tested, it can analyze the never-before-seen data. Hence, ML involves algorithms that learn patterns from data without being explicitly programmed with rules for each task. It gives computers new ways to accomplish tasks previously performed by people and teaches a computer system to make correct predictions when data is entered. It accelerates the development of AI's potential as its most crucial component.

\subsubsection{Deep Learning}
\label{sec:dl}
DL is a subset of AI that uses Artificial Neural Networks (ANNs) to learn specific tasks~\cite{goodfellow2016deep}. ANNs are inspired by biological NNs, which can be seen as a possible explanation for ANNs' flexibility and problem-solving capacity. NN is a complicated program comprising many internal (hidden) layers with configurable parameters - the weight coefficients of artificial neurons comprising each network layer. The network gets a vector of features that characterize the object - data in the form of signals - at the first input layer. They are processed on the inner layers: the input vector is multiplied by the matrix of connections, and the vector of new features created as a result is sent to the next layer. The outcome of signal processing is transferred to the network's output layer.

Each NN comprises many parameters that cannot be altered manually; therefore, the NN is trained automatically with the help of the data being fed to it. Throughout the training process, the weight coefficients of neurons continually change and are adjusted so that the result of calculations and signal processing becomes meaningful. The weight coefficients are updated using a basic optimization approach based on the gradient descent method, which allows modeling of how the NN's signal processing output changes with a slight change in each weight coefficient. The NN weights are modified at each layer, which is why it is termed deep. The more sophisticated the NN, the more layers and neurons it has, and the more computational operations it executes, the better (on average) the output it produces. However, it is more difficult to understand what is going on in its underlying layers.

\subsection{Classification by Learning Type}
\subsubsection{Supervised Learning}
\label{sec:supervised}
Supervised learning is a type of AI comprised of algorithms and methods that learn based on a set of examples containing known "input-output" pairs. A supervised learning algorithm must be fed with samples containing a vector of independent variables (most commonly called features) and a vector of learnable values (target output) that the model should try to produce during training. The difference between the target and actual outputs produced by the model is called a residual error. At the heart of each supervised learning algorithm is the residual error minimization process (also known as objective function optimization), which teaches the model with the help of right (small error) and wrong (significant error) examples. This residual error is used to calculate model parameter corrections at each training iteration. Classification and regression are two main types of problems solved with the help of supervised learning (see Table~\ref{table:algorithms}). The class label is used as a target variable in the former type, while in the latter, a numeric value is the desired output. Supervised learning has a wide range of applications in various fields. It is commonly used in text and image analysis, object recognition, predictive analysis, and modeling. The main supervised learning algorithms used in the literature are listed in Fig.~\ref{table:MLDL}. 

\tikzset{every picture/.style={line width=0.5pt}} 

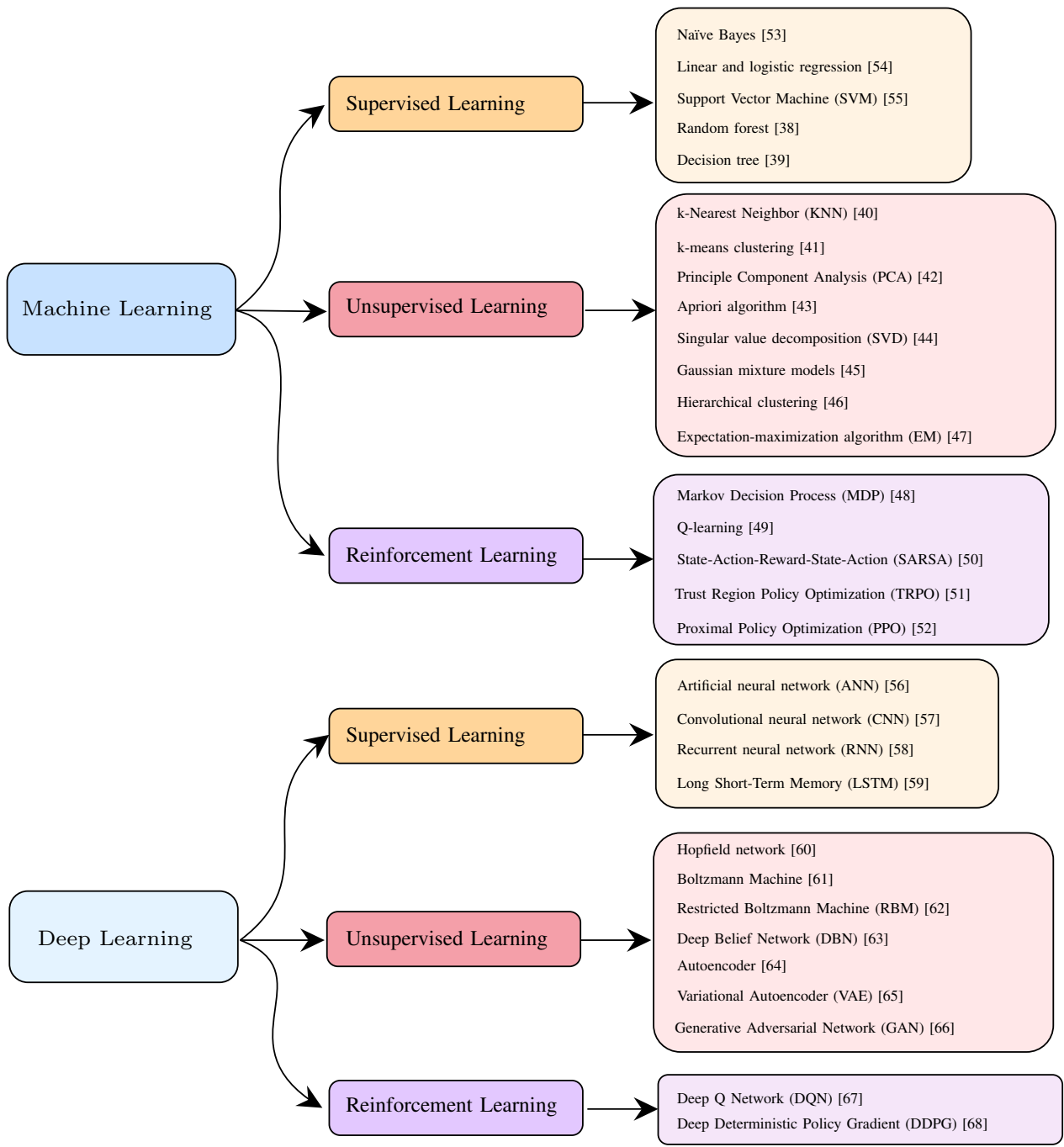
\begin {figure*}
\begin{adjustbox}{width=\textwidth}
\begin{tikzpicture}[x=0.75pt,y=0.75pt,yscale=-1,xscale=1]
\path (50,400); 

\draw  [fill={rgb, 255:red, 200; green, 226; blue, 255 }  ,fill opacity=1 ] (90,134) .. controls (90,129.58) and (93.58,126) .. (98,126) -- (182,126) .. controls (186.42,126) and (190,129.58) .. (190,134) -- (190,158) .. controls (190,162.42) and (186.42,166) .. (182,166) -- (98,166) .. controls (93.58,166) and (90,162.42) .. (90,158) -- cycle ;

\draw  [fill={rgb, 255:red, 228; green, 243; blue, 255 }  ,fill opacity=1 ] (91,409) .. controls (91,404.58) and (94.58,401) .. (99,401) -- (183,401) .. controls (187.42,401) and (191,404.58) .. (191,409) -- (191,433) .. controls (191,437.42) and (187.42,441) .. (183,441) -- (99,441) .. controls (94.58,441) and (91,437.42) .. (91,433) -- cycle ;

\draw  [fill={rgb, 255:red, 255; green, 213; blue, 154 }  ,fill opacity=1 ] (231,48.8) .. controls (231,46.15) and (233.15,44) .. (235.8,44) -- (337.2,44) .. controls (339.85,44) and (342,46.15) .. (342,48.8) -- (342,63.2) .. controls (342,65.85) and (339.85,68) .. (337.2,68) -- (235.8,68) .. controls (233.15,68) and (231,65.85) .. (231,63.2) -- cycle ;

\draw  [fill={rgb, 255:red, 227; green, 200; blue, 255 }  ,fill opacity=1 ] (231,246.8) .. controls (231,244.15) and (233.15,242) .. (235.8,242) -- (337.2,242) .. controls (339.85,242) and (342,244.15) .. (342,246.8) -- (342,261.2) .. controls (342,263.85) and (339.85,266) .. (337.2,266) -- (235.8,266) .. controls (233.15,266) and (231,263.85) .. (231,261.2) -- cycle ;

\draw  [fill={rgb, 255:red, 244; green, 159; blue, 169 }  ,fill opacity=1 ] (231,138.8) .. controls (231,136.15) and (233.15,134) .. (235.8,134) -- (337.2,134) .. controls (339.85,134) and (342,136.15) .. (342,138.8) -- (342,153.2) .. controls (342,155.85) and (339.85,158) .. (337.2,158) -- (235.8,158) .. controls (233.15,158) and (231,155.85) .. (231,153.2) -- cycle ;

\draw  [fill={rgb, 255:red, 255; green, 213; blue, 154 }  ,fill opacity=1 ] (231,325.8) .. controls (231,323.15) and (233.15,321) .. (235.8,321) -- (337.2,321) .. controls (339.85,321) and (342,323.15) .. (342,325.8) -- (342,340.2) .. controls (342,342.85) and (339.85,345) .. (337.2,345) -- (235.8,345) .. controls (233.15,345) and (231,342.85) .. (231,340.2) -- cycle ;

\draw  [fill={rgb, 255:red, 244; green, 159; blue, 169 }  ,fill opacity=1 ] (230,414.8) .. controls (230,412.15) and (232.15,410) .. (234.8,410) -- (336.2,410) .. controls (338.85,410) and (341,412.15) .. (341,414.8) -- (341,429.2) .. controls (341,431.85) and (338.85,434) .. (336.2,434) -- (234.8,434) .. controls (232.15,434) and (230,431.85) .. (230,429.2) -- cycle ;

\draw  [fill={rgb, 255:red, 227; green, 200; blue, 255 }  ,fill opacity=1 ] (231,488.8) .. controls (231,486.15) and (233.15,484) .. (235.8,484) -- (337.2,484) .. controls (339.85,484) and (342,486.15) .. (342,488.8) -- (342,503.2) .. controls (342,505.85) and (339.85,508) .. (337.2,508) -- (235.8,508) .. controls (233.15,508) and (231,505.85) .. (231,503.2) -- cycle ;

\draw    (190,146.5) .. controls (229.2,117.1) and (190.61,87.7) .. (226.7,58.3) ;
\draw [shift={(229,56.5)}, rotate = 143.13] [fill={rgb, 255:red, 0; green, 0; blue, 0 }  ][line width=0.08]  [draw opacity=0] (10.72,-5.15) -- (0,0) -- (10.72,5.15) -- (7.12,0) -- cycle    ;
\draw    (192,422.5) .. controls (230.42,437.28) and (182.48,468.06) .. (225.95,494.78) ;
\draw [shift={(228,496)}, rotate = 209.88] [fill={rgb, 255:red, 0; green, 0; blue, 0 }  ][line width=0.08]  [draw opacity=0] (10.72,-5.15) -- (0,0) -- (10.72,5.15) -- (7.12,0) -- cycle    ;
\draw    (192,422.5) .. controls (231.2,393.1) and (192.61,363.7) .. (228.7,334.3) ;
\draw [shift={(231,332.5)}, rotate = 143.13] [fill={rgb, 255:red, 0; green, 0; blue, 0 }  ][line width=0.08]  [draw opacity=0] (10.72,-5.15) -- (0,0) -- (10.72,5.15) -- (7.12,0) -- cycle    ;
\draw    (190,146.5) .. controls (233.34,156.84) and (185.48,225.88) .. (227.04,252.8) ;
\draw [shift={(229,254)}, rotate = 210.02] [fill={rgb, 255:red, 0; green, 0; blue, 0 }  ][line width=0.08]  [draw opacity=0] (10.72,-5.15) -- (0,0) -- (10.72,5.15) -- (7.12,0) -- cycle    ;
\draw    (190,146.5) -- (227,146.96) ;
\draw [shift={(230,147)}, rotate = 180.72] [fill={rgb, 255:red, 0; green, 0; blue, 0 }  ][line width=0.08]  [draw opacity=0] (10.72,-5.15) -- (0,0) -- (10.72,5.15) -- (7.12,0) -- cycle    ;
\draw    (192,422.5) -- (226,422.5) ;
\draw [shift={(229,422.5)}, rotate = 180] [fill={rgb, 255:red, 0; green, 0; blue, 0 }  ][line width=0.08]  [draw opacity=0] (10.72,-5.15) -- (0,0) -- (10.72,5.15) -- (7.12,0) -- cycle    ;
\draw  [fill={rgb, 255:red, 255; green, 243; blue, 227 }  ,fill opacity=1 ] (374,25) .. controls (374,18.92) and (378.92,14) .. (385,14) -- (501,14) .. controls (507.08,14) and (512,18.92) .. (512,25) -- (512,79.5) .. controls (512,85.58) and (507.08,90.5) .. (501,90.5) -- (385,90.5) .. controls (378.92,90.5) and (374,85.58) .. (374,79.5) -- cycle ;
\draw  [fill={rgb, 255:red, 255; green, 243; blue, 227 }  ,fill opacity=1 ] (374,308.85) .. controls (374,303.68) and (378.18,299.5) .. (383.35,299.5) -- (514.65,299.5) .. controls (519.82,299.5) and (524,303.68) .. (524,308.85) -- (524,355.15) .. controls (524,360.32) and (519.82,364.5) .. (514.65,364.5) -- (383.35,364.5) .. controls (378.18,364.5) and (374,360.32) .. (374,355.15) -- cycle ;
\draw  [fill={rgb, 255:red, 255; green, 231; blue, 232 }  ,fill opacity=1 ] (374,109.5) .. controls (374,101.77) and (380.27,95.5) .. (388,95.5) -- (535,95.5) .. controls (542.73,95.5) and (549,101.77) .. (549,109.5) -- (549,196.5) .. controls (549,204.23) and (542.73,210.5) .. (535,210.5) -- (388,210.5) .. controls (380.27,210.5) and (374,204.23) .. (374,196.5) -- cycle ;
\draw  [fill={rgb, 255:red, 255; green, 231; blue, 232 }  ,fill opacity=1 ] (373,387.19) .. controls (373,380.73) and (378.23,375.5) .. (384.69,375.5) -- (536.31,375.5) .. controls (542.77,375.5) and (548,380.73) .. (548,387.19) -- (548,459.81) .. controls (548,466.27) and (542.77,471.5) .. (536.31,471.5) -- (384.69,471.5) .. controls (378.23,471.5) and (373,466.27) .. (373,459.81) -- cycle ;
\draw  [fill={rgb, 255:red, 245; green, 230; blue, 249 }  ,fill opacity=1 ] (373,229.21) .. controls (373,223.3) and (377.8,218.5) .. (383.71,218.5) -- (535.29,218.5) .. controls (541.2,218.5) and (546,223.3) .. (546,229.21) -- (546,282.29) .. controls (546,288.2) and (541.2,293) .. (535.29,293) -- (383.71,293) .. controls (377.8,293) and (373,288.2) .. (373,282.29) -- cycle ;
\draw  [fill={rgb, 255:red, 245; green, 230; blue, 249 }  ,fill opacity=1 ] (375,485.89) .. controls (375,483.46) and (376.96,481.5) .. (379.39,481.5) -- (547.61,481.5) .. controls (550.04,481.5) and (552,483.46) .. (552,485.89) -- (552,507.61) .. controls (552,510.04) and (550.04,512) .. (547.61,512) -- (379.39,512) .. controls (376.96,512) and (375,510.04) .. (375,507.61) -- cycle ;
\draw    (342,55.5) -- (370,55.5) ;
\draw [shift={(373,55.5)}, rotate = 180] [fill={rgb, 255:red, 0; green, 0; blue, 0 }  ][line width=0.08]  [draw opacity=0] (10.72,-5.15) -- (0,0) -- (10.72,5.15) -- (7.12,0) -- cycle    ;
\draw    (343,146.5) -- (371,146.5) ;
\draw [shift={(374,146.5)}, rotate = 180] [fill={rgb, 255:red, 0; green, 0; blue, 0 }  ][line width=0.08]  [draw opacity=0] (10.72,-5.15) -- (0,0) -- (10.72,5.15) -- (7.12,0) -- cycle    ;
\draw    (342,255.5) -- (370,255.5) ;
\draw [shift={(373,255.5)}, rotate = 180] [fill={rgb, 255:red, 0; green, 0; blue, 0 }  ][line width=0.08]  [draw opacity=0] (10.72,-5.15) -- (0,0) -- (10.72,5.15) -- (7.12,0) -- cycle    ;
\draw    (342,332.5) -- (370,332.5) ;
\draw [shift={(373,332.5)}, rotate = 180] [fill={rgb, 255:red, 0; green, 0; blue, 0 }  ][line width=0.08]  [draw opacity=0] (10.72,-5.15) -- (0,0) -- (10.72,5.15) -- (7.12,0) -- cycle    ;
\draw    (341,422.5) -- (369,422.5) ;
\draw [shift={(372,422.5)}, rotate = 180] [fill={rgb, 255:red, 0; green, 0; blue, 0 }  ][line width=0.08]  [draw opacity=0] (10.72,-5.15) -- (0,0) -- (10.72,5.15) -- (7.12,0) -- cycle    ;
\draw    (344,496.5) -- (372,496.5) ;
\draw [shift={(375,496.5)}, rotate = 180] [fill={rgb, 255:red, 0; green, 0; blue, 0 }  ][line width=0.08]  [draw opacity=0] (10.72,-5.15) -- (0,0) -- (10.72,5.15) -- (7.12,0) -- cycle    ;

\draw (382,63) node [anchor=north west][inner sep=0.75pt]  [font=\tiny] [align=left] {Random forest \cite{breiman2001random}};
\draw (382,77) node [anchor=north west][inner sep=0.75pt]  [font=\tiny] [align=left] {Decision tree \cite{quinlan1986induction}};
\draw (382,100) node [anchor=north west][inner sep=0.75pt]  [font=\tiny] [align=left] {k-Nearest Neighbor (KNN) \cite{cover1967nearest}};
\draw (382,115) node [anchor=north west][inner sep=0.75pt]  [font=\tiny] [align=left] {k-means clustering \cite{hartigan1979algorithm}};
\draw (382,128) node [anchor=north west][inner sep=0.75pt]  [font=\tiny] [align=left] {Principle Component Analysis (PCA) \cite{wold1987principal}};
\draw (382,141) node [anchor=north west][inner sep=0.75pt]  [font=\tiny] [align=left] {Apriori algorithm \cite{hegland2007apriori}};
\draw (382,155) node [anchor=north west][inner sep=0.75pt]  [font=\tiny] [align=left] {Singular value decomposition (SVD) \cite{golub1971singular}};
\draw (382,169) node [anchor=north west][inner sep=0.75pt]  [font=\tiny] [align=left] {Gaussian mixture models \cite{reynolds2009gaussian}};
\draw (382,183) node [anchor=north west][inner sep=0.75pt]  [font=\tiny] [align=left] {Hierarchical clustering \cite{johnson1967hierarchical}};
\draw (382,198) node [anchor=north west][inner sep=0.75pt]  [font=\tiny] [align=left] {Expectation-maximization algorithm (EM) \cite{moon1996expectation}};
\draw (382,224) node [anchor=north west][inner sep=0.75pt]  [font=\tiny] [align=left] {Markov Decision Process (MDP) \cite{puterman1990markov}};
\draw (382,238) node [anchor=north west][inner sep=0.75pt]  [font=\tiny] [align=left] {Q-learning \cite{watkins1992q}};
\draw (382,252) node [anchor=north west][inner sep=0.75pt]  [font=\tiny] [align=left] {State-Action-Reward-State-Action (SARSA) \cite{sutton1995generalization}};
\draw (381,267) node [anchor=north west][inner sep=0.75pt]  [font=\tiny] [align=left] {Trust Region Policy Optimization (TRPO) \cite{schulman2015trust}};
\draw (382,282) node [anchor=north west][inner sep=0.75pt]  [font=\tiny] [align=left] {Proximal Policy Optimization (PPO) \cite{schulman2017proximal}};
\draw (382,22) node [anchor=north west][inner sep=0.75pt]  [font=\tiny] [align=left] {Na\"ive Bayes \cite{rish2001empirical}};
\draw (382,36) node [anchor=north west][inner sep=0.75pt]  [font=\tiny] [align=left] {Linear and logistic regression \cite{1023071268603}};
\draw (382,50) node [anchor=north west][inner sep=0.75pt]  [font=\tiny] [align=left] {Support Vector Machine (SVM) \cite{boser1992training}};
\draw (382,307) node [anchor=north west][inner sep=0.75pt]  [font=\tiny] [align=left] {Artificial neural network (ANN) \cite{jain1996artificial}};
\draw (382,322) node [anchor=north west][inner sep=0.75pt]  [font=\tiny] [align=left] {Convolutional neural network (CNN) \cite{krizhevsky2017imagenet}};
\draw (382,335) node [anchor=north west][inner sep=0.75pt]  [font=\tiny] [align=left] {Recurrent neural network (RNN) \cite{mikolov2010recurrent}};
\draw (382,350) node [anchor=north west][inner sep=0.75pt]  [font=\tiny] [align=left] {Long Short-Term Memory (LSTM) \cite{hochreiter1997long}};
\draw (382,379) node [anchor=north west][inner sep=0.75pt]  [font=\tiny] [align=left] {Hopfield network \cite{paik1992image} };
\draw (382,392) node [anchor=north west][inner sep=0.75pt]  [font=\tiny] [align=left] {Boltzmann Machine \cite{salakhutdinov2010efficient}};
\draw (382,405) node [anchor=north west][inner sep=0.75pt]  [font=\tiny] [align=left] {Restricted Boltzmann Machine (RBM) \cite{fischer2012introduction}};
\draw (382,418) node [anchor=north west][inner sep=0.75pt]  [font=\tiny] [align=left] {Deep Belief Network (DBN) \cite{hinton2009deep}};
\draw (382,430) node [anchor=north west][inner sep=0.75pt]  [font=\tiny] [align=left] {Autoencoder \cite{bank2020autoencoders}};
\draw (382,443) node [anchor=north west][inner sep=0.75pt]  [font=\tiny] [align=left] {Variational Autoencoder (VAE) \cite{doersch2016tutorial} };
\draw (381,457) node [anchor=north west][inner sep=0.75pt]  [font=\tiny] [align=left] {Generative Adversarial Network (GAN) \cite{goodfellow2020generative}};
\draw (382,488) node [anchor=north west][inner sep=0.75pt]  [font=\tiny] [align=left] {Deep Q Network (DQN) \cite{mnih2013playing}};
\draw (382,499) node [anchor=north west][inner sep=0.75pt]  [font=\tiny] [align=left] {Deep Deterministic Policy Gradient (DDPG) \cite{silver2014deterministic}};
\draw (237,140) node [anchor=north west][inner sep=0.75pt]  [font=\small] [align=left] {{\scriptsize Unsupervised Learning}};
\draw (237,51) node [anchor=north west][inner sep=0.75pt]  [font=\small] [align=left] {{\scriptsize Supervised Learning}};
\draw (237,249) node [anchor=north west][inner sep=0.75pt]  [font=\small] [align=left] {{\scriptsize Reinforcement Learning}};
\draw (237,328) node [anchor=north west][inner sep=0.75pt]  [font=\small] [align=left] {{\scriptsize Supervised Learning}};
\draw (237,417) node [anchor=north west][inner sep=0.75pt]  [font=\small] [align=left] {{\scriptsize Unsupervised Learning}};
\draw (237,490) node [anchor=north west][inner sep=0.75pt]  [font=\small] [align=left] {{\scriptsize Reinforcement Learning}};
\draw (95,141) node [anchor=north west][inner sep=0.75pt]  [font=\small] [align=left] {{\fontfamily{helvet}\selectfont {\scriptsize Machine Learning}}};
\draw (102,417) node [anchor=north west][inner sep=0.75pt]  [font=\small] [align=left] {{\fontfamily{helvet}\selectfont {\scriptsize Deep Learning}}};

\end{tikzpicture}

\end{adjustbox}
\caption{Different AI techniques and algorithms surveyed in this paper and their respective classification.}
\label{table:MLDL}
\vspace{-0.1cm}
\end{figure*}

\subsubsection{Unsupervised Learning}
\label{sec:unsupervised}
Unsupervised learning is a subset of AI that does not use an objective function to correct the parameters of a trained model. In other words, in unsupervised training models, it is not necessary to have predefined model outputs, and so the output error of the model on the training set is not calculated. Instead, information about the current state of the model parameters and examples of the training set are used. For instance, the Euclidean distance between the example's feature vector and the neuron's weight vector can control the correction of model parameters during training. The main application of unsupervised learning is building models for clustering and association. Since the cluster data structure and the underlying relation between data points are not known in advance but determined during the model's training, it is impossible to use any target values. The main unsupervised learning algorithms used in the literature are listed in Fig.~\ref{table:MLDL}. 

\subsubsection{Reinforcement Learning}
\label{sec:rl}
RL is a branch of AI that studies the behavior of intelligent agents operating in a particular environment and making decisions~\cite{sutton2018reinforcement}. The environment's response to the decisions made is reinforcement signals, based on which the agent is trained. Therefore, such learning is a particular case of supervised learning, where the teacher is the environment. A reinforcement system is any set of rules that can change the state of a model over time. In this system, the agents aim to maximize their performance in a particular environment. A reward the agent receives signifies that the action contributed to achieving the goal sooner, while a penalty is given when the agent fails to produce the right move. The problem narrows down to choosing a sequence of steps (or actions) necessary to maximize the reward. Each decision brings new information to the agent, leading to the development of new strategies and solutions, which in turn lead to greater rewards. Some techniques focus on maximizing immediate compensation (called "exploitation"), while others tend to prioritize finding as much information about the environment as possible (called "exploration"). The majority of RL algorithms balance these two paradigms. Similarly to supervised and unsupervised learning, RL is an iterative process. The strategy improves as training progresses, and the agent receives more feedback. The exact way this happens solely depends on the RL algorithm. Furthermore, a list of the main RL algorithms is presented in Fig.~\ref{table:MLDL}. The key differences between supervised, unsupervised, and RL algorithms are presented in Table~\ref{table:algorithms}.

\begin{figure*}[t!]
    \centering
    \includegraphics[width=0.9\linewidth]{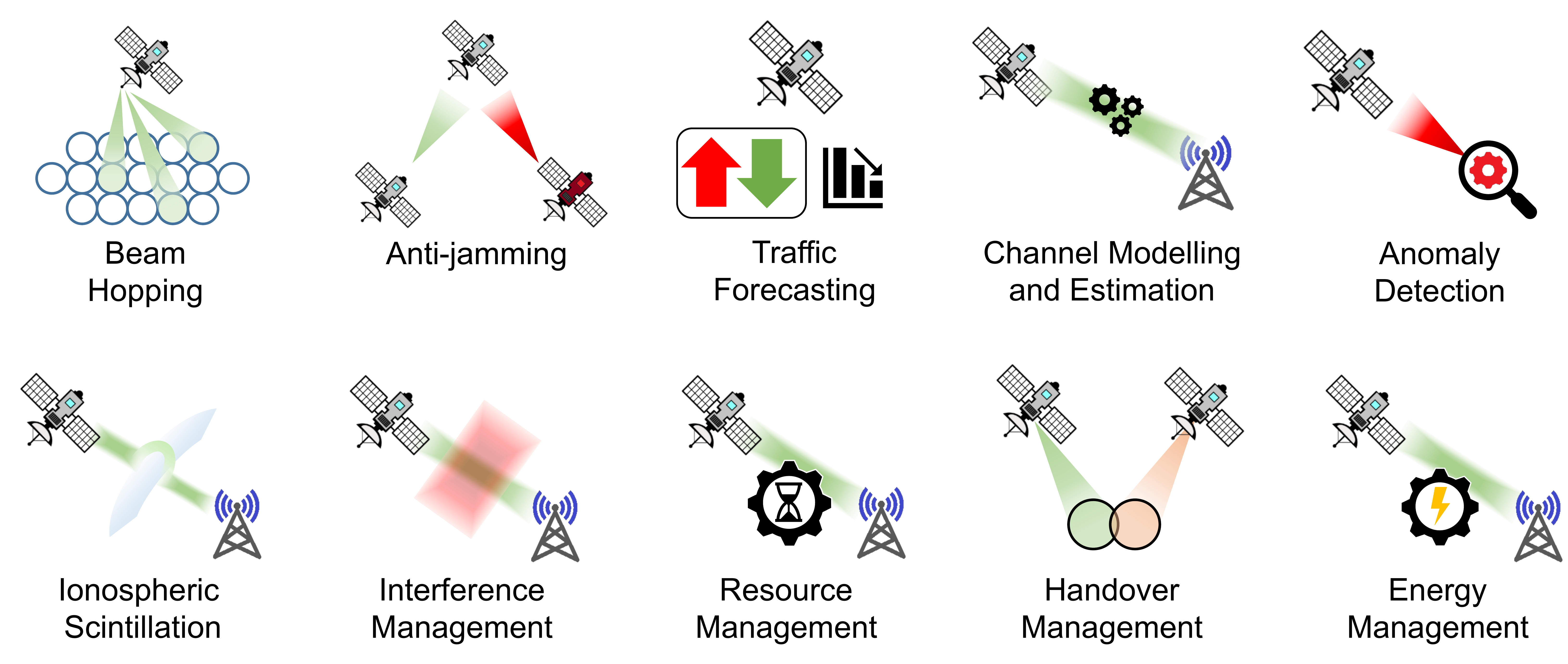}
    \caption{Challenges in satellite communications addressed by AI and discussed in this survey.}
    \label{fig:sat}
    \vspace{-0.2cm}
\end{figure*}
\section{AI for Satellite Communications}
\label{sec:satellite}
Satellite communications provide many advantages to enhance the user experience, such as extended coverage, good transmission quality, high communication capacity, convenient and rapid networking, and seamless global communications. On the other hand, the financial implications of deploying and maintaining satellite infrastructure require developing robust resource management and time-sharing systems. Several papers in the literature aimed at proposing solutions for the limitations of satellite networks~\cite{Wang2023Joint}. Recently, AI tools have been considered to deal with the massive data satellite networks generate and to provide efficient solutions to enable fast integration of satellites in the SAGIN architecture. In this section, we aim to survey the primary papers that use tools from AI to address challenges and optimize satellite communications. Namely, we focus on the AI algorithms that tackle challenges such as beam hopping, anti-jamming, traffic forecasting, channel modeling, anomaly detection, interference management, ionospheric scintillation, energy management, and resource management. Table~\ref{tab:sat1} and Table~\ref{tab:sat2} summarize most works in the literature that use AI to address the challenges of satellite communications. Fig.~\ref{fig:sat} visualizes the main challenges of satellite communications considered in this survey.

\subsection{Beam Hopping}\label{sec:sat_beam}
Beam hopping emerged as a promising solution to manage diverse traffic demands by enabling satellites to reallocate capacity between beams in response to ground requests~\cite{kodheli2021satellite}. This is accomplished using time-division multiplexing with a single frequency instead of the earlier method of isolating spot beams by "color (frequency divided by polarization). Beam hopping is used to employ flexibility in dealing with unpredictable and time-varying traffic requirements in the satellite coverage region. It mainly uses a small number of active beams simultaneously to dynamically illuminate a specific cell based on the traffic demand. 

The main challenge of beam hopping is the choice of the illumination pattern, that is, which and for how long each beam must be selected~\cite{Han2023Beam}. Furthermore, the need for high-precision position control of the laser beam in the pointing, acquisition, and tracking system is yet another challenge for efficient beam hopping systems~\cite{sun2010}. Another critical factor is the optimization of the system parameters, such as the number of serving beams, transmit power, beam directions, and sizes~\cite{al2020traffic}. Several works in the literature attempt to propose solutions for the challenges of beam hopping, and most methods have relied on classical optimization algorithms. For instance, the authors in~\cite{Lyu2023Beam} aim at minimizing the number of satellite beam positions subject to a predefined requirement on the radius of a beam. In~\cite{sharma2012}, the authors discuss cognitive techniques for enhancing spectral efficiency in satellite communications. In~\cite{sun2010}, the authors research adaptive control algorithms for jitter in laser beam pointing and tracking systems. 

While optimizing algorithms yields favorable outcomes in enhancing flexibility and reducing delays in beam hopping systems~\cite{Li2023Pattern}, certain challenges persist. As the number of beams increases, there is a significant expansion in the search space, making it inherently challenging to devise the optimal parameters amid numerous local optima. This issue becomes particularly pronounced in the case of complex satellite systems employing extensive beamforming capabilities, with hundreds or even thousands of individual beams, as computational demands of conventional optimization algorithms can quickly escalate, exceeding acceptable time constraints and hindering real-world implementation. Furthermore, classical optimization algorithms necessitate adjustments when there are moderate changes in the scenario, which, in turn, increases the computational complexity. This complexity becomes impractical, especially when managing resources onboard, where adaptability is crucial.

To overcome the limitations of traditional optimization techniques, several recent works leverage tools from AI and ML to enhance the performance of beam hopping schemes. For instance, the authors in~\cite{lei2020deep} apply a DL technique to optimize the beam hopping illumination pattern. A Fully connected neural network (FC-NN)-based method is proposed to create a synergy of optimization and learning. The significant advantage of the technique is in the computation time, which is substantially low in contrast to other iterative methods. Furthermore, the authors use the cardinality of beam hopping patterns as a feature to achieve improved performance and accuracy. A similar algorithm for finding the efficient beam hopping illumination pattern design is presented in~\cite{lei2020beam}. Having identified the proper features necessary for good predictions, namely asymmetric traffic demands, the authors propose a learning-and-optimization algorithm combining DL and optimization benefits. Experimental validation shows that FC-NN can efficiently limit the search space of optimization problems in case studies involving multi-beam-defined GEO satellites. 

RL is yet another AI tool frequently used to improve the throughput of beam hopping systems in satellite networks. In~\cite{han2021dynamic}, researchers focus on multi-beam satellite systems and aim to establish the traffic model of forward links. Thus, a deep reinforcement learning (DRL)-based algorithm is proposed to optimize the allocation of resources for beam hopping. The results show that the proposed method can reduce the transmission delay and increase the throughput. Similar results can be achieved by predicting long-term utilization rates. For instance, the authors in~\cite{hu2019deep} propose a DRL-based approach that considers traffic demands in spatial and temporal domains, antenna radiation patterns, and interference. Demonstrating the efficacy of the proposed DRL-based method, the study reveals notable reductions in transmission delay (up to $52.2\%$) and increases in system throughput (up to $11.4\%$) compared to existing methodologies. 

An overview of different beam hopping algorithms in LEO satellite constellation networks is presented in~\cite{li2021overview}. Simulation results compare the presented algorithms in terms of performance and complexity metrics, concluding that DRL-based algorithms are the fastest to deliver results comparable with hybrid simulated annealing/particle swarm optimization (PSO) and genetic algorithm (GA)-based methods. The only problem identified is the poor robustness of the DRL models, which can be solved by pairing them with GAs. In this case, the GA, acting as an evolution algorithm, can optimize the hyperparameters and architecture of the DRL models, thereby enhancing their overall performance. Another study involving GAs is presented in~\cite{zhang2019dynamic}, in which researchers show that a DRL algorithm can solve the problem of the "curse of dimensionality" in satellite communications. The authors note, however, that while the multi-objective DRL method can increase the throughput, maintain the fairness of an individual cell, and decrease the delay, the GA can achieve comparable results but with $110$~times less complexity.

Another DRL-based approach called double-loop learning is adopted in~\cite{hu2020dynamic} to ensure both delay reduction and throughput maximization while maintaining the fairness of beam hopping selection for each cell. The simulations show that the suggested technique may pursue many objectives at the same time and intelligently allocate resources based on user requirements and channel characteristics. The work in~\cite{lin2022dynamic} focuses on achieving real-time beam pattern illumination and bandwidth allocation to satisfy non-uniform and time-varying traffic requests. In this case, deep Q-learning is adopted to produce high generalization as traffic demand grows.
\begin{table*}[]
\caption{Summary of AI-Aided Satellite Communications solutions}
\resizebox{\textwidth}{!}{%
\begin{tabular}{|llllll|}
\hline
\multicolumn{1}{|l|}{\textbf{Publication}} & \multicolumn{1}{l|}{\textbf{Year}} & \multicolumn{1}{l|}{\textbf{Objective}} & \multicolumn{1}{l|}{\textbf{AI type}} & \multicolumn{1}{l|}{\textbf{AI algorithm}} & \textbf{Performance metrics} \\ \hline
\multicolumn{6}{|c|}{\textbf{Beam Hopping}} \\ \hline
\multicolumn{1}{|l|}{\cite{zhang2019dynamic}} & \multicolumn{1}{l|}{2019} & \multicolumn{1}{l|}{\begin{tabular}[c]{@{}l@{}}Multi-objective optimization \\ for multi-beam dynamic \\ beam hopping\end{tabular}} & \multicolumn{1}{l|}{Reinforcement Learning} & \multicolumn{1}{l|}{Deep Q-Network} & Throughput, delay \\ \hline
\multicolumn{1}{|l|}{\cite{hu2019deep}} & \multicolumn{1}{l|}{2019} & \multicolumn{1}{l|}{\begin{tabular}[c]{@{}l@{}}Optimization policy for beam\\  hopping illumination plan\end{tabular}} & \multicolumn{1}{l|}{Reinforcement Learning} & \multicolumn{1}{l|}{Deep Q-Network} & Delay reduction, system throughput \\ \hline
\multicolumn{1}{|l|}{\cite{lei2020deep}} & \multicolumn{1}{l|}{2020} & \multicolumn{1}{l|}{\begin{tabular}[c]{@{}l@{}}Optimize the beam-hopping \\ illumination pattern\end{tabular}} & \multicolumn{1}{l|}{Supervised Learning} & \multicolumn{1}{l|}{Deep Neural Network} & \begin{tabular}[c]{@{}l@{}}Offered capacity to requested \\ demand ratio (OCDR)\end{tabular} \\ \hline
\multicolumn{1}{|l|}{\cite{lei2020beam}} & \multicolumn{1}{l|}{2020} & \multicolumn{1}{l|}{\begin{tabular}[c]{@{}l@{}}Learning-and-optimization \\ algorithm for efficient beam \\ hopping illumination pattern design\end{tabular}} & \multicolumn{1}{l|}{Supervised Learning} & \multicolumn{1}{l|}{Deep Neural Network} & OCDR \\ \hline
\multicolumn{1}{|l|}{\cite{hu2020dynamic}} & \multicolumn{1}{l|}{2020} & \multicolumn{1}{l|}{Fairness-centric beam hopping strategy} & \multicolumn{1}{l|}{Reinforcement Learning} & \multicolumn{1}{l|}{Deep Q-Network} & \begin{tabular}[c]{@{}l@{}}Delay reduction, system throughput, \\ fairness of each cell\end{tabular} \\ \hline
\multicolumn{1}{|l|}{\cite{li2021overview}} & \multicolumn{1}{l|}{2021} & \multicolumn{1}{l|}{Optimal beam-hopping algorithm} & \multicolumn{1}{l|}{Reinforcement Learning} & \multicolumn{1}{l|}{Deep Q-Network} & Satisfied user rate \\ \hline
\multicolumn{1}{|l|}{\cite{han2021dynamic}} & \multicolumn{1}{l|}{2021} & \multicolumn{1}{l|}{\begin{tabular}[c]{@{}l@{}}Traffic model to optimize the allocation\\  of resources for beam hopping\end{tabular}} & \multicolumn{1}{l|}{Reinforcement Learning} & \multicolumn{1}{l|}{Deep Q-Network} & Throughput, delay \\ \hline
\multicolumn{1}{|l|}{\cite{lin2022dynamic}} & \multicolumn{1}{l|}{2022} & \multicolumn{1}{l|}{\begin{tabular}[c]{@{}l@{}}Real-time beam pattern and \\ bandwidth allocation\end{tabular}} & \multicolumn{1}{l|}{Reinforcement Learning} & \multicolumn{1}{l|}{Deep Q-learning} & Long-term throughput and delay fairness \\ \hline
\multicolumn{6}{|c|}{\textbf{Anti-jamming}} \\ \hline
\multicolumn{1}{|l|}{\cite{lee2019synchronization}} & \multicolumn{1}{l|}{2019} & \multicolumn{1}{l|}{\begin{tabular}[c]{@{}l@{}}Synchronization technique\\  for anti-jamming purposes\end{tabular}} & \multicolumn{1}{l|}{Supervised Learning} & \multicolumn{1}{l|}{Long Short-Term Memory} & MSE of uplink hop timing estimation \\ \hline
\multicolumn{1}{|l|}{\cite{li2019performance}} & \multicolumn{1}{l|}{2019} & \multicolumn{1}{l|}{\begin{tabular}[c]{@{}l@{}}Jamming method to study the\\  response of the anti-jamming techniques\end{tabular}} & \multicolumn{1}{l|}{Reinforcement Learning} & \multicolumn{1}{l|}{Q-learning} & Normalized throughput \\ \hline
\multicolumn{1}{|l|}{\cite{liu2019pattern}} & \multicolumn{1}{l|}{2019} & \multicolumn{1}{l|}{\begin{tabular}[c]{@{}l@{}}Jamming identification and \\ channel selection\end{tabular}} & \multicolumn{1}{l|}{\begin{tabular}[c]{@{}l@{}}Reinforcement Learning \\ and Supervised Learning\end{tabular}} & \multicolumn{1}{l|}{\begin{tabular}[c]{@{}l@{}}Q-learning and Convolutional\\  Neural Network\end{tabular}} & Cumulative throughput \\ \hline
\multicolumn{1}{|l|}{\cite{han2020spatial}} & \multicolumn{1}{l|}{2020} & \multicolumn{1}{l|}{Spatial anti-jamming technique} & \multicolumn{1}{l|}{Reinforcement Learning} & \multicolumn{1}{l|}{Deep Q-Network} & Routing cost \\ \hline
\multicolumn{1}{|l|}{\cite{han2020dynamic}} & \multicolumn{1}{l|}{2020} & \multicolumn{1}{l|}{\begin{tabular}[c]{@{}l@{}}Reducing energy consumption\\  in jamming scenarios\end{tabular}} & \multicolumn{1}{l|}{Reinforcement Learning} & \multicolumn{1}{l|}{Q-learning} & Energy efficiency \\ \hline
\multicolumn{1}{|l|}{\cite{wei2021optimal}} & \multicolumn{1}{l|}{2021} & \multicolumn{1}{l|}{\begin{tabular}[c]{@{}l@{}}Defend against sweep jamming\\  attacks\end{tabular}} & \multicolumn{1}{l|}{Reinforcement Learning} & \multicolumn{1}{l|}{\begin{tabular}[c]{@{}l@{}}Multi-step prediction \\ Bellman iterative equation\end{tabular}} & Utility of the communication system \\ \hline
\multicolumn{1}{|l|}{\cite{li2021satellite}} & \multicolumn{1}{l|}{2021} & \multicolumn{1}{l|}{Anti-jamming by blind separation} & \multicolumn{1}{l|}{Unsupervised learning} & \multicolumn{1}{l|}{Artificial Bee Colony} & Bit error rate \\ \hline
\multicolumn{1}{|l|}{\cite{yan2023cross}} & \multicolumn{1}{l|}{2023} & \multicolumn{1}{l|}{\begin{tabular}[c]{@{}l@{}}Channel and path selection for \\ in jamming scenarios\end{tabular}} & \multicolumn{1}{l|}{Reinforcement Learning} & \multicolumn{1}{l|}{Q-learning} & Payoff, time delay, convergence \\ \hline
\multicolumn{6}{|c|}{\textbf{Traffic Forecasting}} \\ \hline
\multicolumn{1}{|l|}{\cite{ziluan2018short}} & \multicolumn{1}{l|}{2018} & \multicolumn{1}{l|}{Short-term traffic forecasting} & \multicolumn{1}{l|}{Supervised Learning} & \multicolumn{1}{l|}{Deep Neural Network} & \begin{tabular}[c]{@{}l@{}}Forecasting accuracy, training time, \\ and robustness\end{tabular} \\ \hline
\multicolumn{1}{|l|}{\cite{na2018distributed}} & \multicolumn{1}{l|}{2018} & \multicolumn{1}{l|}{\begin{tabular}[c]{@{}l@{}}Traffic prediction for routing \\ decisions\end{tabular}} & \multicolumn{1}{l|}{Supervised Learning} & \multicolumn{1}{l|}{Extreme Learning Machine} & \begin{tabular}[c]{@{}l@{}}Link utilization, delay, \\ packet loss rate accuracy\end{tabular} \\ \hline
\multicolumn{1}{|l|}{\cite{bie2019combined}} & \multicolumn{1}{l|}{2019} & \multicolumn{1}{l|}{Traffic forecasting} & \multicolumn{1}{l|}{Supervised Learning} & \multicolumn{1}{l|}{Extreme Learning Machine} & \begin{tabular}[c]{@{}l@{}}Forecasting accuracy and speed, \\ complexity\end{tabular} \\ \hline
\multicolumn{1}{|l|}{\cite{jones2020short}} & \multicolumn{1}{l|}{2020} & \multicolumn{1}{l|}{Short-term traffic forecasting} & \multicolumn{1}{l|}{Supervised Learning} & \multicolumn{1}{l|}{XG Boost, Recurrent Neural Network} & \begin{tabular}[c]{@{}l@{}}Unmet terminal demand and total\\  power consumption\end{tabular} \\ \hline
\multicolumn{1}{|l|}{\cite{yang2020noval}} & \multicolumn{1}{l|}{2020} & \multicolumn{1}{l|}{\begin{tabular}[c]{@{}l@{}}Predicting spatial and \\ temporal traffic features\end{tabular}} & \multicolumn{1}{l|}{Supervised Learning} & \multicolumn{1}{l|}{\begin{tabular}[c]{@{}l@{}}Graph  Convolutional Neural Network\\  and Gated Recurrent Unit\end{tabular}}

& Traffic RMSE and accuracy \\ \hline
\multicolumn{1}{|l|}{\cite{rajagopal2021optimal}} & \multicolumn{1}{l|}{2021} & \multicolumn{1}{l|}{\begin{tabular}[c]{@{}l@{}}Distributed routing model \\ for traffic forecasting\end{tabular}} & \multicolumn{1}{l|}{Supervised Learning} & \multicolumn{1}{l|}{\begin{tabular}[c]{@{}l@{}}Extreme Learning Machine (ELM) and\\  multitask beetle antenna search (MBAS)\end{tabular}} & \begin{tabular}[c]{@{}l@{}}Average delay, packet loss ratio (PLR),\\  and queuing delay\end{tabular} \\ \hline
\multicolumn{1}{|l|}{\cite{huang2021ipr}} & \multicolumn{1}{l|}{2021} & \multicolumn{1}{l|}{Packet routing framework} & \multicolumn{1}{l|}{Reinforcement Learning} & \multicolumn{1}{l|}{Long Short-Term Memory and Deep Q-Networks} & Average delivery time, average hops \\ \hline
\multicolumn{1}{|l|}{\cite{li2021research}} & \multicolumn{1}{l|}{2021} & \multicolumn{1}{l|}{\begin{tabular}[c]{@{}l@{}}Traffic forecasting based \\ on transfer learning\end{tabular}} & \multicolumn{1}{l|}{Supervised Learning} & \multicolumn{1}{l|}{Gated Recurrent Unit Neural Network} & Traffic RMSE \\ \hline
\multicolumn{6}{|c|}{\textbf{Channel Modeling and Estimation}} \\ \hline
\multicolumn{1}{|l|}{\cite{ates2019path}} & \multicolumn{1}{l|}{2019} & \multicolumn{1}{l|}{Channel parameters estimation} & \multicolumn{1}{l|}{Supervised Learning} & \multicolumn{1}{l|}{Convolutional Neural Network} & \begin{tabular}[c]{@{}l@{}}Accuracy of predicted path loss exponent \\ and standard deviation of shadowing\end{tabular} \\ \hline
\multicolumn{1}{|l|}{\cite{ahmadien2020predicting}} & \multicolumn{1}{l|}{2020} & \multicolumn{1}{l|}{Path loss distribution prediction} & \multicolumn{1}{l|}{Supervised Learning} & \multicolumn{1}{l|}{Convolutional Neural Network} & MSE of predicted path loss \\ \hline
\multicolumn{1}{|l|}{\cite{zhang2021}} & \multicolumn{1}{l|}{2021} & \multicolumn{1}{l|}{CSI prediction} & \multicolumn{1}{l|}{Supervised Learning} & \multicolumn{1}{l|}{Long Short-Term Memory} & Normalized MSE of predicted CSI \\ \hline
\multicolumn{1}{|l|}{\cite{zhang2022sat}} & \multicolumn{1}{l|}{2022} & \multicolumn{1}{l|}{Downlink CSI estimation} & \multicolumn{1}{l|}{Supervised Learning} & \multicolumn{1}{l|}{Deep Neural Network} & Normalized MSE of predicted CSI \\ \hline
\multicolumn{6}{|c|}{\textbf{Anomaly Detection}} \\ \hline
\multicolumn{1}{|l|}{\cite{ibrahim2018machine}} & \multicolumn{1}{l|}{2018} & \multicolumn{1}{l|}{\begin{tabular}[c]{@{}l@{}}Telemetry mining for LEO \\ satellite data\end{tabular}} & \multicolumn{1}{l|}{Supervised Learning} & \multicolumn{1}{l|}{\begin{tabular}[c]{@{}l@{}}Auto-Regressive Integrated Moving \\ Average, Multilayer Perceptron,\\  Recurrent Neural Network, Deep Long\\  Short-Term Memory Recurrent \\ Neural Network, Deep Gated \\ Recurrent Unit Recurrent \\ Neural Network\end{tabular}} & \begin{tabular}[c]{@{}l@{}}RMSE, MAE of predicted load \\ current values and voltage on\\  power bus values\end{tabular} \\ \hline
\multicolumn{1}{|l|}{\cite{wan2019study}} & \multicolumn{1}{l|}{2019} & \multicolumn{1}{l|}{Telemetry data compression} & \multicolumn{1}{l|}{Supervised Learning} & \multicolumn{1}{l|}{\begin{tabular}[c]{@{}l@{}}Classification Probability calculation - \\ Window Step optimization (CP-WS)\end{tabular}} & Compression ratio \\ \hline
\multicolumn{1}{|l|}{\cite{ramadan2020tensor}} & \multicolumn{1}{l|}{2020} & \multicolumn{1}{l|}{Tensor-based anomaly detection} & \multicolumn{1}{l|}{Supervised Learning} & \multicolumn{1}{l|}{\begin{tabular}[c]{@{}l@{}}Support Vector Machine, Deep Neural\\  Network\end{tabular}} & Detected anomalies \\ \hline
\multicolumn{1}{|l|}{\cite{peng2020machine}} & \multicolumn{1}{l|}{2020} & \multicolumn{1}{l|}{Orbit prediction} & \multicolumn{1}{l|}{Supervised Learning} & \multicolumn{1}{l|}{Support Vector Machine} & Error of predicted orbit positions \\ \hline
\multicolumn{1}{|l|}{\cite{shehab2020recurrent}} & \multicolumn{1}{l|}{2020} & \multicolumn{1}{l|}{Telemetry data compression} & \multicolumn{1}{l|}{Supervised Learning} & \multicolumn{1}{l|}{\begin{tabular}[c]{@{}l@{}}Long Short-Term Memory Recurrent\\  Neural Network\end{tabular}} & Compression ratio \\ \hline
\multicolumn{1}{|l|}{\cite{mahmoud2021different}} & \multicolumn{1}{l|}{2021} & \multicolumn{1}{l|}{Telemetry data compression} & \multicolumn{1}{l|}{Supervised Learning} & \multicolumn{1}{l|}{Long Short-Term Memory} & \begin{tabular}[c]{@{}l@{}}Prediction gain and entropy, \\ compression ratio\end{tabular} \\ \hline
\multicolumn{6}{|c|}{\textbf{Ionospheric Scintillation}} \\ \hline
\multicolumn{1}{|l|}{\cite{rezende2010survey}} & \multicolumn{1}{l|}{2010} & \multicolumn{1}{l|}{Ionospheric scintillation prediction} & \multicolumn{1}{l|}{Supervised Learning} & \multicolumn{1}{l|}{Decision tree} & MSE of predicted scintillation data \\ \hline
\multicolumn{1}{|l|}{\cite{jiao2017performance}} & \multicolumn{1}{l|}{2017} & \multicolumn{1}{l|}{\begin{tabular}[c]{@{}l@{}}Ionospheric phase and amplitude \\ scintillation detection\end{tabular}} & \multicolumn{1}{l|}{Supervised Learning} & \multicolumn{1}{l|}{Support Vector Machine} & Prediction accuracy \\ \hline
\multicolumn{1}{|l|}{\cite{jiao2017automatic}} & \multicolumn{1}{l|}{2017} & \multicolumn{1}{l|}{Ionospheric scintillation detection} & \multicolumn{1}{l|}{Supervised Learning} & \multicolumn{1}{l|}{\begin{tabular}[c]{@{}l@{}}Support Vector Machine and \\ Gaussian Support Vector Machine\end{tabular}} & \begin{tabular}[c]{@{}l@{}}Receiver operating characteristic \\ curve and confusion matrix\end{tabular} \\ \hline
\multicolumn{1}{|l|}{\cite{linty2018detection}} & \multicolumn{1}{l|}{2018} & \multicolumn{1}{l|}{\begin{tabular}[c]{@{}l@{}}Early detection of amplitude \\ ionospheric scintillation events\end{tabular}} & \multicolumn{1}{l|}{Supervised Learning} & \multicolumn{1}{l|}{Decision tree} & \begin{tabular}[c]{@{}l@{}}Confusion matrix, accuracy, \\ precision, recall, F-score\end{tabular} \\ \hline
\multicolumn{1}{|l|}{\cite{imam2020distinguishing}} & \multicolumn{1}{l|}{2020} & \multicolumn{1}{l|}{\begin{tabular}[c]{@{}l@{}}Distinguishing between multipath\\  and ionospheric scintillation\end{tabular}} & \multicolumn{1}{l|}{Supervised Learning} & \multicolumn{1}{l|}{Bagged Decision tree} & \begin{tabular}[c]{@{}l@{}}Classification accuracy, confusion \\ matrix, miss rate, false alarms rate\end{tabular} \\ \hline
\end{tabular}%
}
\label{tab:sat1}
\end{table*}

\begin{table*}[t!]
\caption{Summary of AI-Aided Satellite Communications solutions (Continued)}
\resizebox{\textwidth}{!}{%
\begin{tabular}{|llllll|}
\hline
\multicolumn{1}{|l|}{\textbf{Publication}} & \multicolumn{1}{l|}{\textbf{Year}} & \multicolumn{1}{l|}{\textbf{Objective}} & \multicolumn{1}{l|}{\textbf{AI type}} & \multicolumn{1}{l|}{\textbf{AI algorithm}} & \textbf{Performance metrics} \\ \hline
\multicolumn{6}{|c|}{\textbf{Interference Management}} \\ \hline
\multicolumn{1}{|l|}{\cite{henarejos2019deep}} & \multicolumn{1}{l|}{2019} & \multicolumn{1}{l|}{\begin{tabular}[c]{@{}l@{}}Interference detection and \\ classification\end{tabular}} & \multicolumn{1}{l|}{Supervised Learning} & \multicolumn{1}{l|}{Long Short-Term Memory} & \begin{tabular}[c]{@{}l@{}}Probability Density Function and \\ Cumulative Density Function of \\ the MSE vector for signals with and \\ without interference\end{tabular} \\ \hline
\multicolumn{1}{|l|}{\cite{pellaco2019spectrum}} & \multicolumn{1}{l|}{2019} & \multicolumn{1}{l|}{\begin{tabular}[c]{@{}l@{}}Short-term and long-term \\ interference detection\end{tabular}} & \multicolumn{1}{l|}{Supervised Learning} & \multicolumn{1}{l|}{Long Short-Term Memory} & Maximum MSE on predicted spectrum \\ \hline
\multicolumn{1}{|l|}{\cite{liang2021realizing}} & \multicolumn{1}{l|}{2021} & \multicolumn{1}{l|}{\begin{tabular}[c]{@{}l@{}}Intelligent spectrum management\\  for satellite and ground networks\end{tabular}} & \multicolumn{1}{l|}{\begin{tabular}[c]{@{}l@{}}Supervised Learning \\ and Reinforcement Learning\end{tabular}} & \multicolumn{1}{l|}{\begin{tabular}[c]{@{}l@{}}Software-defined network,\\  Deep Neural Network, Deep Q-Networks\end{tabular}} & Optimal resource distribution \\ \hline
\multicolumn{1}{|l|}{\cite{yun2023dynamic}} & \multicolumn{1}{l|}{2023} & \multicolumn{1}{l|}{\begin{tabular}[c]{@{}l@{}}Frequency allocation to\\  maximize throughput \\ and minimize interference\end{tabular}} & \multicolumn{1}{l|}{\begin{tabular}[c]{@{}l@{}}Reinforcement Learning\end{tabular}} & \multicolumn{1}{l|}{\begin{tabular}[c]{@{}l@{}}Upper confidence bound \\ and Deep Q-Network\end{tabular}} & Throughput and collision rate \\ \hline
\multicolumn{1}{|l|}{\cite{Cho2023Multi}} & \multicolumn{1}{l|}{2023} & \multicolumn{1}{l|}{\begin{tabular}[c]{@{}l@{}}Interference-aware \\ channel allocation\end{tabular}} & \multicolumn{1}{l|}{\begin{tabular}[c]{@{}l@{}}Reinforcement Learning\end{tabular}} & \multicolumn{1}{l|}{\begin{tabular}[c]{@{}l@{}}Deep Q-Network\end{tabular}} & SINR and interference power \\ \hline
\multicolumn{6}{|c|}{\textbf{Resource Management}} \\ \hline
\multicolumn{1}{|l|}{\cite{liu2018deep}} & \multicolumn{1}{l|}{2018} & \multicolumn{1}{l|}{\begin{tabular}[c]{@{}l@{}}Dynamic channel allocation algorithm\\  for efficient resource utilization\end{tabular}} & \multicolumn{1}{l|}{Reinforcement Learning} & \multicolumn{1}{l|}{\begin{tabular}[c]{@{}l@{}}Convolutional Neural Network \\ and Deep Q-Network\end{tabular}} & \begin{tabular}[c]{@{}l@{}}Service blocking probability,\\  spectrum resources under-utilization\end{tabular} \\ \hline
\multicolumn{1}{|l|}{\cite{liao2020distributed}} & \multicolumn{1}{l|}{2020} & \multicolumn{1}{l|}{Radio resources scheduling} & \multicolumn{1}{l|}{Reinforcement Learning} & \multicolumn{1}{l|}{Multi-agent Deep Reinforcement Learning} & Transmission efficiency, broadband coverage \\ \hline
\multicolumn{1}{|l|}{\cite{jiang2020reinforcement}} & \multicolumn{1}{l|}{2020} & \multicolumn{1}{l|}{Capacity management} & \multicolumn{1}{l|}{Reinforcement Learning} & \multicolumn{1}{l|}{Q-learning} & Inter-layer capacity, utility \\ \hline
\multicolumn{1}{|l|}{\cite{huang2023}} & \multicolumn{1}{l|}{2023} & \multicolumn{1}{l|}{Distributed routing scheme} & \multicolumn{1}{l|}{Reinforcement Learning} & \multicolumn{1}{l|}{Q-learning} & \begin{tabular}[c]{@{}l@{}}Data packets delivery loss, average\\  end-to-end delay\end{tabular} \\ \hline
\multicolumn{1}{|l|}{\cite{Zhao2023Flexible}} & \multicolumn{1}{l|}{2023} & \multicolumn{1}{l|}{\begin{tabular}[c]{@{}l@{}}Spectrum efficiency optimization \\ to meet traffic demands \end{tabular}} & \multicolumn{1}{l|}{\begin{tabular}[c]{@{}l@{}}Self-supervise learning \\ and reinforcement learning\end{tabular}} & \multicolumn{1}{l|}{Proximal Policy Optimization} & \begin{tabular}[c]{@{}l@{}}Spectrum efficiency, average satisfaction index\end{tabular} \\ \hline
\multicolumn{1}{|l|}{\cite{Deng2023Dynamic}} & \multicolumn{1}{l|}{2023} & \multicolumn{1}{l|}{\begin{tabular}[c]{@{}l@{}}Joint sub-channel and power \\ allocation\end{tabular}} & \multicolumn{1}{l|}{Reinforcement Learning} & \multicolumn{1}{l|}{\begin{tabular}[c]{@{}l@{}}Twin-delayed Deep \\ Deterministic Policy Gradient \end{tabular}} & \begin{tabular}[c]{@{}l@{}}Average sum log spectral efficiency\end{tabular} \\ \hline
\multicolumn{1}{|l|}{\cite{Hassan2023Satellite}} & \multicolumn{1}{l|}{2023} & \multicolumn{1}{l|}{\begin{tabular}[c]{@{}l@{}}Offloading, computing and \\bandwidth allocation\end{tabular}} & \multicolumn{1}{l|}{Reinforcement Learning} & \multicolumn{1}{l|}{Proximal Policy Optimization} & \begin{tabular}[c]{@{}l@{}}Service time and service price\end{tabular} \\ \hline
\multicolumn{6}{|c|}{\textbf{Handover Management}} \\ \hline
\multicolumn{1}{|l|}{\cite{he2020load}} & \multicolumn{1}{l|}{2020} & \multicolumn{1}{l|}{Satellite handover strategy} & \multicolumn{1}{l|}{Reinforcement Learning} & \multicolumn{1}{l|}{Multi-agent Q-learning} & Average number of handovers, blocking rate \\ \hline
\multicolumn{1}{|l|}{\cite{zhang2020}} & \multicolumn{1}{l|}{2020} & \multicolumn{1}{l|}{Handover strategy} & \multicolumn{1}{l|}{Supervised Learning} & \multicolumn{1}{l|}{Convolutional Neural Network} & Average handover number, signal strength \\ \hline
\multicolumn{1}{|l|}{\cite{huang2020efficient}} & \multicolumn{1}{l|}{2020} & \multicolumn{1}{l|}{\begin{tabular}[c]{@{}l@{}}Algorithm to reduce the handover\\  problem to a classification problem\\  solved using DNN\end{tabular}} & \multicolumn{1}{l|}{Supervised Learning} & \multicolumn{1}{l|}{Deep Neural Network} & Radio link failure rate, ping-pong rate \\ \hline
\multicolumn{1}{|l|}{\cite{wu2021lb}} & \multicolumn{1}{l|}{2021} & \multicolumn{1}{l|}{\begin{tabular}[c]{@{}l@{}}Load-balancing based method \\ for handover decision\end{tabular}} & \multicolumn{1}{l|}{Reinforcement Learning} & \multicolumn{1}{l|}{Double Deep Q-Network} & \begin{tabular}[c]{@{}l@{}}Average number of handovers,\\  mobile users access, load imbalance, \\ network throughput\end{tabular} \\ \hline
\multicolumn{1}{|l|}{\cite{Galli2023Playing}} & \multicolumn{1}{l|}{2023} & \multicolumn{1}{l|}{\begin{tabular}[c]{@{}l@{}}Resource allocation \\ for handover decision\end{tabular}} & \multicolumn{1}{l|}{Reinforcement Learning} & \multicolumn{1}{l|}{Cooperative multi-armed bandit} & \begin{tabular}[c]{@{}l@{}}Execution time and average regret\end{tabular} \\ \hline
\multicolumn{6}{|c|}{\textbf{Energy Management}} \\ \hline
\multicolumn{1}{|l|}{\cite{luis2019deep}} & \multicolumn{1}{l|}{2019} & \multicolumn{1}{l|}{\begin{tabular}[c]{@{}l@{}}Power allocation in multi-beam \\ satellite systems\end{tabular}} & \multicolumn{1}{l|}{Reinforcement Learning} & \multicolumn{1}{l|}{Proximal Policy Optimization} & \begin{tabular}[c]{@{}l@{}}Normalized throughput, unmet system \\ demand, power consumption\end{tabular} \\ \hline
\multicolumn{1}{|l|}{\cite{tsuchida2020efficient}} & \multicolumn{1}{l|}{2020} & \multicolumn{1}{l|}{Power allocation in LEO satellites} & \multicolumn{1}{l|}{Reinforcement Learning} & \multicolumn{1}{l|}{Q-learning} & Battery lifetime \\ \hline
\multicolumn{1}{|l|}{\cite{zhao2020deep}} & \multicolumn{1}{l|}{2020} & \multicolumn{1}{l|}{\begin{tabular}[c]{@{}l@{}}Energy-efficient channel allocation \\ in Satellite IoT\end{tabular}} & \multicolumn{1}{l|}{Reinforcement Learning} & \multicolumn{1}{l|}{Deep Q-Network} & Energy consumption, satisfaction rate \\ \hline
\multicolumn{1}{|l|}{\cite{cui2020latency}} & \multicolumn{1}{l|}{2020} & \multicolumn{1}{l|}{\begin{tabular}[c]{@{}l@{}}Offloading and resource allocation\\  joint optimization\end{tabular}} & \multicolumn{1}{l|}{Reinforcement Learning} & \multicolumn{1}{l|}{Deep Q-Network} & Latency and energy costs \\ \hline
\multicolumn{1}{|l|}{\cite{Gao2021Energy}} & \multicolumn{1}{l|}{2021} & \multicolumn{1}{l|}{\begin{tabular}[c]{@{}l@{}}Energy-constrained \\ online scheduling\end{tabular}} & \multicolumn{1}{l|}{Reinforcement Learning} & \multicolumn{1}{l|}{Data-driven bandit} & \begin{tabular}[c]{@{}l@{}}Average regret and energy consumption\end{tabular} \\ \hline
\multicolumn{1}{|l|}{\cite{Hu2023Joint}} & \multicolumn{1}{l|}{2023} & \multicolumn{1}{l|}{\begin{tabular}[c]{@{}l@{}}Energy management through \\ efficient caching strategy\end{tabular}} & \multicolumn{1}{l|}{Reinforcement Learning} & \multicolumn{1}{l|}{Q-learning} & \begin{tabular}[c]{@{}l@{}}Energy consumption and average delay\end{tabular} \\ \hline
\multicolumn{1}{|l|}{\cite{Zhang2023Satellite}} & \multicolumn{1}{l|}{2023} & \multicolumn{1}{l|}{\begin{tabular}[c]{@{}l@{}}Offloading strategy for \\ reduced energy consumption\end{tabular}} & \multicolumn{1}{l|}{Reinforcement Learning} & \multicolumn{1}{l|}{Deep Deterministic Policy Gradients} & \begin{tabular}[c]{@{}l@{}}Average time delay and energy consumption\end{tabular} \\ \hline
\multicolumn{1}{|l|}{\cite{Li2023Multi}} & \multicolumn{1}{l|}{2023} & \multicolumn{1}{l|}{\begin{tabular}[c]{@{}l@{}}Energy management through \\ cache design\end{tabular}} & \multicolumn{1}{l|}{Reinforcement Learning} & \multicolumn{1}{l|}{Deep Deterministic Policy Gradients} & \begin{tabular}[c]{@{}l@{}}Energy efficiency and hit rate\end{tabular} \\ \hline
\end{tabular}%
}
\label{tab:sat2}
\vspace{-0.2cm}
\end{table*}

\subsection{Anti-jamming}\label{sec:sat_jamming}
Satellite jamming is an electronic anti-satellite attack that disrupts signals sent and received by satellites by generating noise at the same radio frequency as the satellite's antennas use~\cite{ahmad2022security}. Jamming attacks are mainly initiated within a satellite network to diminish the data throughput and deteriorate the user experience. Thus, efficient anti-jamming techniques should be designed to prevent jamming and maintain secure satellite communications. Several papers focus on reducing jamming attacks by developing effective anti-jamming solutions that rely on classical methods~\cite{shen2013, quintanadiaz2021, ahmad2022security, Weerackody2021Satellite}. For instance, the authors in~\cite{Weerackody2021Satellite} investigate using satellite diversity to overcome the effect of jamming. However, many classical anti-jamming methods cannot counter smart jamming attacks that can adapt and evolve through interaction and learning. AI is, therefore, a mandatory tool to provide intelligent and adaptive solutions for detecting and mitigating jamming attacks in real-time while improving interference mitigation. 

Several papers consider using different AI algorithms to optimize anti-jamming methods in satellite networks. For instance, the synchronization of the frequency hopping signal technique, which is frequently used for its anti-jamming properties, can be achieved with the help of a long short-term memory (LSTM) network, as described in~\cite{lee2019synchronization}. The proposed method for synchronizing frequency hopping signals significantly reduces synchronization time, thus enhancing anti-jamming abilities. Furthermore, an artificial bee colony (ABC) is proposed in~\cite{li2021satellite} to solve the blind separation problem between jamming and communication signals and find the sub-optimal solution. 

Among the different AI algorithms, RL has proven its efficiency in learning the optimal communication policy without having prior knowledge of the jamming conditions and the radio channel model in a dynamic environment. In~\cite{wei2021optimal}, the authors present a multi-step prediction Markov decision process (MDP) and build up a multi-step prediction Bellman iterative equation to defend against sweep jamming attacks while considering the inherent transmission latency of an actual communication system. Probabilities of state transition are calculated to maximize the communication system's utility. The optimal anti-jamming attack technique is investigated under the premise that the jammer can learn the strategy of the communication system. A combination of convolutional neural network (CNN)-based jamming pattern recognition and Q-learning-based online channel selection is used in~\cite{liu2019pattern} for effective anti-jamming communication. The authors conclude that introducing the channel switching cost improves the anti-jamming algorithm while capturing the trade-off between throughput and communication overhead. Another work~\cite{yan2023cross} utilizes the same RL-type Q-learning technique to choose the channel in the high-intensity jamming scenario. At the same time, a deep Q-network (DQN) determines the signal route from source to destination.

In addition to RL, DRL is used repeatedly to address the anti-jamming challenge. For instance, the work in~\cite{han2020spatial} combines game theory with DRL to learn the anti-jamming policy in a dynamic internet of satellite environment. Specifically, a Stackelberg game models the interactions between the jammers and the satellites. A DRL-based routing algorithm is used to solve the routing selection issue while preserving an available routing subset, and another Q-learning-based algorithm is used to adapt the anti-jamming strategy based on the selected routing subset. Simulations demonstrate that the proposed method has a lower routing cost and greater performance than existing algorithms and approaches the Stackelberg equilibrium. A similar work in~\cite{han2020dynamic} aims to reduce energy usage in a jamming scenario. The distributed dynamic anti-jamming system consists of a hierarchical Stackelberg game, a coalition formation game, and a Q-learning algorithm to obtain the anti-jamming policy. In~\cite{li2019performance}, the authors tackle the anti-jamming problem from the point of view of the jammer. They investigate scenarios where DRL-based users with varying communication modes combat DRL-based jammers using different jamming methods. The simulation results suggest that the proposed DRL-based jamming may effectively limit the performance of the DRL-based anti-jamming.
\vspace{-0.2cm}
\subsection{Traffic Forecasting}\label{sec:sat_traffic}
Traffic forecasting involves predicting future traffic patterns and demands in satellite networks to ensure efficient resource allocation, congestion control, and network planning. Existing terrestrial traffic forecasting models suffer from several problems, such as high computational complexity, which makes them unsuitable for satellites with limited onboard computing resources~\cite{yaacoub2019key}. In the literature, several works attempt to perform traffic forecasting in satellite networks using classical methods. For instance, a traffic prediction-based dynamic routing technique is proposed in~\cite{Yan2015TPDR} for LEO/GEO satellite networks and evaluated regarding the end-to-end delay and the packet loss rate. As satellite traffic is self-similar and demonstrates long-range dependence, the proposed traffic forecasting models must be more adaptive and achieve sufficient accuracy. AI can, therefore, be helpful because it can reduce the complexity and provide intelligent and automated solutions for accurate and efficient traffic forecasting in satellite communications. AI techniques, such as ML algorithms, can analyze historical traffic data, identify patterns, and predict future traffic demands, enabling better resource allocation and network planning~\cite{abderrahim2020}.

Supervised learning is widely used in performing traffic forecasting in satellite networks. In~\cite{rajagopal2021optimal}, a new hybridization of extreme learning machine and multitask beetle antenna search (MBAS-ELM) algorithm-based distributed routing is developed for LEO satellite networks. Practical and trustworthy routing for LEO satellite networks is challenging because of changing topology, connection modifications, and uneven communication load. The suggested model selects routes based on traffic forecasts concerning the level of traffic circulation on the Earth. The results are tested using different simulation times and data sensing rates. The collected results demonstrate that the suggested MBAS-ELM model outperforms previous techniques.

A combined forecasting model for satellite network traffic based on the extreme learning machine (ELM) is proposed in~\cite{bie2019combined}. This algorithm aims to empirically divide long-range-dependent self-similar satellite network traffic into several short-range dependent components. This allows for faster and more precise forecasting with reduced complexity. A similar work relying on ELM can be found in~\cite{na2018distributed}. The work in~\cite{jones2020short} proposes a two-stage approach using XG Boost and recurrent neural network (RNN) algorithms. This work compares the performance of these algorithms, concluding that applying either one at default settings shows an improvement up to $9.5\%$ for XG Boost and $9.2\%$ for RNNs. In~\cite{ziluan2018short}, short-term traffic loads are predicted using principal component analysis (PCA) and a generalized regression NN. The presented results demonstrate that the proposed method outperforms previous state-of-the-art algorithms regarding forecasting accuracy, training time, and robustness, making it the go-to approach for real-time traffic forecasting in satellite networks. 

The authors of~\cite{li2021research} combine a transfer learning-based gated recurrent unit (GRU) NN and particle filter online training algorithm to predict traffic faster and more accurately with insufficient online traffic data. The work done in~\cite{yang2020noval} depicts satellite network traffic's spatial and temporal features using a graph convolutional network (GCN) and GRU model with improved forecast accuracy. Apart from supervised learning, RL once again proves its usability in performance optimization tasks. For instance, the work in~\cite{huang2021ipr} introduces a DRL-based framework to improve the performance of satellite network routing algorithms. The fully distributed framework is created to mitigate the issues that emerge due to centralized training and execution found in other methods. 

\subsection{Channel Modeling and Estimation}\label{sec:sat_channel}
Accurate channel models are essential for evaluating the performance of satellite systems and, in turn, improving coverage for existing deployments. Channel models can also predict signal propagation within planned deployment scenarios, enabling efficient planning and assessment before actual deployment. Channel modeling has been traditionally done with ray tracing simulations and extensive outdoor experiments to estimate the values of the channel parameters. However, with the need for real experiments in satellite networks and the difficulty of applying ray tracing, new tools are required for efficient channel modeling and estimation. Furthermore, with the integration of massive multiple-input multiple-output (MIMO) techniques in satellite systems, obtaining effective channel state information (CSI) to construct efficient hybrid beamforming mechanisms is extremely challenging due to the dynamic nature of these systems, long delay times, and low payloads~\cite{ma2020}. Thus, several works in the literature propose schemes for efficient satellite channel modeling and estimation without relying on AI. For example, using ray tracing simulations, the authors in~\cite{ma2020} characterize the satellite-terrestrial channel in a high-speed railway environment. In~\cite{kumar2023}, the authors investigate the performance of satellite communication systems using the space shift keying modulation technique over shadowed-rician land mobile satellite links. In~\cite{al2020traffic}, a traffic simulator is developed for multi-beam satellite communication systems to address the challenge of setting optimal system parameters. 

AI techniques, on the other side, can process real-time environmental information, predict channel conditions, and optimize modulation and coding schemes to maximize system throughput~\cite{zhang2022weather}. AI can also enhance the accuracy and efficiency of channel modeling, leading to improved performance and reliability in satellite communications. Early attempts to use AI for channel modeling and estimation in terrestrial networks rely on traditional ML algorithms such as support vector machines (SVM)~\cite{Uccellari2016use} and decision trees~\cite{Oroza2017Machine}. However, DL algorithms are more effective for channel modeling and estimation in satellite networks. For instance, using a deep CNN, the authors in~\cite{ates2019path} estimate channel characteristics (path loss exponent and standard deviation of shadowing) from 2D satellite photos. In this work, the authors present a computationally efficient and trustworthy alternative to ray tracing simulations, with experimental findings demonstrating that at $900$~MHz, the prediction accuracy is equal to or higher than $76\%$. The work done in~\cite{ahmadien2020predicting} presents a similar approach for predicting path loss distributions from 2D satellite pictures. The authors conclude that accurate path loss prediction can be achieved in real-time and for various frequencies and heights. 

As establishing efficient instantaneous CSI is challenging due to the changing environment and substantial transmission delays in satellite networks, The authors in~\cite{zhang2021} offer a DL-based prediction technique to address this issue using correlations in different channels via a collection of LSTM units. The predictor is trained offline before being utilized online to extract channel information and forecast future CSI in LEO satellite settings. The findings indicate that the proposed solution efficiently alleviates channel aging issues in LEO satellite massive MIMO systems. In~\cite{zhang2022sat}, authors propose using deep neural networks (DNNs) to realize downlink CSI acquirement and hybrid beamforming design. The proposed schemes can predict future downlink CSI from observed uplink CSI without estimating and generate beamformers from predicted downlink CSI without complex optimization. The findings indicate that the offered methods can successfully support LEO systems with the massive MIMO technology.
\vspace{-0.2cm}
\subsection{Anomaly Detection}\label{sec:sat_anomaly}
Due to the harsh space environment and the exposition to heat, vacuum, and radiation, satellite communications are vulnerable to failures and disconnections. Thus, anomaly detection via telemetry mining, which involves detecting outliers or anomalies in satellite time series data to identify signal distortion, attitude instability, adjacent objects' proximity, and equipment failure, is essential for fault diagnosis and maintaining reliable satellite connections. Due to the vast amount of telemetry data, noise and measurement errors, and the complex and high-dimensional datasets, detecting anomalies via manual inspection is not feasible~\cite{schlag2018,jin2023anomaly}. Several papers in the literature address these problems in satellite telemetry mining without relying on AI techniques. For example, the authors in~\cite{jin2023anomaly} propose a cluster-based method for anomaly detection in satellite telemetry data using an extended dominant sets clustering algorithm. This paper provides valuable insights into the challenges and potential solutions for telemetry mining in satellite communications. 

However, AI techniques can effectively handle large volumes of data and complex patterns in satellite telemetry data. AI techniques can also improve the accuracy and efficiency of anomaly detection by learning from historical data and adapting to changing patterns in the telemetry data~\cite{fourati_artificial_2021}. The work in~\cite{ibrahim2018machine} is one of the early works promoting ML use for LEO satellite telemetry mining. Here, the performance of various ML methods such as autoregressive integrated moving average (ARIMA), multilayer perceptron, RNN, deep LSTM RNN, and deep GRU RNN are compared in predicting metrics such as battery temperature, power bus voltage, and load current from spacecraft telemetry data. The results show that ARIMA outperforms the other techniques and provides higher accuracy regarding the root mean square error (RMSE) and the mean absolute error (MAE).

Another approach for anomaly detection is proposed in~\cite{ramadan2020tensor}. Tensor-based anomaly detection (TAD) is a recent direction to detect and identify the corruption in the telemetry data of a satellite. Standard spectral-based methods like PCA can detect anomalies, but ML-based TAD methods, such as SVM and NNs, are shown to provide better results. This approach is reasonable as accumulated data in satellites usually have a tensor structure, e.g., space-time measurements. Orbit prediction of various resident space objects (RSOs) is crucial for avoiding satellite collisions, according to~\cite{peng2020machine}. The results show that the SVM model can enhance the orbit position estimation in most situations. 

Telemetry data compression is often overlooked, with anomaly forecasting being paid more attention to. However, the authors in~\cite{wan2019study} describe a self-learning technique called classification probability calculation - window step optimization for obtaining class characteristics and deciding on individual parameters compression. Simulation results show that the algorithm correctly classifies simulation and real mission data into the appropriate base class, significantly reducing data and computational complexity. Work done in~\cite{mahmoud2021different} employs a two-stage lossless algorithm based on LSTM to achieve the same goal. LSTM coupled with RNN can be a basis of a similar two-stage lossless data compression algorithm, as presented in~\cite{shehab2020recurrent}.

\subsection{Ionospheric Scintillation}\label{sec:sat_iono}
One of the main challenges that hinder the efficient use and implementation of satellite communications is ionospheric scintillation. Ionospheric scintillation is characterized by rapid fading or distortion of radio signals as they pass through the ionosphere~\cite{yang2015correlation}. This can occur due to changes in the density of electrons in the atmosphere, affecting radio waves' transmission. In satellite communications, ionospheric scintillation can lead to signal dropouts and degraded performance of communication systems, resulting in reduced data throughput and reliability~\cite{yang2015correlation}. This poses a significant challenge for satellite-based services such as TV broadcasting and mobile phone networks, as service disruptions can occur. Several papers address the problems associated with ionospheric scintillation in satellite communications. For instance, the work in~\cite{yang2015correlation} analyzes the correlation between the rate of total electron content index (ROTI) and scintillation indices using global positioning system (GPS) data collected in Hong Kong. The authors find a correlation coefficient of about $0.6$ between ROTI and scintillation indices when data from all GPS satellites are used together. In another context, the work in~\cite{vani2017} develops an ionospheric scintillation monitor receivers query tool, a visual exploration and analysis tool for ionospheric scintillation monitoring data. This tool allows for extracting relevant information from the monitoring data generated by ionospheric scintillation monitoring stations. 
In~\cite{rodrigues2019scint}, the authors propose a low-cost GPS-based sensor for detecting and monitoring ionospheric irregularities through detecting amplitude scintillation. 

To assess the impact of the scintillation on the signals, researchers rely on simple techniques involving wavelet methods~\cite{fu1999real}, analysis of statistical properties with histograms~\cite{Romero2016novel}, and adaptive frequency-time techniques~\cite{Miriyala2015Robust}. However, using these simple methods to evaluate the scintillation impact is only sometimes accurate due to its complexity. Additionally, it can be challenging to distinguish signal distortions caused by other factors like multi-path interference. Recently, AI has been used for scintillation detection in satellite communications thanks to its appealing features. AI can provide advanced signal processing techniques, adaptive/robust methods, and parameter estimation to mitigate the effects of ionospheric scintillation on GNSS receivers~\cite{vila-valls2020}. It can also contribute to developing models for describing ionospheric scintillation's effects on GPS receivers, improving the estimation of tracking loop error~\cite{moraes2014}. 

One of the earlier works that applies an ML algorithm to predict the ionospheric scintillation is presented in~\cite{rezende2010survey}. A decision tree-based algorithm indicates the degree of scintillation despite the significant fluctuation of the ionospheric conditions that influence the emergence of such anomalies. Similar work is done in~\cite{linty2018detection}, in which experiment findings suggest that a decision tree-based strategy can outperform traditional approaches, achieving $98\%$ detection accuracy and faster processing, allowing for early scintillation detection. A variation of the decision tree-based algorithm, namely the bagged decision tree, is proposed in~\cite{imam2020distinguishing} to differentiate multipath and ionospheric scintillation in monitoring data. The model classifies the data as scintillated, multipath impacted, or clean GNSS signal with $96\%$ accuracy. Authors in~\cite{jiao2017performance} conclude that an SVM-based amplitude detector is enough at low latitudes, but a phase-scintillation detector is required at high latitudes. A similar approach is adopted in~\cite{jiao2017automatic} where SVM and Gaussian SVM solve the frequency domain ionospheric scintillation detection problem. 
\vspace{-0.2cm}
\subsection{Interference Management}\label{sec:sat_interference}
Interference management is a broad issue affecting most wireless networks' efficiency and has traditionally been addressed using signal processing techniques~\cite{sharma2012}. However, these techniques have limitations in managing interference at a large scale. One problem associated with interference management in satellite communications is the increasing demand for high-speed data rates for satellite multimedia and broadcasting services, coupled with spectrum scarcity in satellite bands~\cite{sharma2012}. This challenge has led to the exploration of new techniques for enhancing spectral efficiency in satellite communication. For example, cognitive communication is proposed as a promising solution~\cite{sharma2012}. Cognitive techniques such as underlay, overlay, interweave, and database-related methods have been studied to improve the efficiency of satellite communication systems~\cite{di2019}. While several papers in the literature aim to mitigate the impact of interference in satellite communications, AI techniques have been proven to enhance the effectiveness of interference management algorithms in large-scale satellite networks~\cite{sharma2012}.  

In~\cite{henarejos2019deep}, the authors propose using DL algorithms, namely LSTM, to mitigate the impact of interference in satellite networks. The work mainly illustrates interference detection and classification performance with interference with different power levels. Auto-encoding techniques are used to decide whether interference is present or not. The results show high accuracy values of the proposed algorithms in low signal-to-interference ratio regimes where interference is more substantial. Researchers also use the LSTM algorithm in~\cite{pellaco2019spectrum} to detect in real-time both short-term and long-term interference in the spectrum of the signal received from the satellite. The proposed method generally applies across a discrete collection of various signal spectra and can pinpoint interference in time and frequency. The work in~\cite{liang2021realizing} proposes a DRL-based framework that exploits software-defined networks and AI to manage spectrum in integrated satellite and terrestrial networks. Spectrum sharing becomes especially relevant to accommodate a broadcasting network within a constrained range. Even though there are some methods, such as cognitive radio, to enable the dynamic sharing of the spectrum between satellite and terrestrial networks, the proposed DRL-based spectrum management techniques show promising results that outperform traditional interference management techniques.

DRL methods for satellite interference management are further explored in~\cite{yun2023dynamic}. This work explores the challenge of allocating resources in LEO satellite networks during dynamic inter-satellite interference. The objective is to optimize throughput while effectively handling interference. The authors suggest employing learning-based frequency allocation strategies, incorporating upper confidence bound (UCB) variants and DRL techniques. Through simulations with various reward and constraint configurations, the findings indicate that DQN outperforms other techniques, particularly in scenarios where prior knowledge of interfering satellites is lacking. Another study~\cite{Cho2023Multi} uses a multi-agent DRL algorithm designed for interference-conscious channel allocation. This algorithm addresses the challenge of frequency sharing between uplink satellite networks and terrestrial services. The primary goal is to identify the most favorable channel index set, minimize interference to the victim system, and meet the signal-to-interference-and-noise ratio (SINR) requirements of the network uplink. The results indicate that the proposed DQN algorithm surpasses the performance of the existing graph coloring algorithm, demonstrating its effectiveness in single-satellite and multi-satellite non-terrestrial network scenarios.
\vspace{-0.3cm}
\subsection{Resource Management}\label{sec:sat_resource}
Resource management involves allocating and scheduling resources such as bandwidth, power, frequency spectrum, transponders, and antennas to ensure efficient and effective utilization of available resources~\cite{kisseleff2021}. In addition to improving overall performance by allowing better control over network quality metrics such as latency and throughput, resource management and allocation enable satellite systems to maximize their efficiency by reducing interference. Additionally, it allows for more reliable communication links essential for mission-critical applications such as military operations or emergency response services. Resource management also plays a vital role in helping to reduce costs associated with operating a satellite system by ensuring that all resources are used efficiently~\cite{kisseleff2021}. 

Several specific problems are associated with resource management in satellite communications, and many papers in the literature propose solutions. The majority of the proposed solutions rely on complex traditional optimization techniques. For instance, the authors in~\cite{abdu2022} propose a demand and interference-aware adaptive resource management approach for GEO high throughput satellite systems. They formulate a multi-objective optimization problem to minimize power consumption and system bandwidth usage while matching the offered capacity with the demand per beam. Another paper~\cite{gost2021} investigates the resource management problem for virtual network function placement in a decentralized LEO satellite network. It proposes a decentralized approach where each satellite manages neighboring sub-network resources and provides computing services for satellite applications. 

With recent AI advancements, satellite communications can be optimized for resource allocation to improve efficiency and reduce costs. Leveraging AI addresses the rising demand for high data rates and flexible radio resource assignment while enabling enhanced spectral efficiencies and cost reductions per bit~\cite{kisseleff2021}. One of the main problems related to efficient resource management is dynamic channel allocation (DCA), which aims to maximize spectrum utilization. Unlike existing works on DCA, which ignore the intrinsic temporal correlation among the sequential channel allocation decisions, the research done in~\cite{liu2018deep} proposes a DRL-based algorithm to solve the problem and models the DCA optimization problem as an MDP. A CNN further extracts useful features from the image-like fashion reformulated system state. Simulation findings suggest that the proposed method may reduce blocking probability while increasing carried traffic and spectrum efficiency. Researchers in~\cite{Zhao2023Flexible} also tackle spectrum efficiency and meeting traffic demands challenges in satellite communications. However, their solution involves a two-stage algorithm incorporating self-supervised learning and DRL to solve spectrum congestion and power consumption issues. Spectrum efficiency is studied in another work~\cite{liao2020distributed}, in which a cooperative multi-agent DRL framework is employed in the bandwidth allocation problem. The experimental findings demonstrate that this strategy can improve transmission efficiency while remaining flexible. 

A Q-learning-based dynamic distributed routing system is presented in~\cite{huang2023} to achieve minimal end-to-end delay and low network traffic overhead burden. The experimental findings show that the scheme can discover the initial routing strategy and provide long-term quality of service (QoS) optimization. Another Q-learning-based approach is adopted in~\cite{jiang2020reinforcement} to solve the capacity management challenge in a three-layer heterogeneous satellite network comprising GEO, MEO, and LEO satellites. The paper presents a low-complexity approach for immediate and long-term optimum capacity allocation to maximize system utility. Another work~\cite{Deng2023Dynamic} explores the application of an enhanced DRL algorithm, namely twin-delayed deep deterministic policy gradient (DDPG), for joint sub-channel and power allocation within multi-beam GEO satellite communication systems. Simulation results reveal that the proposed approach demonstrates substantial improvement compared to baseline schemes. 

A different angle on the resource management problem in satellite communications is adopted in~\cite{Hassan2023Satellite}. The authors suggest using satellite networks for offloading and computation services in intelligent transportation systems (ITS). 
The proposed design leverages a network of distributed MEC nodes comprising LEO and cube satellites. By employing a collaborative multi-agent proximal policy optimization (PPO) DRL framework with an attention mechanism, this architecture facilitates intelligent offloading, computation, and bandwidth allocation decisions optimized for dynamic network conditions. Comprehensive simulations demonstrate enhanced performance compared to baseline scenarios.

\subsection{Handover Management}\label{sec:sat_handover}
To maintain rapid and uninterrupted global connectivity, effective management of mobility aspects in satellite networks, such as satellite handovers, is crucial~\cite{radhakrishnan2016}. The impact of handovers in satellite networks is more prominent than in terrestrial networks due to the high mobility of satellites and their continuous orbiting around the Earth. Several solutions are proposed in the literature to optimize handover strategies in satellite networks and provide seamless service to users. For instance, in many prior research works, handover decisions are typically made using one or more predefined criteria. Specifically, the elevation angle, remaining service time, and the number of available channels are mainly considered~\cite{Zhou2023Handover,Hourani2023Session}. These criteria are linked to signal strength, handover frequency, and satellite workload~\cite{Wang2023Seamless}. The authors in~\cite{yang2016seamless} propose a seamless handover mechanism based on software-defined satellite networking and conduct physical layer simulations to evaluate its performance. Leveraging the competitive dynamics of potential games,~\cite{wu2019handover} proposes a novel handover strategy for LEO satellite networks. This strategy promotes an equilibrium where each satellite handles an appropriate share of the workload, optimizing overall network performance. Another paper~\cite{zhao2022interlink} proposes a digital twin-assisted storage strategy for satellite-terrestrial networks to address the problems of inconsistent service and reduced link utilization during satellite handover. However, these conventional methods fail to achieve comprehensive optimization, given the emergence of larger and faster satellite constellations. Consequently, there remains a need for novel AI-driven handover strategies to ensure uninterrupted connectivity, minimize interruptions, and improve the QoS and user experience. 

One of the most frequently utilized techniques is Q-learning~\cite{he2020load}, which falls under the category of RL. For instance, the research presented in~\cite{he2020load} offers a multi-agent Q-learning satellite handover approach to decrease average satellite handovers while meeting each satellite's load limitation. In the presented RL algorithm, specific attributes of the user are designated as its state, and the handover process is defined as its action. Subsequently, the user can autonomously determine whether to execute a handover based on its state. Simulation results reveal that the suggested technique beats local handover solutions in average satellite handover and user blocking rate. Another work~\cite{wu2021lb} treats load balancing and handover strategy problems jointly by incorporating the load coefficient as one of the handover choice factors. Double DQN achieves this aim and better performance in the handover decision, with mobile user access becoming more balanced. Researchers in~\cite{Galli2023Playing} explore efficiently handling handover events within 5G network infrastructures. In their work, satellites employ a combinatorial multi-armed bandit approach to devise a resource allocation game capable of making decisions over time amid uncertain conditions. This approach enables the dynamic allocation of resources, considering interference with other channels. The simulations show a linear trend in running time and attaining a sub-optimal solution within approximately $20$-$30$~rounds.

Apart from RL, an alternative method involves employing DNN for handover optimization. For instance, the work in~\cite{huang2020efficient} investigates simplifying the handover decision problem to a classification problem based on individual users' SINR changes in 5G networks. The resulting classification problem is then solved using DNN, yielding an algorithm that outperforms traditional methods in terms of radio connection failure rate and ping-pong rate. A different work focuses on minimizing the average number of handovers while ensuring the signal strength~\cite{zhang2020}. Using a CNN, it models the handover process as a directed graph and determines the underlying regularity of different users' optimal handover tactics. Numerical simulation demonstrates that the suggested handover approach may efficiently minimize the number of handovers while maintaining signal strength.
\begin{figure*}
    \centering
    \includegraphics[width=0.9\linewidth]{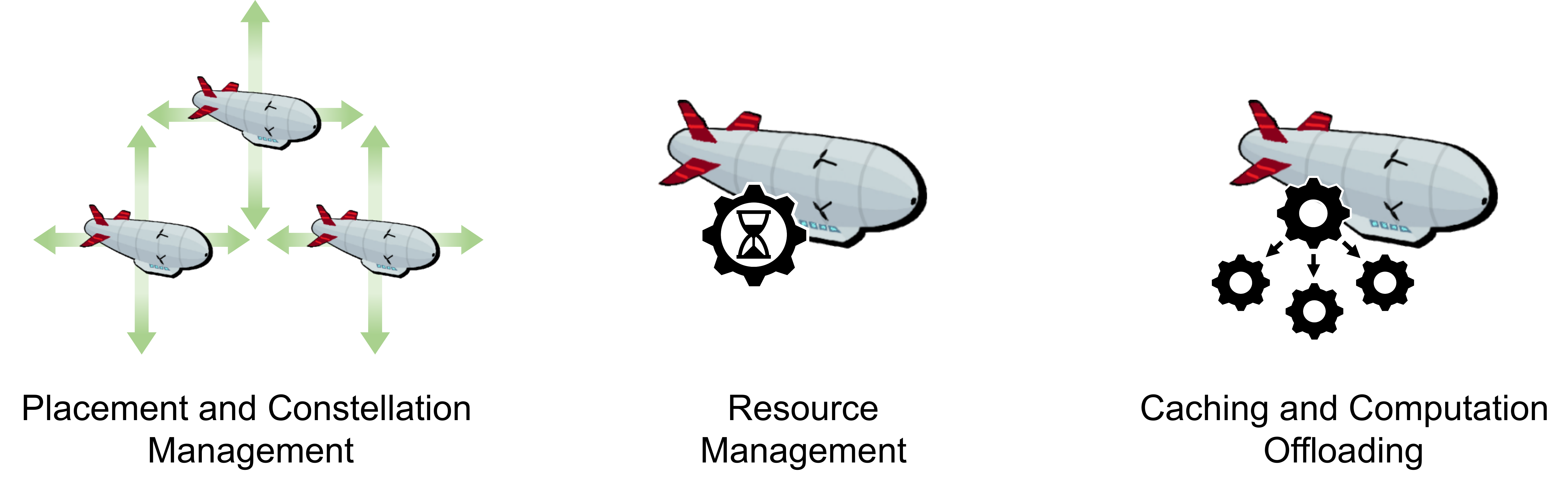}
    \caption{Challenges in HAPS communications addressed by AI and discussed in this survey.}
    \label{fig:haps}
    \vspace{-0.2cm}
\end{figure*}

\subsection{Energy Management}\label{sec:sat_energy}
Energy management strategies are critical in satellite communications due to the limited capacity of onboard batteries. They are meant to maximize power availability for mission operations, extend the system's lifetime, reduce operational costs, ensure reliable operation, and improve overall system efficiency~\cite{saleem2019}. Several classical solutions are proposed in the literature to mitigate the energy constraint and efficiently allocate resources in satellite systems~\cite{Ye2023Multi,Tang2023New,Bao2023Towards}. An interesting approach is presented in~\cite{li2018distributed} and focuses on satellite communication networks where satellite terminals are equipped with energy harvesting devices. In this work, the authors propose a distributed random access scheme considering energy constraints to maximize the average long-term network throughput. The paper adopts a game-theoretic method to approximate the optimal solution and proves the existence and uniqueness of the symmetric Nash equilibrium. 

AI algorithms, on the other hand, offer advanced optimization algorithms that can be used for various energy management solutions. For instance, the work presented in~\cite{luis2019deep} describes a DRL-based technique for power allocation in multi-beam satellite systems. The suggested architecture models the problem as continuous state and action spaces, and the PPO method is utilized to optimize the allocation policy for the least amount of unmet system demand and power usage. Primarily focused on extending battery life,~\cite{tsuchida2020efficient} employs a Q-learning algorithm to dynamically redistribute tasks among overloaded satellites and those experiencing lighter workloads within their vicinity. In~\cite{zhao2020deep}, the authors present a DQN-based approach coupled with a sliding block scheme that simplifies the modeling of dynamic features of the LEO satellites to achieve $67.86\%$ power consumption savings compared to conventional methods. The DQN-based technique is also used in~\cite{cui2020latency} to devise an integrated method for user association and offloading decisions within MEC-augmented satellite networks. This method incorporates an optimal resource allocation mechanism, enabling informed joint decisions on user connectivity and computational task placement for enhanced network performance. The proposed methodology is shown to offer better long-term latency and energy costs. The DQN is particularly useful in solving the energy consumption optimization problem in satellite communications due to its model-free nature, ability to handle sequential decision-making, approximation of Q-values with NNs, experience replay, trade-off between exploration and exploitation, continuous learning and adaptation, and scalability.

Another work~\cite{Gao2021Energy} explores the domain of energy-constrained online scheduling within satellite-terrestrial integrated networks. This study introduces a collaborative task scheduling and resource allocation data-driven bandit optimization-based strategy designed to minimize task latency. This scheme incorporates elements of online learning, online control, and offline historical information. Comparative analysis against baseline schemes demonstrates its superior regret value and energy consumption performance. The challenges of edge caching and energy conservation in non-geostationary orbit (NGSO) satellite constellations are addressed in~\cite{Hu2023Joint}, where the authors aim to formulate an efficient data sharing strategy among NGSO satellites to prevent the loss of cached data advantages caused by device mobility. The study utilizes a spreading dynamics model to establish the relationship between satellite speed, constellation structure, data size, and hit probability, determining a decisive threshold for effective data density. The proposed optimization strategy, incorporating improvements to water-filling and RL algorithms, effectively reduces energy consumption and service delay compared to traditional strategies in various scenarios. 

Focusing on integrated terrestrial-satellite networks employing non-orthogonal multiple access (NOMA) protocols, another work~\cite{Li2023Multi} proposes a novel approach for joint resource allocation and cache design. This approach leverages a multi-agent DDPG algorithm to optimize resource utilization and network content delivery efficiency. The offloading strategy for energy consumption optimization is presented in~\cite{Zhang2023Satellite}. The presented algorithm incorporates collaborative computing among satellites, enabling users to delegate tasks and alleviate local computing demands. Additionally, DDPG-based algorithms are cited for enhancing offloading decisions and optimizing resource allocation. This suggested method enhances rewards and diminishes energy consumption compared to conventional optimization techniques.
\vspace{-0.3cm}
\subsection{Summary and Lessons Learnt}\label{sec:sat_summary}
Several AI methods are commonly employed in optimizing satellite communications, with DL and DRL techniques standing out prominently. Specifically, beam hopping systems frequently use FC-NN and DRL for efficient illumination pattern design and resource allocation. These AI techniques offer advantages in terms of computation time and adaptability, addressing the challenges posed by the increasing complexity of beam hopping as the number of beams grows. Additionally, anti-jamming techniques benefit significantly from DRL, providing intelligent and adaptive solutions to counter smart jamming attacks in real-time. The combination of game theory, RL, and DRL-based approaches proves effective in optimizing anti-jamming policies, showcasing improved performance and adaptability in dynamic environments. DRL algorithms also take the spotlight in resource and energy management and handover strategy problems due to their ability to autonomously find the optimum solution based on only the data received from interacting with the environment. DL and ML algorithms take the spotlight in channel modeling and estimation, particularly for predicting channel conditions and optimizing modulation and coding schemes, and anomaly detection for satellite telemetry data, which is essential for fault diagnosis. TAD, SVM models, and ANNs showcase the effectiveness of AI in telemetry mining, providing efficient solutions for processing large volumes of telemetry data and detecting anomalies challenging for traditional methods. DL methods have also proved to be more effective in ionospheric scintillation and interference management challenges due to the availability of ground-truth data. 

The prevalence of DL, DRL, and ML in addressing specific challenges in satellite communications is attributed to their adaptability, efficiency, and improved performance compared to traditional methods. These AI techniques offer solutions beyond classical optimization algorithms, providing the necessary intelligence and adaptability to tackle the complexities and dynamic nature of satellite communication systems effectively. Other AI methods not explicitly mentioned in this section that can be useful include GAs and swarm intelligence (SI), which have shown effectiveness in various communication-related optimization problems and could contribute to addressing challenges in satellite communications. Researchers might also succeed in filling the research gap by applying the existing methods to other challenges reflecting the complexity and dynamic nature of satellite communications. The list might include security concerns, satellite maintenance optimization, and mitigation of environmental factors.  

\section{AI for High Altitude Platforms Communications}
\label{sec:haps}

\begin{table*}[]
\caption{Summary of AI-Aided HAPS Communications solutions}
\resizebox{\textwidth}{!}{%
\begin{tabular}{|llllll|}
\hline
\multicolumn{1}{|l|}{\textbf{Publication}} & \multicolumn{1}{l|}{\textbf{Year}} & \multicolumn{1}{l|}{\textbf{Objective}} & \multicolumn{1}{l|}{\textbf{AI type}} & \multicolumn{1}{l|}{\textbf{AI algorithm}} & \textbf{Performance metrics} \\ \hline
\multicolumn{6}{|c|}{\textbf{Placement and Constellation Management}} \\ \hline
\multicolumn{1}{|l|}{\cite{dong2016}} & \multicolumn{1}{l|}{2016} & \multicolumn{1}{l|}{\begin{tabular}[c]{@{}l@{}}Optimizing HAPS \\ network constellation\end{tabular}} & \multicolumn{1}{l|}{Supervised Learning} & \multicolumn{1}{l|}{Artificial immune system} & \begin{tabular}[c]{@{}l@{}}Network capacity per cost \\ under the Quality of Service \\ metrics\end{tabular} \\ \hline
\multicolumn{1}{|l|}{\cite{anicho2019comparative}} & \multicolumn{1}{l|}{2019} & \multicolumn{1}{l|}{\begin{tabular}[c]{@{}l@{}}Unmanned coordination for \\ communications area \\ coverage\end{tabular}} & \multicolumn{1}{l|}{Reinforcement Learning} & \multicolumn{1}{l|}{\begin{tabular}[c]{@{}l@{}}Deep Q-Network, \\ Swarm Intelligence \end{tabular}} & Individual and global coverage \\ \hline
\multicolumn{1}{|l|}{\cite{qiu2020}} & \multicolumn{1}{l|}{2020} & \multicolumn{1}{l|}{Multiple HAPS placement} & \multicolumn{1}{l|}{Reinforcement Learning} & \multicolumn{1}{l|}{\begin{tabular}[c]{@{}l@{}}Deep Q-Network, Prioritized \\ Experience replay\end{tabular}} & Coverage rate \\ \hline
\multicolumn{1}{|l|}{\cite{anicho2021}} & \multicolumn{1}{l|}{2021} & \multicolumn{1}{l|}{Multiple HAPS coordination} & \multicolumn{1}{l|}{\begin{tabular}[c]{@{}l@{}}Reinforcement Learning, \\ Unsupervised Learning\end{tabular}} & \multicolumn{1}{l|}{\begin{tabular}[c]{@{}l@{}}Deep Q-Network, \\ Swarm Intelligence\end{tabular}} & User coverage \\ \hline
\multicolumn{6}{|c|}{\textbf{Resource Management}} \\ \hline
\multicolumn{1}{|l|}{\cite{xiao2019location}} & \multicolumn{1}{l|}{2019} & \multicolumn{1}{l|}{\begin{tabular}[c]{@{}l@{}}Next moment location \\ prediction\end{tabular}} & \multicolumn{1}{l|}{Supervised Learning} & \multicolumn{1}{l|}{Long Short-Term Memory} & \begin{tabular}[c]{@{}l@{}}Prediction accuracy, \\ capacity loss\end{tabular} \\ \hline
\multicolumn{1}{|l|}{\cite{guan_intelligent_2019}} & \multicolumn{1}{l|}{2019} & \multicolumn{1}{l|}{\begin{tabular}[c]{@{}l@{}}Intelligent wireless channel \\ allocation\end{tabular}} & \multicolumn{1}{l|}{Reinforcement Learning} & \multicolumn{1}{l|}{Deep Q-Network} & Channel allocation accuracy \\ \hline
\multicolumn{1}{|l|}{\cite{Wu2019}} & \multicolumn{1}{l|}{2019} & \multicolumn{1}{l|}{Channel resource allocation} & \multicolumn{1}{l|}{Reinforcement Learning} & \multicolumn{1}{l|}{Deep Q-Network} & \begin{tabular}[c]{@{}l@{}}Outage and blocking rates, \\ and grade of service\end{tabular} \\ \hline
\multicolumn{1}{|l|}{\cite{guan2020}} & \multicolumn{1}{l|}{2020} & \multicolumn{1}{l|}{Intelligent beamforming} & \multicolumn{1}{l|}{Unsupervised Learning} & \multicolumn{1}{l|}{\begin{tabular}[c]{@{}l@{}}Genetic Algorithm, Particle \\ Swarm Optimization\end{tabular}} & \begin{tabular}[c]{@{}l@{}}Coverage area, user \\ interference\end{tabular} \\ \hline
\multicolumn{1}{|l|}{\cite{Jo2022DeepQT}} & \multicolumn{1}{l|}{2022} & \multicolumn{1}{l|}{Transmission power control} & \multicolumn{1}{l|}{Reinforcement Learning} & \multicolumn{1}{l|}{Deep Q-Network} & \begin{tabular}[c]{@{}l@{}}Outage probability, \\ interference\end{tabular} \\ \hline
\multicolumn{1}{|l|}{\cite{wada2022dynamic}} & \multicolumn{1}{l|}{2022} & \multicolumn{1}{l|}{Dynamic antenna control} & \multicolumn{1}{l|}{Reinforcement Learning} & \multicolumn{1}{l|}{Fuzzy Q-learning} & \begin{tabular}[c]{@{}l@{}}System throughput, SINR \\ improvement\end{tabular} \\ \hline
\multicolumn{1}{|l|}{\cite{yang2023deep}} & \multicolumn{1}{l|}{2023} & \multicolumn{1}{l|}{Antenna parameters control} & \multicolumn{1}{l|}{Reinforcement Learning} & \multicolumn{1}{l|}{Deep Q-Network} & \begin{tabular}[c]{@{}l@{}}Number of low throughput \\ users\end{tabular} \\ \hline
\multicolumn{1}{|l|}{\cite{Zhang2023HAP}} & \multicolumn{1}{l|}{2023} & \multicolumn{1}{l|}{Communication resource allocation} & \multicolumn{1}{l|}{Reinforcement Learning} & \multicolumn{1}{l|}{Proximal Policy Optimization} & \begin{tabular}[c]{@{}l@{}}Age of information, \\ data rate\end{tabular} \\ \hline
\multicolumn{1}{|l|}{\cite{seid2023multi}} & \multicolumn{1}{l|}{2023} & \multicolumn{1}{l|}{Maximizing HAPS utility} & \multicolumn{1}{l|}{Reinforcement Learning} & \multicolumn{1}{l|}{\begin{tabular}[c]{@{}l@{}}Multi-agent Deep \\Deterministic Policy Gradient\end{tabular}} & \begin{tabular}[c]{@{}l@{}}Utility, throughput, delay, \\fairness, convergence rate\end{tabular} \\ \hline
\multicolumn{6}{|c|}{\textbf{Caching and Computation Offloading}} \\ \hline
\multicolumn{1}{|l|}{\cite{lakew2021}} & \multicolumn{1}{l|}{2021} & \multicolumn{1}{l|}{\begin{tabular}[c]{@{}l@{}}HAPS partial offloading \\ scheme and communication \\ resource allocation\end{tabular}} & \multicolumn{1}{l|}{Reinforcement Learning} & \multicolumn{1}{l|}{\begin{tabular}[c]{@{}l@{}}Deep Deterministic \\ Policy Gradient\end{tabular}} & \begin{tabular}[c]{@{}l@{}}Total delay, energy \\ consumption\end{tabular} \\ \hline
\multicolumn{1}{|l|}{\cite{ren2022}} & \multicolumn{1}{l|}{2022} & \multicolumn{1}{l|}{\begin{tabular}[c]{@{}l@{}}Caching and computation \\ offloading\end{tabular}} & \multicolumn{1}{l|}{Reinforcement Learning} & \multicolumn{1}{l|}{Multi-agent Deep Q-Network} & \begin{tabular}[c]{@{}l@{}}Bandwidth and computing \\ resources\end{tabular} \\ \hline
\multicolumn{1}{|l|}{\cite{truong2022mec}} & \multicolumn{1}{l|}{2022} & \multicolumn{1}{l|}{\begin{tabular}[c]{@{}l@{}}HAPS offloading scheme in \\ MEC-enhanced aerial serving \\ network\end{tabular}} & \multicolumn{1}{l|}{Reinforcement Learning} & \multicolumn{1}{l|}{\begin{tabular}[c]{@{}l@{}}Deep Deterministic \\ Policy Gradient\end{tabular}} & \begin{tabular}[c]{@{}l@{}}Total task latency, energy \\ consumption\end{tabular} \\ \hline
\multicolumn{1}{|l|}{\cite{truong2022hamec}} & \multicolumn{1}{l|}{2022} & \multicolumn{1}{l|}{\begin{tabular}[c]{@{}l@{}}HAPS offloading scheme in \\ MEC system and RSMA \\ environment\end{tabular}} & \multicolumn{1}{l|}{Reinforcement Learning} & \multicolumn{1}{l|}{\begin{tabular}[c]{@{}l@{}}Deep Deterministic \\ Policy Gradient\end{tabular}} & \begin{tabular}[c]{@{}l@{}}System latency, power \\ consumption\end{tabular} \\ \hline
\multicolumn{1}{|l|}{\cite{zhang2023distributed}} & \multicolumn{1}{l|}{2023} & \multicolumn{1}{l|}{\begin{tabular}[c]{@{}l@{}}Distributed computation \\ offloading\end{tabular}} & \multicolumn{1}{l|}{Reinforcement Learning} & \multicolumn{1}{l|}{Soft Actor-Critic} & \begin{tabular}[c]{@{}l@{}}Transmission and computation \\ times\end{tabular} \\ \hline
\end{tabular}%
}
\label{tab:HAPS}
\vspace{-0.2cm}
\end{table*}
Innovations in technology have fueled the advancement of HAPS systems, aiming to enhance broadband communication accessibility. HAPSs operate in the stratosphere and can effectively cover a large area or supplement existing broadband services. Beyond data collection, HAPS systems offer exciting possibilities for collaborative computing and distributed ML. Their interconnected nature and high altitude provide a unique platform for decentralized intelligence, where processing power and learning algorithms can be distributed across the network to analyze data in real-time and facilitate collaborative decision-making. However, challenges of HAPS communications, such as deployment and constellation management, mobility and energy constraints, resource allocation, and security considerations, remain key barriers to fully exploiting HAPS potential. In this section, we survey existing papers that utilize AI to efficiently address the challenges in deploying HAPS-assisted networks. This includes HAPS placement and topology management, resource management, caching, and computation offloading. Table~\ref{tab:HAPS} presents a summary of all works utilizing AI for addressing the challenges of HAPS communications, while Fig.~\ref{fig:haps} visualizes the challenges of HAPS communications highlighted in this survey.

\subsection{Placement and Constellation Management}\label{sec:haps_placement}
A significant obstacle in deploying HAPS systems lies in developing robust and efficient self-organizing network architectures.  Conventional self-organizing networks incorporate key functionalities like self-configuration, optimization, and self-healing capabilities. These become crucial in aerial networks, given their higher dynamism compared to fixed cellular networks; in the former, elements' positions might shift over time due to various factors such as alterations in user needs, atmospheric variations, coverage demands, battery status, or sudden shifts in network traffic. Thus, managing HAPS constellations has always been regarded as indispensable to ensure reliable communication by optimizing coverage, minimizing interference, and maximizing system capacity~\cite{arum2020}. For instance, the work in~\cite{xu2017coverage} presents a geometry-based HAPS channel model for optimizing HAPS deployment, focusing on Line-of-sight (LoS) transmission probability and path loss. The study also provides an algorithm for maximizing deployment efficiency that has been validated through simulations. Another study~\cite{alsharoa2020improvement} introduces a new wireless scheme that integrates satellite, airborne, and terrestrial networks to improve user throughput. The focus is on resource allocation and HAPS placement optimization through a two-stage approach, which includes approximated and low-complexity solutions. The study also demonstrates its advantages through numerical results. A layered architecture integrating HAPS at different altitudes into cellular networks is presented in another work~\cite{ahmadi2017}. It emphasizes using self-organizing network features for optimal HAPS placement to enhance coverage and capacity, with initial simulations indicating improved service for users. 

\vspace{-0.22cm}
Thanks to the advantages of AI techniques, they have been regarded as key enablers in designing and optimizing HAPS constellations and providing self-organizing and self-healing capabilities. AI provides significant advantages over traditional optimization approaches due to its ability to continuously adapt and improve based on real-time data and complex environmental factors. This allows for dynamic decision-making that considers changing weather conditions, traffic patterns, and network demands, ultimately leading to more efficient and resilient network operations - a challenge for static rule-based classical methods to achieve. Several papers in the literature attempt to optimize HAPS placement and constellation by relying on different AI algorithms. For instance, the performance of RL and SI algorithms in coordinating a swarm of HAPS for communication area coverage is compared in~\cite{anicho2019comparative}. Due to its simple rules-based logic, the SI method demonstrates higher convergence speed and stable coverage. On the other hand, the RL method achieves greater peak user coverage figures by using a dynamic epsilon-greedy approach and a decreasing learning rate, although it causes some coverage gaps due to the exploration strategy of each HAPS. Since RL-based approaches do not depend on feedback loops and cross-agents, they exhibit inherent coordination resilience. Therefore, it can be concluded that SI-based techniques are more efficient and reliable for building coordination algorithms but provide less ideal coverage outcomes, whereas RL algorithms produce superior peaks in coverage. The authors of a similar study~\cite{anicho2021} examine the efficiency of RL and SI in coordinating multiple unmanned HAPSs. The research builds upon existing work on both algorithms, intending to address the challenge posed by the continuous state space through partitioning this high-dimensional space. The findings indicate that SI continues to outperform RL across important performance metrics like mean overall user coverage and convergence rates despite RL demonstrating higher average peak user coverage. Nevertheless, its unpredictable coverage dip negates this advantage, rendering SI more appropriate.

In~\cite{dong2016}, an artificial immune algorithm enhances network capacity while optimizing a HAPS network constellation. The study assesses QoS limitations alongside user demand metrics, like signal-to-noise ratio (SNR), Bit Error Rate (BER), and bits per second coverage. In another scenario, NNs are utilized to mitigate the issue of frequent hand-offs that users at the cell boundary may face due to HAPS mobility. The proposed HAPS constellation aims to efficiently balance cost considerations while meeting end-user demands and ensuring QoS through isolation, rate integrity, and information availability. This particular constellation pattern can be leveraged for designing multiple HAPS constellations. It is demonstrated that a DRL approach can optimize the positioning of multiple HAPS in urgent situations requiring wireless connectivity, with limited coverage~\cite{qiu2020}. The complexity arises from irregular coverage due to site-specific blockage in 3D space. A two-tier design is proposed, including an initial model based on LoS and an advanced design considering LoS/non-LoS channel states. This utilizes double DQN and prioritized experience replay algorithms for decision-making, leading to significantly improved coverage rates compared to benchmark methods like basic DQN and K-means algorithms.

\subsection{Resource Management}\label{sec:haps_resource}
As in all wireless communication systems, radio resource management (RRM) is a crucial aspect of ensuring the performance of HAPS-assisted networks. RRM in HAPS networks involves efficient channel allocation, beamforming and antenna management, energy efficiency and power control, and communication resource allocation. Efficient channel allocation is crucial due to the limited spectrum availability. Moreover, meticulous energy management strategies are needed since HAPS systems depend heavily on limited-capacity batteries and variable solar energy. Antenna control involves adapting beamforming and steering strategies to maintain stable connections with ground users. Lastly, communication resource allocation addresses the dynamic distribution of bandwidth and computational resources to meet the demands of various applications and users. Traditional approaches to address these issues often involve intricate optimization and decision-making processes. For instance, the work in~\cite{thornton2003resman} suggests a method for optimizing an array of antennas on a HAPS for cellular coverage by predicting co-channel interference based on curve-fit approximations for radiation patterns of elliptic beams and estimating optimum beam widths for each cell of a regular hexagonal layout. However, due to the complexity and scale of HAPS networks, advanced AI techniques are needed to develop more intelligent solutions. AI has the advantage of fast adaptation and learning from past experiences and mistakes, making it crucial for any HAPS management system. In the rest of this subsection, we survey the main papers that use AI to address antenna control, channel allocation, power control, and communication resource allocation in HAPS-assisted networks.

\subsubsection{Antenna Control}
According to~\cite{guan2020}, various systems' dynamic and interdependent properties in the communication link present a challenge for HAPS resource allocation. Modifying resource allocation for one service user interferes with the resource allocation of other users, necessitating readjustment. The authors propose an intelligent beamforming algorithm based on PSO to address this challenge by effectively managing power through game theory principles, thus dealing with interference between service users and expanding coverage zones. The importance of intelligent beamforming is also studied in~\cite{xiao2019location}, where it is argued that non-precise beamforming can lead to an increased capacity loss. Therefore, having an accurate HAPS location available at all times by incorporating a location prediction model based on LSTM and trained on the available two-dimensional angle data becomes crucial. Dynamic beam control is also studied in a multi-cell configuration to enhance system throughput and address environmental factors by optimizing antenna parameters using fuzzy Q-learning~\cite{wada2022dynamic}. Another work~\cite{yang2023deep} presents a DRL-based technique for dynamic antenna control in a HAPS system. It addresses the challenge of reducing low-throughput users caused by HAPS movement due to wind and outperforms conventional evolution algorithms and RL methods, particularly in non-uniform user distribution scenarios. 

\subsubsection{Channel Allocation}
An intelligent DQN-based method for dynamic channel allocation in the HAPS 5G communication system is proposed in~\cite{guan_intelligent_2019}, which can effectively improve the system's overall performance by assessing channel quality and service priority. The system can be paired with massive MIMO technology to address issues present in current ground communication systems, offering higher accuracy in channel allocation compared to existing solutions at varying traffic levels. The work in~\cite{Wu2019} presents an algorithm that integrates Q-learning with an ANN to enable HAPS 5G systems to autonomously learn from their environment and effectively allocate channel resources.

\subsubsection{Power Control and Communication Resource Allocation}
In~\cite{Jo2022DeepQT}, the authors propose a multi-agent DQN-based algorithm for transmission power control to minimize HAPS interference with existing systems while maintaining high-speed data communication. This approach effectively avoids overestimating action values without compromising performance and produces results comparable to those of the optimal exhaustive search algorithm in various conditions. Additionally, the authors develop a double DQN to prevent action value overestimation. Researchers in~\cite{Zhang2023HAP} discuss the challenges of providing timely communication services in rural areas through HAPS networks, particularly for real-time applications like smart agriculture and digital forestry. The conducted work highlights the importance of balancing resource allocation among various services, both freshness-aware and non-freshness-aware, and explores free space optics backhaul in multi-layer HAP networks. The study introduces static and DRL-based resource allocation schemes. It demonstrates that the dynamic PPO method significantly outperforms heuristic algorithms, improving performance by nearly $2.5$~times compared to ant colony optimization methods. Another crucial scenario involving HAPS resource allocation is presented in a recent paper~\cite{seid2023multihaps}. The paper introduces an intelligent distributed multi-agent DDPG-based resource allocation scheme for HAPS-enabled Internet of Vehicles networks. The proposed scheme optimizes the association and resource allocation strategies of vehicles and other mobile devices to maximize the utility of HAPSs.
\begin{figure*}
    \centering
    \includegraphics[width=0.75\linewidth]{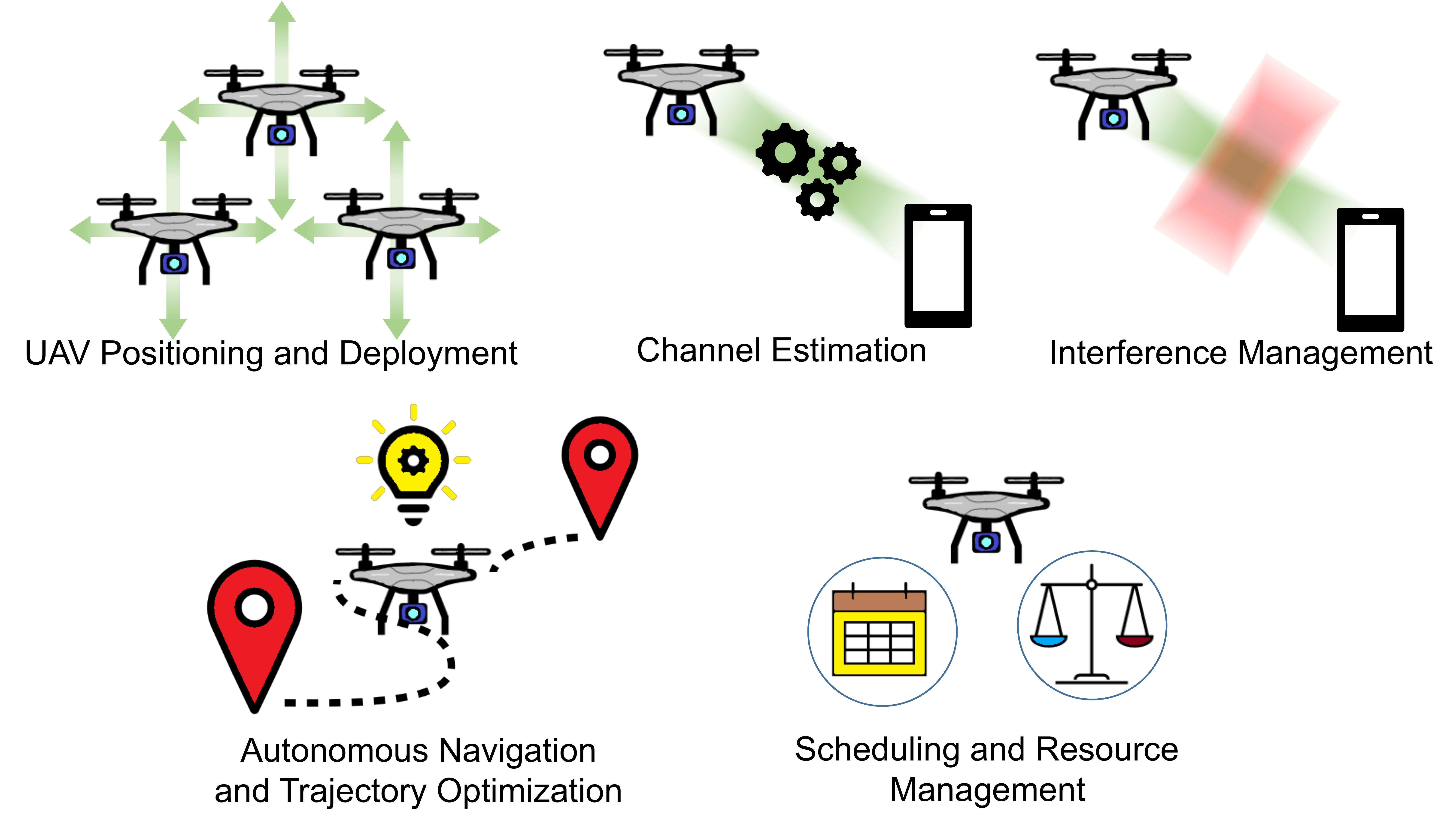}
    \caption{Challenges in UAV communications addressed by AI and discussed in this survey.}
    \label{fig:uav}
    \vspace{-0.3cm}
\end{figure*}

\subsection{Caching and Computation Offloading}\label{sec:haps_caching}
Leveraging their superior computational resources compared to user terminals, HAPS can function as agile aerial data centers, facilitating efficient offloading of intensive processing tasks. This capability, combined with their high altitude vantage point enabling robust LoS connections, minimizes the risk of disconnections during offloading, fostering seamless and reliable data transfer. However, increased latency and response delays are critical issues to be addressed when using HAPSs as aerial data centers~\cite{kurt2021vision}. This can be achieved by optimizing how frequently and where to store content to cache in HAPS-assisted networks. However, determining what content to cache and where to place it in the dynamic HAPS environments can be complex due to varying user demands and mobility patterns. Thus, optimizing the distribution of computation offloading between HAPSs and other network components is crucial for balancing computational load, minimizing energy consumption, and ensuring low latency. Efficient content caching and computation offloading can significantly improve QoS, reduce latency by bypassing congested terrestrial networks, and support applications such as remote sensing, emergency response, and IoT connectivity. 

AI and ML techniques can be used to develop inventive algorithms for intelligent decision-making to accommodate dynamic network conditions and user requirements and meet the challenges of efficient caching and computation offloading in HAPS networks. A study of caching and computation offloading by HAPS in collaboration with terrestrial nodes and intelligent vehicles is presented in~\cite{ren2022}. The multi-agent DQN-based solution targets optimal bandwidth and computing resource distribution, emphasizing the importance of caching at network boundaries and highlighting the advantages of HAPS for reducing delays. Another paper aims to minimize the total transmission and computation times in a HAPS-assisted Internet of vehicles network~\cite{zhang2023distributed}. The developed optimization problem is formulated and solved using the soft actor-critic (SAC) algorithm, with simulation results demonstrating time reduction through the application of HAPS. Additionally, it shows significant superiority compared to existing approaches. A similar scenario, where HAPS serves as a computing server that can take offloaded tasks from aerial vehicles such as drones and UAVs, is depicted in~\cite{truong2022mec}. A partial offloading strategy based on a DRL method reduces total task latency and energy consumption costs. 

A related work~\cite{truong2022hamec} introduces a HAPS-mounted MEC system in a rate-splitting multiple access environment, enabling efficient task offloading for aerial users. It optimally designs key parameters for response latency and energy consumption minimization using a DDPG algorithm with parameter noises, showing superior performance compared to benchmark schemes in simulations. In~\cite{lakew2021}, a DDPG-based algorithm is used to achieve similar objectives. The paper investigates the use of HAPS in B5G wireless networks to improve user device capacity and computing capabilities, focusing on partial task offloading and communication resource allocation. Employing a DDPG algorithm within an MDP framework, the study seeks to maximize the accomplishment of user tasks while minimizing the energy footprint. It demonstrates that their approach outperforms other methods in simulation experiments.

\vspace{-0.2cm}
\subsection{Summary and Lessons Learnt}\label{sec:haps_summary}
AI methods are crucial in addressing numerous challenges in HAPS communications, including placement and constellation management, resource management, and caching and computation offloading. These challenges involve complex optimization and decision-making processes that AI techniques can greatly enhance. AI methods such as RL and SI algorithms are employed to optimize HAPS constellation coordination for placement and constellation management. RL approaches, while being more complex, offer superior coverage peaks, making them suitable for dynamic environments and applications requiring peak performance. SI-based techniques are more efficient and reliable but may offer less ideal coverage outcomes. Resource management in HAPS communications, including channel allocation, antenna control, power regulation, and communication resource assignment, can benefit from AI techniques such as PSO and DRL. These approaches facilitate intelligent beamforming, dynamic antenna control, and autonomous resource distribution while providing substantial advantages for adapting to intricate and ever-changing HAPS network environments. Caching and computation offloading challenges in HAPS communications are addressed with DRL techniques, including DQNs, PPO, and DDPG algorithms. These methods optimize content caching and computation offloading, reducing latency and energy consumption while enhancing the quality of service. In addition to the AI methods mentioned, other techniques such as GAs, Bayesian optimization, and evolutionary algorithms could also be useful for HAPS communications. The flexibility and adaptability of AI make it a valuable tool in addressing the evolving challenges of HAPS communications, ensuring efficient and reliable connectivity for various applications and user demands. The presented AI algorithms can extend the range of issues addressed in the literature regarding interference management, scalability, real-time applications, link stability, security and privacy, environmental impact, and weather resilience. 
\section{AI for Unmanned Aerial Vehicles Communications}
\label{sec:uavs}
UAVs are considered an essential component of the SAGIN architecture, characterized by enhanced flexibility, reduced costs, and increased capabilities. The research community has shown a growing interest in considering UAVs to extend the coverage of terrestrial networks by transmitting information to and from edge users located in remote locations. On the other hand, the efficient integration of UAVs in SAGIN architectures requires addressing several challenges, such as UAV positioning and deployment, trajectory optimization, channel estimation, autonomous navigation, scheduling and resource management, and interference management~\cite{mozaffari2019}. Recently, AI tools have been considered to provide efficient solutions to enable fast integration of UAVs in the SAGIN architecture. In this section, we survey existing papers trying to use AI to address the challenges of UAV-assisted wireless networks. Fig.~\ref{fig:uav} presents the challenges of UAV communications tackled in this survey.

\begin{table*}[]
\caption{Summary of AI-Aided UAV Communications Solutions for the Positioning and Deployment Challenge}
\resizebox{\textwidth}{!}{%
\begin{tabular}{|llllll|}
\hline
{\textbf{Publication}} & \multicolumn{1}{l|}{\textbf{Year}} & \multicolumn{1}{l|}{\textbf{Objective}} & \multicolumn{1}{l|}{\textbf{AI type}} & \multicolumn{1}{l|}{\textbf{AI algorithm}} & \textbf{Performance metrics} \\ \hline
\multicolumn{6}{|c|}{\textbf{UAVs Positioning and Deployment}} \\ \hline
\multicolumn{1}{|l|}{\cite{chen_2017}} & \multicolumn{1}{l|}{2017} & \multicolumn{1}{l|}{Radio map reconstruction} & \multicolumn{1}{l|}{Unsupervised Learning} & \multicolumn{1}{l|}{\begin{tabular}[c]{@{}l@{}}KNN-based iterative \\ clustering and regression \\ algorithm\end{tabular}} & Reconstruction RMSE \\ \hline
\multicolumn{1}{|l|}{\cite{jailton_2017}} & \multicolumn{1}{l|}{2017} & \multicolumn{1}{l|}{Flight path planning} & \multicolumn{1}{l|}{Supervised Learning} & \multicolumn{1}{l|}{\begin{tabular}[c]{@{}l@{}}Generalized Regression \\ Neural Network\end{tabular}} & Throughput, delay \\ \hline
\multicolumn{1}{|l|}{\cite{zhang_2018}} & \multicolumn{1}{l|}{2018} & \multicolumn{1}{l|}{\begin{tabular}[c]{@{}l@{}}Congestion and hotspot \\ prediction\end{tabular}} & \multicolumn{1}{l|}{Unsupervised Learning} & \multicolumn{1}{l|}{K-means clustering} & \begin{tabular}[c]{@{}l@{}}Downlink transmission \\ and mobility power\end{tabular} \\ \hline
\multicolumn{1}{|l|}{\cite{xiao_2019}} & \multicolumn{1}{l|}{2019} & \multicolumn{1}{l|}{Arrival angle prediction} & \multicolumn{1}{l|}{Supervised Learning} & \multicolumn{1}{l|}{Recurrent Neural Network} & \begin{tabular}[c]{@{}l@{}}Accuracy of the predicted \\ angle\end{tabular} \\ \hline
\multicolumn{1}{|l|}{\cite{liu2020fast}} & \multicolumn{1}{l|}{2020} & \multicolumn{1}{l|}{Fast deployment strategy} & \multicolumn{1}{l|}{Supervised Learning} & \multicolumn{1}{l|}{Deep Neural Network} & Sum-rate (bps/Hz) \\ \hline
\multicolumn{1}{|l|}{\cite{zhang2021joint}} & \multicolumn{1}{l|}{2021} & \multicolumn{1}{l|}{\begin{tabular}[c]{@{}l@{}}3D deployment and power \\ allocation\end{tabular}} & \multicolumn{1}{l|}{Reinforcement Learning} & \multicolumn{1}{l|}{\begin{tabular}[c]{@{}l@{}}Deep Deterministic Policy \\ Gradient\end{tabular}} & Throughput \\ \hline
\multicolumn{1}{|l|}{\cite{Cao2023UAV}} & \multicolumn{1}{l|}{2023} & \multicolumn{1}{l|}{\begin{tabular}[c]{@{}l@{}}Optimal association and \\ dynamic deployment\end{tabular}} & \multicolumn{1}{l|}{Reinforcement Learning} & \multicolumn{1}{l|}{\begin{tabular}[c]{@{}l@{}}Multi-agent Soft \\ Actor-Critic\end{tabular}} & \begin{tabular}[c]{@{}l@{}}Throughput, coverage\end{tabular} \\ \hline
\multicolumn{1}{|l|}{\cite{Tsipi2023Machine}} & \multicolumn{1}{l|}{2023} & \multicolumn{1}{l|}{\begin{tabular}[c]{@{}l@{}}Scheme for rapid UAV \\ deployment\end{tabular}} & \multicolumn{1}{l|}{Unsupervised Learning} & \multicolumn{1}{l|}{\begin{tabular}[c]{@{}l@{}}K-means clustering\end{tabular}} & \begin{tabular}[c]{@{}l@{}}SINR\end{tabular} \\ \hline
\multicolumn{1}{|l|}{\cite{Xu2023Joint}} & \multicolumn{1}{l|}{2023} & \multicolumn{1}{l|}{\begin{tabular}[c]{@{}l@{}}Joint deployment and \\ resource allocation\end{tabular}} & \multicolumn{1}{l|}{Reinforcement Learning} & \multicolumn{1}{l|}{\begin{tabular}[c]{@{}l@{}}Personalized Federated \\ Reinforcement Learning\end{tabular}} & \begin{tabular}[c]{@{}l@{}}Long-term network \\ throughput and user privacy\end{tabular} \\ \hline
\multicolumn{1}{|l|}{\cite{Zhang2023Deployment}} & \multicolumn{1}{l|}{2023} & \multicolumn{1}{l|}{\begin{tabular}[c]{@{}l@{}}Optimal deployment of UAVs \\ and BSs\end{tabular}} & \multicolumn{1}{l|}{Reinforcement Learning} & \multicolumn{1}{l|}{\begin{tabular}[c]{@{}l@{}}Twin-delayed deep \\ deterministic policy gradient\end{tabular}} & \begin{tabular}[c]{@{}l@{}}Deployment cost\end{tabular} \\ \hline
\multicolumn{1}{|l|}{\cite{Mostafa2023Machine}} & \multicolumn{1}{l|}{2023} & \multicolumn{1}{l|}{\begin{tabular}[c]{@{}l@{}}UAV deployment for \\ cellular network offloading\end{tabular}} & \multicolumn{1}{l|}{Reinforcement Learning} & \multicolumn{1}{l|}{\begin{tabular}[c]{@{}l@{}}Q-learning\end{tabular}} & \begin{tabular}[c]{@{}l@{}}Number of UAVs\end{tabular} \\ \hline
\multicolumn{1}{|l|}{\cite{Xu2023Soft}} & \multicolumn{1}{l|}{2023} & \multicolumn{1}{l|}{\begin{tabular}[c]{@{}l@{}}3D deployment and power \\ allocation for UAVs\end{tabular}} & \multicolumn{1}{l|}{Reinforcement Learning} & \multicolumn{1}{l|}{\begin{tabular}[c]{@{}l@{}}Soft Actor-Critic\end{tabular}} & \begin{tabular}[c]{@{}l@{}}Achievable minimum user \\rate\end{tabular} \\ \hline
\multicolumn{1}{|l|}{\cite{Fu2023Joint}} & \multicolumn{1}{l|}{2023} & \multicolumn{1}{l|}{\begin{tabular}[c]{@{}l@{}}3D deployment and power \\ allocation for UAVs\end{tabular}} & \multicolumn{1}{l|}{Reinforcement Learning} & \multicolumn{1}{l|}{\begin{tabular}[c]{@{}l@{}}Proximal Policy Optimization\end{tabular}} & \begin{tabular}[c]{@{}l@{}}System throughput and \\ energy efficiency\end{tabular} \\ \hline
\multicolumn{1}{|l|}{\cite{Wang2023Joint2}} & \multicolumn{1}{l|}{2023} & \multicolumn{1}{l|}{\begin{tabular}[c]{@{}l@{}}Placement optimization\end{tabular}} & \multicolumn{1}{l|}{Reinforcement Learning} & \multicolumn{1}{l|}{\begin{tabular}[c]{@{}l@{}}Double Deep Q-Network \\and Deep Deterministic Policy \\Gradient\end{tabular}} & \begin{tabular}[c]{@{}l@{}}Total time delay\end{tabular} \\ \hline
\multicolumn{1}{|l|}{\cite{Sun2023UAV}} & \multicolumn{1}{l|}{2023} & \multicolumn{1}{l|}{\begin{tabular}[c]{@{}l@{}}Optimal deployment\end{tabular}} & \multicolumn{1}{l|}{Reinforcement Learning} & \multicolumn{1}{l|}{\begin{tabular}[c]{@{}l@{}}Deep Q-Network and \\ Convolutional Neural Network\end{tabular}} & \begin{tabular}[c]{@{}l@{}}Sum throughput and service \\ time of mobile clients\end{tabular} \\ \hline
\multicolumn{1}{|l|}{\cite{Nasr2022Single}} & \multicolumn{1}{l|}{2023} & \multicolumn{1}{l|}{\begin{tabular}[c]{@{}l@{}}Deployment and trajectory \\design\end{tabular}} & \multicolumn{1}{l|}{Reinforcement Learning} & \multicolumn{1}{l|}{\begin{tabular}[c]{@{}l@{}}Soft Actor-Critic\end{tabular}} & \begin{tabular}[c]{@{}l@{}}Downlink rate, convergence speed\end{tabular} \\ \hline
\multicolumn{1}{|l|}{\cite{Zhang2022Joint}} & \multicolumn{1}{l|}{2023} & \multicolumn{1}{l|}{\begin{tabular}[c]{@{}l@{}}Joint deployment, resource\\ and caching optimization\end{tabular}} & \multicolumn{1}{l|}{Reinforcement Learning} & \multicolumn{1}{l|}{\begin{tabular}[c]{@{}l@{}}Deep Deterministic Policy \\Gradient\end{tabular}} & \begin{tabular}[c]{@{}l@{}}Content delivery delay\end{tabular} \\ \hline
\multicolumn{1}{|l|}{\cite{Lee2022Multiagent}} & \multicolumn{1}{l|}{2023} & \multicolumn{1}{l|}{\begin{tabular}[c]{@{}l@{}}Deployment and power \\control\end{tabular}} & \multicolumn{1}{l|}{Reinforcement Learning} & \multicolumn{1}{l|}{\begin{tabular}[c]{@{}l@{}}Multi-agent Q-learning\end{tabular}} & \begin{tabular}[c]{@{}l@{}}Average energy efficiency, \\number of average outage users\end{tabular} \\ \hline
\end{tabular}%
}
\label{tab:UAVpositioning}
\vspace{-0.2cm}
\end{table*}

\subsection{UAV Positioning and Deployment}\label{sec:uav_position}
The optimum placement of UAVs is crucial for enhancing coverage and ensuring reliable communication links between UAVs and other airborne/ground nodes~\cite{zhou2019}. By strategically specifying the UAV locations, substantial gains can be obtained in terms of improved energy efficiency, increased throughput, and reduced latency. Furthermore, optimal UAV placement is needed in multi-UAV scenarios to reduce interference, as spectrum scarcity may necessitate frequency reuse over the spatial domain. Unfortunately, due to the dynamic environment and the limited capabilities of UAVs, finding optimal locations for the UAVs is not an easy task, and several challenges must be resolved. The challenges of optimizing UAV placement become even more intensified in complex and cluttered environments~\cite{khuwaja2019}, where external obstructions block incoming signals, causing degradation of the wireless transmission.

To optimize the placement of UAVs in wireless networks, researchers explored various approaches by employing classical optimization methods. For instance, the authors in~\cite{wang2020} rely on the semi-definite relaxation technique to enhance the physical layer security in UAV networks. The proposed optimization algorithm focuses on adapting the three-dimensional positions of the UAVs to maximize the probability of non-zero secrecy capacity subject to airspace and obstacle constraints. In~\cite{zhong2019}, the authors optimize the deployment of a cluster of UAVs for data collection from randomly distributed sensors. The main objective of the work in~\cite{zhong2019} is to maximize the network's total capacity by employing a local interaction game model and an online learning approach to find Nash equilibrium points. In~\cite{zhang_2020}, the authors aim to predict the users' distribution and the downlink data needed during an increased traffic event using a weighted expectation-maximization algorithm and later use this information to optimize UAV positioning. Additionally, the research in~\cite{peng_2018} focuses on predicting nearby flying objects' positions and categorizing them into groups to aid control and communication protocols using expectation maximization and Bayesian inference.

Due to the dynamic features of UAVs, and as the integration of UAVs in SAGINs increases further the search space and the number of components affecting the UAV connections, classical optimization tools fail to mitigate the added complexity in obtaining the best UAV locations to maximize the overall performance of the network. At the same time, there has been a growing interest in utilizing AI to devise efficient deployment strategies for UAV networks. AI can handle complex and dynamic decision-making processes, adapt to changing scenarios, and learn from data. This leads to more effective and automated strategies for optimizing UAV positioning, reducing the need for human intervention, and minimizing reaction time in response to problems or accidents. Table~\ref{tab:UAVpositioning} lists the main works in the literature that use AI to enhance UAV deployment in wireless networks.

Numerous studies propose the successful deployment of UAVs alongside ground BSs during network traffic overload. When cellular hotspot areas with increased data demand require more BSs, utilizing UAVs through predictive methods can be seen as a cost-effective and efficient approach for on-demand network communication. In~\cite{zhang_2018}, the authors propose an ML framework that predicts possible network congestion and determines optimal locations for UAV deployment to provide wireless service when needed. The results demonstrate that this predictive ML method, based on a K-means clustering algorithm, can significantly reduce power requirements for downlink transmission and mobility compared to deploying UAVs without any ML prediction techniques. Another study~\cite{Zhang2023Deployment} examines optimizing deployment and backhauling topology in tethered UAV-assisted integrated access and backhaul networks, introducing a twin-delayed DDPG-based learning framework to minimize deployment costs. Q-learning is also utilized to efficiently size excess mixed traffic demands on terrestrial BSs and manage subsequent offloading, treating UAVs as temporary BSs~\cite{Mostafa2023Machine}.

Another subset of studies explores the optimization of UAV placement using advanced AI techniques to improve the throughput. The authors in~\cite{liu2020fast} sequentially employ various RL and DL methods to predict the optimal positioning of UAVs in multi-UAV and multi-user scenarios, aiming to maximize throughput, while the work done in~\cite{Sun2023UAV} uses DQN to achieve the same goal. Similarly, a DDPG-based approach for 3D deployment and power allocation is proposed in~\cite{zhang2021joint}, demonstrating superior performance compared to other DQN and GA-based algorithms in increasing system throughput. The work in~\cite{Wang2023Joint2} combines DDPG and double DQN for UAV placement, resource allocation, and computation offloading in the terahertz (THz) band. Another method~\cite{Xu2023Joint} incorporates DRL into an FL framework, allowing UAVs to make real-time decisions based on local observations while achieving a globally optimal solution, with the objective of maximizing network throughput while ensuring user privacy. The SAC-based algorithm can also be utilized for optimizing the 3D deployment of UAVs while minimizing power requirements~\cite{Xu2023Soft}. Prioritizing both fairness and efficiency, the proposed framework strives to maximize the guaranteed minimum user rate, ensuring all users experience equitable resource allocation. Furthermore, this framework exhibits enhanced stability and greater long-term rewards compared to other DRL-based approaches. SAC can also be utilized for the joint initial deployment and trajectory design of UAVs to reduce downlink rate and improve convergence speed~\cite{Nasr2022Single}. In~\cite{Fu2023Joint}, the authors introduce the DRL PPO algorithm, which jointly optimizes UAV deployment and power allocation, outperforming other baselines in system throughput and energy efficiency. A different approach to achieving a trade-off between energy conservation and user connectivity is proposed in~\cite{Lee2022Multiagent}. This work presents a multi-agent Q-learning algorithm to optimize the placement and transmission power of UAV-BSs, aiming to minimize the number of ground users experiencing outages while minimizing network energy consumption.

To address the challenges in optimizing the deployment of UAV-aided communication networks for emergency scenarios, the authors in~\cite{Cao2023UAV} propose an iterative two-stage multi-agent SAC approach. The main objective is to maximize the system throughput and coverage during the deployment duration in emergency communication networks. Another paper~\cite{Tsipi2023Machine} proposes a scheme for quickly deploying a UAV-enabled emergency cellular network in disaster scenarios to support rescue operations. The placement of UAV aerial BSs is achieved through a modified K-means algorithm, and end-user signal quality is improved through joint coordinated multi-point transmission and reception.

Several other papers address various aspects of UAV positioning and deployment challenges. In~\cite{chen_2017}, the authors propose an efficient radio map reconstruction method for autonomous positioning algorithms of UAVs using a k-nearest neighbor (KNN)-based approach. In~\cite{xiao_2019}, an RNN is used to predict the angle arrival for better position estimation of a UAV. The work in~\cite{jailton_2017} tackles generic flying path planning using a generalized regression NN. Finally, the authors in~\cite{Zhang2022Joint} utilize a DDPG-based algorithm to optimize the deployment of UAVs for handling dynamic content requests and user mobility in UAV-assisted AR applications.

In conclusion, the deployment and optimal placement of UAVs in SAGINs represent a challenging yet crucial area of research. The various AI algorithms and techniques discussed in this survey highlight the diverse approaches taken to address main system considerations, such as power allocation and system throughput, by carefully optimizing UAV deployment. Furthermore, applying RL algorithms, such as Q-learning, DDPG, and multi-agent SAC, showcases the potential of adaptable and intelligent UAV-assisted communication systems. 

\subsection{Channel Estimation}\label{sec:uav_channel}

\begin{table*}[]
\caption{Summary of AI-Aided UAV Communications solutions for the Channel Estimation Challenge}
\resizebox{\textwidth}{!}{%
\begin{tabular}{|llllll|}
\hline
{\textbf{Publication}} & \multicolumn{1}{l|}{\textbf{Year}} & \multicolumn{1}{l|}{\textbf{Objective}} & \multicolumn{1}{l|}{\textbf{AI type}} & \multicolumn{1}{l|}{\textbf{AI algorithm}} & \textbf{Performance metrics} \\ \hline
\multicolumn{6}{|c|}{\textbf{Channel Estimation}} \\ \hline
\multicolumn{1}{|l|}{\cite{timoteo_2014}} & \multicolumn{1}{l|}{2014} & \multicolumn{1}{l|}{Path loss prediction} & \multicolumn{1}{l|}{Unsupervised Learning} & \multicolumn{1}{l|}{Support Vector Machine} & RMSE of path loss \\ \hline
\multicolumn{1}{|l|}{\cite{zhang_2018_air}} & \multicolumn{1}{l|}{2018} & \multicolumn{1}{l|}{\begin{tabular}[c]{@{}l@{}}Air-to-air path loss \\ prediction\end{tabular}} & \multicolumn{1}{l|}{Unsupervised Learning} & \multicolumn{1}{l|}{\begin{tabular}[c]{@{}l@{}}Random forest \\and K-Nearest Neighbors\end{tabular}} & \begin{tabular}[c]{@{}l@{}}MAE and RMSE of path \\ loss\end{tabular} \\ \hline
\multicolumn{1}{|l|}{\cite{alsamhi2018predictive}} & \multicolumn{1}{l|}{2018} & \multicolumn{1}{l|}{\begin{tabular}[c]{@{}l@{}}Signal strength and channel \\ propagation estimation\end{tabular}} & \multicolumn{1}{l|}{Supervised Learning} & \multicolumn{1}{l|}{Artificial Neural Network} & \begin{tabular}[c]{@{}l@{}}Signal strength, channel \\ fading, and LoS probability \\ estimation\end{tabular} \\ \hline
\multicolumn{1}{|l|}{\cite{ladosz_2019}} & \multicolumn{1}{l|}{2019} & \multicolumn{1}{l|}{\begin{tabular}[c]{@{}l@{}}Multiple ray tracing \\ simulation for \\ communication channel \\ model\end{tabular}} & \multicolumn{1}{l|}{Supervised Learning} & \multicolumn{1}{l|}{\begin{tabular}[c]{@{}l@{}}Receding Horizon with \\ Deep Neural Network\end{tabular}} & \begin{tabular}[c]{@{}l@{}}Probability of successful \\ communication\end{tabular} \\ \hline
\multicolumn{1}{|l|}{\cite{egi2019}} & \multicolumn{1}{l|}{2019} & \multicolumn{1}{l|}{\begin{tabular}[c]{@{}l@{}}Signal power path loss \\ prediction\end{tabular}} & \multicolumn{1}{l|}{Supervised Learning} & \multicolumn{1}{l|}{Artificial Neural Network} & \begin{tabular}[c]{@{}l@{}}MAPE of signal power \\ path loss\end{tabular} \\ \hline
\multicolumn{1}{|l|}{\cite{goudos_2019}} & \multicolumn{1}{l|}{2019} & \multicolumn{1}{l|}{\begin{tabular}[c]{@{}l@{}}Received signal strength \\ prediction\end{tabular}} & \multicolumn{1}{l|}{Supervised Learning} & \multicolumn{1}{l|}{Artificial Neural Network} & \begin{tabular}[c]{@{}l@{}}MAE, RMSE, MAPE of \\ received signal strength\end{tabular} \\ \hline
\multicolumn{1}{|l|}{\cite{goudos_2019_application}} & \multicolumn{1}{l|}{2019} & \multicolumn{1}{l|}{\begin{tabular}[c]{@{}l@{}}Received signal strength \\ prediction\end{tabular}} & \multicolumn{1}{l|}{\begin{tabular}[c]{@{}l@{}}Supervised and \\ Unsupervised Learning\end{tabular}} & \multicolumn{1}{l|}{\begin{tabular}[c]{@{}l@{}}Ensemble of Support Vector \\ Machines, Gaussian Processes, \\ Artificial Neural Network, the \\ Least Squares Boosting, \\ Bagging Prediction method\end{tabular}} & \begin{tabular}[c]{@{}l@{}}MAE, RMSE, MAPE of \\ received signal strength\end{tabular} \\ \hline
\multicolumn{1}{|l|}{\cite{wang_2019}} & \multicolumn{1}{l|}{2019} & \multicolumn{1}{l|}{\begin{tabular}[c]{@{}l@{}}3D channel modeling for  \\ communication link quality \\ evaluation\end{tabular}} & \multicolumn{1}{l|}{Unsupervised Learning} & \multicolumn{1}{l|}{K-means clustering} & RMSE of path loss \\ \hline
\multicolumn{1}{|l|}{\cite{yang2019machine}} & \multicolumn{1}{l|}{2019} & \multicolumn{1}{l|}{\begin{tabular}[c]{@{}l@{}}Prediction of path loss and \\ delay spread in mm-Wave \\ channels\end{tabular}} & \multicolumn{1}{l|}{Unsupervised Learning} & \multicolumn{1}{l|}{\begin{tabular}[c]{@{}l@{}}Random forest \\and K-Nearest Neighbors\end{tabular}} & \begin{tabular}[c]{@{}l@{}}RMSE of path loss and \\ delay spread\end{tabular} \\ \hline
\multicolumn{1}{|l|}{\cite{xia2022generative}} & \multicolumn{1}{l|}{2020} & \multicolumn{1}{l|}{\begin{tabular}[c]{@{}l@{}}Modelling mm-wave UAV \\ communication\end{tabular}} & \multicolumn{1}{l|}{Supervised Learning} & \multicolumn{1}{l|}{Variational Autoencoder} & MSE of path loss CDF \\ \hline
\multicolumn{1}{|l|}{\cite{zhang2022distributed}} & \multicolumn{1}{l|}{2022} & \multicolumn{1}{l|}{\begin{tabular}[c]{@{}l@{}}Modelling mm-wave UAV \\ communication\end{tabular}} & \multicolumn{1}{l|}{Unsupervised Learning} & \multicolumn{1}{l|}{\begin{tabular}[c]{@{}l@{}}Conditional Generative \\ Adversarial Network\end{tabular}} & \begin{tabular}[c]{@{}l@{}}Probability of learning \\ completion\end{tabular} \\ \hline
\multicolumn{1}{|l|}{\cite{mao2022machine}} & \multicolumn{1}{l|}{2022} & \multicolumn{1}{l|}{\begin{tabular}[c]{@{}l@{}}Modelling mm-wave UAV \\ communication\end{tabular}} & \multicolumn{1}{l|}{\begin{tabular}[c]{@{}l@{}}Supervised and \\ Unsupervised Learning\end{tabular}} & \multicolumn{1}{l|}{\begin{tabular}[c]{@{}l@{}}Artificial Neural Network and \\ Generative Adversarial Network\end{tabular}} & \begin{tabular}[c]{@{}l@{}}Power delay profile, \\ Doppler power spectrum \\ density, cross-correlation \end{tabular} \\ \hline
\multicolumn{1}{|l|}{\cite{Mai2022}} & \multicolumn{1}{l|}{2022} & \multicolumn{1}{l|}{\begin{tabular}[c]{@{}l@{}}UAV-to-ground channel \\ estimation\end{tabular}} & \multicolumn{1}{l|}{\begin{tabular}[c]{@{}l@{}}Supervised Learning\end{tabular}} & \multicolumn{1}{l|}{\begin{tabular}[c]{@{}l@{}}Long Short-Term Memory\end{tabular}} & \begin{tabular}[c]{@{}l@{}}MSE of estimated channel \end{tabular} \\ \hline
\multicolumn{1}{|l|}{\cite{yu2022deep}} & \multicolumn{1}{l|}{2022} & \multicolumn{1}{l|}{\begin{tabular}[c]{@{}l@{}}Dynamic modeling \\ IRS-assisted UAV \\ communication\end{tabular}} & \multicolumn{1}{l|}{\begin{tabular}[c]{@{}l@{}}Supervised and \\ Unsupervised Learning\end{tabular}} & \multicolumn{1}{l|}{\begin{tabular}[c]{@{}l@{}}Deep Neural Network and \\ Bi-directional Long Short-Term \\ Memory\end{tabular}} & NMSE of channel tracking \\ \hline
\multicolumn{1}{|l|}{\cite{Rasheed2023LSTM}} & \multicolumn{1}{l|}{2023} & \multicolumn{1}{l|}{\begin{tabular}[c]{@{}l@{}}Modelling mm-wave UAV \\ communication\end{tabular}} & \multicolumn{1}{l|}{Unsupervised Learning} & \multicolumn{1}{l|}{\begin{tabular}[c]{@{}l@{}}Conditional Generative \\ Adversarial Network\end{tabular}} & \begin{tabular}[c]{@{}l@{}}Average Jensen-Shannon \\ divergence and \\ communication overhead\end{tabular} \\ \hline
\end{tabular}%
}
\label{tab:UAVchannel}
\vspace{-0.2cm}
\end{table*}
To achieve precise alignment of beams between UAV and other SAGIN components, efficient estimates of the aerial wireless channels are required. Unfortunately, channel estimation in UAV communications entails unique challenges due to the high altitude and three-dimensional placement of UAVs~\cite{Zeng2019Accessing}. Furthermore, the aerial channels are subject to the Doppler effect due to the continuous UAV navigation. The placement of UAVs, along with the surrounding environment, also significantly impacts the propagation characteristics of UAV communications. This leads to fading and time-frequency selectivity in dynamic UAV channels~\cite{khuwaja2018chanest}. Moreover, when UAVs are equipped with many antennas, the non-negligible propagation delay across the array aperture causes a beam squint effect, which further burdens the estimation of UAV channels. 

Traditional methods involving signal processing and numerical optimization have conventionally tackled channel estimation in UAV networks~\cite{zhao2023,zhao2021,ban2020}. However, optimization algorithms typically demand significant computational complexity, creating a gap between theoretical design and real-time processing requirements. Therefore, leveraging previous dataset observations and employing AI models become valuable strategies for learning the complex mapping from compressed received pilots to channels. For instance, DNNs approximate optimization problems by identifying suitable parameters to minimize approximation errors. This application of DNNs is anticipated to significantly reduce computational complexity and processing overhead, relying on several layers performing elementary operations like matrix-vector multiplications. Table~\ref{tab:UAVchannel} highlights the main works in the literature that use AI to optimize channel estimation in UAV networks.

The authors in~\cite{xia2022generative} investigate a data-driven generative NN approach for designing millimeter wave (mmWave) air-to-ground channels. The method can assess the directional properties of the channel at both transmitter and receiver ends. The proposed generative model consists of two stages: first, it predicts the link's state and then uses this state as input to a conditional variational auto-encoder that calculates arrival and departure angles, delays, and path losses of all propagation paths. The benefits of using synthetic big data for channel estimation are explored in~\cite{ladosz_2019}, which involves multiple ray tracing simulations with various urban landscapes to create a communication channel model of UAVs acting as a communication relay. This work studies a hybrid channel modeling approach for optimizing communication relay UAV flight paths. The results demonstrate that coupling receding horizon (RH) with the NN outperforms the RH-only approach by approximately $10\%$ increased probability of achieving the necessary communication strength. Other works~\cite{zhang_2018_air, yang2019machine} also utilize ray tracing software to generate samples subsequently used by the random forest and KNN algorithms for channel estimation.

In~\cite{Mai2022}, an LSTM-based technique for air-to-ground transmission is proposed to generate more accurate CSI and greater resilience compared to the least square and minimum mean square error algorithms. The input, forget, and output gates are used to learn the time correlation of UAV channels, while the forget and input gates are utilized to build a memory function. This memory function and the output gate are then used to estimate the current slot CSI. Another study~\cite{Rasheed2023LSTM} also employs LSTM in formulating a channel estimation method based on the LSTM-distributed conditional generative adversarial network (GAN). Several studies utilize ANNs and DNNs to forecast diverse channel parameters, such as signal strength and fading channel conditions~\cite{alsamhi2018predictive, egi2019, goudos_2019, mao2022machine, yu2022deep}. These approaches facilitate adaptive data transmission and lower power consumption. Various ML algorithms employed for the same purpose include ensemble learning methods~\cite{goudos_2019_application}, K-means clustering~\cite{wang_2019}, SVMs~\cite{timoteo_2014}, and GANs~\cite{zhang2022distributed}.

When selecting a method, key considerations include aligning with specific channel parameters, accounting for the amount and quality of training data, evaluating computational complexity, and prioritizing methods with low latency for real-time requirements in channel estimation or prediction. LSTMs and DNNs emerge as robust choices for UAV channel estimation, especially when dealing with complex, time-varying data. K-means clustering can prove useful in simpler scenarios or when data is limited, while GANs offer versatility by aiding data augmentation or exploring potential channel variations.

\subsection{Interference Management}\label{sec:uav_interference}
The high altitude of UAVs results in predominantly strong LoS links between the UAVs and the ground nodes and also with the upper layers of the SAGIN, presenting both advantageous opportunities and accompanying challenges. On one side, LoS links improve communication quality with UAVs and reduce losses due to blockages. Conversely, this setup can introduce significant interference in uplink and downlink communication, posing a more complex challenge in handling interference, particularly in scenarios where terrestrial and aerial nodes coexist. Several studies tackle the issues related to interference management in UAV networks by referring to classical optimization methods to optimize resource and spectrum allocation in addition to UAV placement and path planning. For instance, the authors in~\cite{liu2019multi} present a method for cooperative interference cancellation and sum-rate maximization in multi-beam UAV communication in the cellular uplink. Their study focuses on reducing interference from other users in the cellular network to enhance overall system performance using the successive convex approximation method. Additionally, the authors in~\cite{xie2022} investigate vulnerability assessment of UAV networks in interference scenarios using a coupled-map-lattices-based approach. The objective is to evaluate the susceptibility of crucial nodes in the UAV network, such as the control center, under interference conditions. 

\begin{table*}[]
\caption{Summary of AI-Aided UAV Communications Solutions for the Interference Management Challenge}
\resizebox{\textwidth}{!}{%
\begin{tabular}{|llllll|}
\hline
{\textbf{Publication}} & \multicolumn{1}{l|}{\textbf{Year}} & \multicolumn{1}{l|}{\textbf{Objective}} & \multicolumn{1}{l|}{\textbf{AI type}} & \multicolumn{1}{l|}{\textbf{AI algorithm}} & \textbf{Performance metrics} \\ \hline
\multicolumn{6}{|c|}{\textbf{Interference Management}} \\ \hline
\multicolumn{1}{|l|}{\cite{challita2019}} & \multicolumn{1}{l|}{2019} & \multicolumn{1}{l|}{\begin{tabular}[c]{@{}l@{}}Interference-aware path \\ planning\end{tabular}} & \multicolumn{1}{l|}{Reinforcement Learning} & \multicolumn{1}{l|}{Echo state network} & \begin{tabular}[c]{@{}l@{}}Number of steps, delay, \\ latency, transmit power\end{tabular} \\ \hline
\multicolumn{1}{|l|}{\cite{ghavimi2020energy}} & \multicolumn{1}{l|}{2020} & \multicolumn{1}{l|}{\begin{tabular}[c]{@{}l@{}}Interference-aware \\ energy-efficient scheme\end{tabular}} & \multicolumn{1}{l|}{Reinforcement Learning} & \multicolumn{1}{l|}{Deep Q-Network} & \begin{tabular}[c]{@{}l@{}}Energy and spectral \\ efficiency, interference \\ to terrestrial users\end{tabular} \\ \hline
\multicolumn{1}{|l|}{\cite{viana2022convolutional}} & \multicolumn{1}{l|}{2022} & \multicolumn{1}{l|}{\begin{tabular}[c]{@{}l@{}}Detecting interference attacks \\ on UAVs\end{tabular}} & \multicolumn{1}{l|}{Supervised Learning} & \multicolumn{1}{l|}{Convolutional Neural Network} & \begin{tabular}[c]{@{}l@{}}Precision, recall and F1 \\ score in identifying \\ jamming attacks, accuracy \\ vs attacker power and speed\end{tabular} \\ \hline
\multicolumn{1}{|l|}{\cite{tang2022deep}} & \multicolumn{1}{l|}{2022} & \multicolumn{1}{l|}{\begin{tabular}[c]{@{}l@{}}UAV deployment and receiver \\ jamming strategy\end{tabular}} & \multicolumn{1}{l|}{\begin{tabular}[c]{@{}l@{}}Supervised Learning and \\ Reinforcement Learning\end{tabular}} & \multicolumn{1}{l|}{\begin{tabular}[c]{@{}l@{}}Deep Neural Network and \\ Deep Q-Network\end{tabular}} & Secrecy rate \\ \hline
\multicolumn{1}{|l|}{\cite{vaezi2023deep}} & \multicolumn{1}{l|}{2023} & \multicolumn{1}{l|}{\begin{tabular}[c]{@{}l@{}}Interference management \\ without CSI information\end{tabular}} & \multicolumn{1}{l|}{\begin{tabular}[c]{@{}l@{}}Reinforcement Learning\end{tabular}} & \multicolumn{1}{l|}{\begin{tabular}[c]{@{}l@{}}Deep Q-learning\end{tabular}} & Sum-rate, SINR \\ \hline
\multicolumn{1}{|l|}{\cite{hashesh2023aiaided}} & \multicolumn{1}{l|}{2023} & \multicolumn{1}{l|}{\begin{tabular}[c]{@{}l@{}}Height optimization for \\ interference management\end{tabular}} & \multicolumn{1}{l|}{\begin{tabular}[c]{@{}l@{}}Supervised Learning\end{tabular}} & \multicolumn{1}{l|}{\begin{tabular}[c]{@{}l@{}}Support Vector regression, \\ Linear regression and \\ Artificial Neural Network\end{tabular}} & \multicolumn{1}{l|}{\begin{tabular}[c]{@{}l@{}}Outage probability, bit \\ error rate \end{tabular}}\\ \hline
\multicolumn{1}{|l|}{\cite{li2023radio}} & \multicolumn{1}{l|}{2023} & \multicolumn{1}{l|}{\begin{tabular}[c]{@{}l@{}}Radio resource management \\ for interference management\end{tabular}} & \multicolumn{1}{l|}{\begin{tabular}[c]{@{}l@{}}Reinforcement Learning\end{tabular}} & \multicolumn{1}{l|}{\begin{tabular}[c]{@{}l@{}}Double Dueling Deep Q-Network, \\Twin Delayed Deep Deterministic \\Policy Gradient\end{tabular}} & \multicolumn{1}{l|}{\begin{tabular}[c]{@{}l@{}}Sum-rate, outage probability,\\ collision probability\end{tabular}}\\ \hline
\end{tabular}%
}
\label{tab:UAVinterference}
\end{table*}
At the same time, AI techniques can be particularly effective in addressing interference-related issues by efficiently processing large amounts of data to detect and/or avoid such interference. AI can adapt to dynamic environments, optimize communication parameters, reduce signaling overhead, and learn from data. This enables efficient and reliable communication even in complex multi-UAV scenarios. Table~\ref{tab:UAVinterference} highlights several recent papers that use AI tools to devise efficient interference management techniques in UAV networks.

One way to mitigate the degradation effect of interference in UAV networks is through careful flight path planning. Focusing on cellular-connected UAV networks,~\cite{challita2019} introduces an innovative path-planning algorithm that considers potential interference between UAVs and cellular infrastructure to optimize network performance. They utilize echo state network cells and a DRL algorithm to minimize time-dependent utility functions by transferring each network state observation to an action. Simulation results demonstrate improved wireless latency and ground user rate, with fewer steps required compared to a heuristic baseline. The optimal altitude of the UAVs varies based on terrestrial network density and user data rate requirements. Height optimization aiming to minimize interference is also investigated in~\cite{hashesh2023aiaided}. The method proposed, based on support vector regression, linear regression, and ANNs, is assessed using metrics like outage probability and BER and demonstrates encouraging outcomes in enhancing the performance and effectiveness of UAV networks. A similar approach is followed in~\cite{ghavimi2020energy}, where the authors aim to improve the energy efficiency of UAVs while minimizing interference on ground users. Simulations show that the deep Q-learning technique enhances UAV energy and spectrum efficiency while minimizing disruption to terrestrial users.

In UAV communication, interference may also result from jamming attacks on the exposed link. In~\cite{viana2022convolutional}, the authors propose a CNN-based method for detecting such jamming attacks in UAV networks, making it possible for UAVs to mitigate the interference caused by them. To address the low processing capability of UAVs, they suggest incorporating a DNN with a self-attention layer that can accurately identify attacks with as little as $2$~dBm power. Simulations indicate that it is easier to detect attacks with three or more attackers, fewer users, and power levels exceeding $10$~dBm, particularly considering the 3D distance between the small cell and the authenticated UAV for improved recognition accuracy. Examining the UAV deployment and receiver jamming method simultaneously for optimal security while considering channel uncertainty is another approach to interference management~\cite{tang2022deep}. This study introduces a data-trained DNN and DQN to approximate the best UAV deployment, enhancing transmission security performance. This algorithm enables UAVs to serve as secure relays for forwarding sensitive information.

Another DRL-driven approach is presented in~\cite{vaezi2023deep}, which addresses the challenge of inter-cell interference in three-dimensional cellular networks, particularly with UAVs. The proposed solution based on deep Q-learning effectively mitigates interference without requiring channel information. The paper also discusses methods to scale the algorithms efficiently and decentralize them using multi-agent RL, facilitating the growth of civilian UAVs. Another study~\cite{li2023radio} suggests a hybrid double duelling DQN-TD3 approach to address inter-cell interference and improve wireless transmission quality for UAVs. The proposed algorithm demonstrates superior performance compared to existing methods regarding sum rate, outage probability, and collision probability.
\vspace{-0.2cm}
\subsection{Autonomous Navigation and Trajectory Optimization}\label{sec:uav_trajectory}

\begin{table*}[]
\caption{Summary of AI-Aided UAV Communications Solutions for the Autonomous Navigation Challenge}
\resizebox{\textwidth}{!}{%
\begin{tabular}{|llllll|}
\hline
{\textbf{Publication}} & \multicolumn{1}{l|}{\textbf{Year}} & \multicolumn{1}{l|}{\textbf{Objective}} & \multicolumn{1}{l|}{\textbf{AI type}} & \multicolumn{1}{l|}{\textbf{AI algorithm}} & \textbf{Performance metrics} \\ \hline
\multicolumn{6}{|c|}{\textbf{Autonomous Navigation and Trajectory Optimization}} \\ \hline
\multicolumn{1}{|l|}{\cite{imanberdiyev2016}} & \multicolumn{1}{l|}{2016} & \multicolumn{1}{l|}{\begin{tabular}[c]{@{}l@{}}High level control method \\ for autonomous navigation \\ of UAVs\end{tabular}} & \multicolumn{1}{l|}{Reinforcement Learning} & \multicolumn{1}{l|}{\begin{tabular}[c]{@{}l@{}}Decision trees-based \\Reinforcement Learning\end{tabular}} & \begin{tabular}[c]{@{}l@{}}Result trajectory and action \\ reward\end{tabular} \\ \hline
\multicolumn{1}{|l|}{\cite{liu2018efficient}} & \multicolumn{1}{l|}{2018} & \multicolumn{1}{l|}{\begin{tabular}[c]{@{}l@{}}UAV communication \\ coverage through navigation\end{tabular}} & \multicolumn{1}{l|}{Reinforcement Learning} & \multicolumn{1}{l|}{Deep Q-Network} & \begin{tabular}[c]{@{}l@{}}Connectivity and energy \\ consumption\end{tabular} \\ \hline
\multicolumn{1}{|l|}{\cite{pham2018cooperative}} & \multicolumn{1}{l|}{2018} & \multicolumn{1}{l|}{\begin{tabular}[c]{@{}l@{}}Multi-agent algorithm for \\ cooperative learning of the \\ environment\end{tabular}} & \multicolumn{1}{l|}{Reinforcement Learning} & \multicolumn{1}{l|}{Q-learning} & \begin{tabular}[c]{@{}l@{}}Number of steps until \\ conversion\end{tabular} \\ \hline
\multicolumn{1}{|l|}{\cite{liu2019trajectory}} & \multicolumn{1}{l|}{2019} & \multicolumn{1}{l|}{\begin{tabular}[c]{@{}l@{}}Trajectory design and power \\ control based on users' \\ mobility information\end{tabular}} & \multicolumn{1}{l|}{Reinforcement Learning} & \multicolumn{1}{l|}{\begin{tabular}[c]{@{}l@{}}Q-learning and Echo \\ state network\end{tabular}} & \begin{tabular}[c]{@{}l@{}}Prediction accuracy, \\ throughput\end{tabular} \\ \hline
\multicolumn{1}{|l|}{\cite{liu2019opttraj}} & \multicolumn{1}{l|}{2019} & \multicolumn{1}{l|}{Trajectory design} & \multicolumn{1}{l|}{Reinforcement Learning} & \multicolumn{1}{l|}{Double Q-learning} & Number of satisfied users \\ \hline
\multicolumn{1}{|l|}{\cite{munoz_2019}} & \multicolumn{1}{l|}{2019} & \multicolumn{1}{l|}{\begin{tabular}[c]{@{}l@{}}Trajectory design with \\ obstacles avoidance\end{tabular}} & \multicolumn{1}{l|}{Reinforcement Learning} & \multicolumn{1}{l|}{Double Deep Q-network} & \begin{tabular}[c]{@{}l@{}}Obstacle avoidance success \\ rate\end{tabular} \\ \hline
\multicolumn{1}{|l|}{\cite{wang2019}} & \multicolumn{1}{l|}{2019} & \multicolumn{1}{l|}{\begin{tabular}[c]{@{}l@{}}UAV navigation in \\ large-scale complex \\ environments\end{tabular}} & \multicolumn{1}{l|}{Reinforcement Learning} & \multicolumn{1}{l|}{\begin{tabular}[c]{@{}l@{}}Recurrent Deterministic \\ Policy Gradient\end{tabular}} & \begin{tabular}[c]{@{}l@{}}Normalized return, success \\ rate, stray rate, crash rate, \\ average flight distance\end{tabular} \\ \hline
\multicolumn{1}{|l|}{\cite{zhu2019deep}} & \multicolumn{1}{l|}{2019} & \multicolumn{1}{l|}{\begin{tabular}[c]{@{}l@{}}Transmission control and \\ flight planning\end{tabular}} & \multicolumn{1}{l|}{Reinforcement Learning} & \multicolumn{1}{l|}{\begin{tabular}[c]{@{}l@{}}Deep Deterministic \\ Policy Gradient\end{tabular}} & \begin{tabular}[c]{@{}l@{}}Total throughput, total \\ throughput per unit energy\end{tabular} \\ \hline
\multicolumn{1}{|l|}{\cite{saxena2019optimal}} & \multicolumn{1}{l|}{2019} & \multicolumn{1}{l|}{\begin{tabular}[c]{@{}l@{}}Optimal traffic-aware \\ trajectories\end{tabular}} & \multicolumn{1}{l|}{Reinforcement Learning} & \multicolumn{1}{l|}{Proximal Policy Optimization} & \begin{tabular}[c]{@{}l@{}}Average user throughput and \\ traffic loads\end{tabular} \\ \hline
\multicolumn{1}{|l|}{\cite{khamidehi2019}} & \multicolumn{1}{l|}{2019} & \multicolumn{1}{l|}{Trajectory optimization} & \multicolumn{1}{l|}{Reinforcement Learning} & \multicolumn{1}{l|}{Q-learning} & Average sum-rate of the users \\ \hline
\multicolumn{1}{|l|}{\cite{huang2020}} & \multicolumn{1}{l|}{2020} & \multicolumn{1}{l|}{\begin{tabular}[c]{@{}l@{}}UAV navigation through \\ massive MIMO\end{tabular}} & \multicolumn{1}{l|}{Reinforcement Learning} & \multicolumn{1}{l|}{Deep Q-Network} & Area coverage score \\ \hline
\multicolumn{1}{|l|}{\cite{li2020}} & \multicolumn{1}{l|}{2020} & \multicolumn{1}{l|}{\begin{tabular}[c]{@{}l@{}}UAV ground target tracking \\ under obstacle environments\end{tabular}} & \multicolumn{1}{l|}{Reinforcement Learning} & \multicolumn{1}{l|}{\begin{tabular}[c]{@{}l@{}}Deep Deterministic Policy \\ Gradient and Long Short Term \\ Memory\end{tabular}} & \begin{tabular}[c]{@{}l@{}}Collision avoidance failure \\ rate\end{tabular} \\ \hline
\multicolumn{1}{|l|}{\cite{bouhamed2020aut}} & \multicolumn{1}{l|}{2020} & \multicolumn{1}{l|}{Autonomous path planning} & \multicolumn{1}{l|}{Reinforcement Learning} & \multicolumn{1}{l|}{\begin{tabular}[c]{@{}l@{}}Deep Deterministic Policy \\ Gradient\end{tabular}} & Task completion rate \\ \hline
\multicolumn{1}{|l|}{\cite{li2020trajectory}} & \multicolumn{1}{l|}{2020} & \multicolumn{1}{l|}{Trajectory design} & \multicolumn{1}{l|}{Reinforcement Learning} & \multicolumn{1}{l|}{Proximal Policy Optimization} & \begin{tabular}[c]{@{}l@{}}Instantaneous and average \\ sum rates\end{tabular} \\ \hline
\multicolumn{1}{|l|}{\cite{li2020uav}} & \multicolumn{1}{l|}{2020} & \multicolumn{1}{l|}{Path planning} & \multicolumn{1}{l|}{Reinforcement Learning} & \multicolumn{1}{l|}{\begin{tabular}[c]{@{}l@{}}Deep Deterministic Policy \\ Gradient\end{tabular}} & \begin{tabular}[c]{@{}l@{}}Distance to target point, \\ average reward\end{tabular} \\ \hline
\multicolumn{1}{|l|}{\cite{zeng2021simultaneous}} & \multicolumn{1}{l|}{2021} & \multicolumn{1}{l|}{Navigation and radio mapping} & \multicolumn{1}{l|}{Reinforcement Learning} & \multicolumn{1}{l|}{Duelling Double Deep Q-Network} & \begin{tabular}[c]{@{}l@{}}MSE and MAE of learned \\ radio maps, UAV flight time \\ necessary for data acquisition\end{tabular} \\ \hline
\multicolumn{1}{|l|}{\cite{guo2021}} & \multicolumn{1}{l|}{2021} & \multicolumn{1}{l|}{Collision-free navigation} & \multicolumn{1}{l|}{Reinforcement Learning} & \multicolumn{1}{l|}{Layered Recurrent Q network} & \begin{tabular}[c]{@{}l@{}}Collision number, path length, \\ success rate\end{tabular} \\ \hline
\multicolumn{1}{|l|}{\cite{mei20223d}} & \multicolumn{1}{l|}{2022} & \multicolumn{1}{l|}{Trjectory and phase shift design} & \multicolumn{1}{l|}{Reinforcement Learning} & \multicolumn{1}{l|}{\begin{tabular}[c]{@{}l@{}}Double Deep Q-network and \\ Deep Deterministic Policy \\ Gradient\end{tabular}} & \begin{tabular}[c]{@{}l@{}}Throughput, propulsion energy, \\ energy efficiency\end{tabular} \\ \hline
\multicolumn{1}{|l|}{\cite{zhang2022}} & \multicolumn{1}{l|}{2022} & \multicolumn{1}{l|}{Path planning} & \multicolumn{1}{l|}{Reinforcement Learning} & \multicolumn{1}{l|}{\begin{tabular}[c]{@{}l@{}}Twin Delayed Deep Deterministic \\ Policy Gradient\end{tabular}} & \begin{tabular}[c]{@{}l@{}}Success rate, crash rate, lost \\ rate, average reward\end{tabular} \\ \hline
\multicolumn{1}{|l|}{\cite{zhang2022game}} & \multicolumn{1}{l|}{2022} & \multicolumn{1}{l|}{\begin{tabular}[c]{@{}l@{}}Pursuit-evasion in the obstacled \\ environment\end{tabular}} & \multicolumn{1}{l|}{Reinforcement Learning} & \multicolumn{1}{l|}{\begin{tabular}[c]{@{}l@{}}Multi-agent Deep Deterministic \\ Policy Gradient\end{tabular}} & \begin{tabular}[c]{@{}l@{}}Average reward, success rate, \\ total steps\end{tabular} \\ \hline
\multicolumn{1}{|l|}{\cite{nguyen20223d}} & \multicolumn{1}{l|}{2023} & \multicolumn{1}{l|}{Optimal trajectory planning} & \multicolumn{1}{l|}{Reinforcement Learning} & \multicolumn{1}{l|}{Duelling Deep Q-learning} & \begin{tabular}[c]{@{}l@{}}Expected reward and \\ throughput\end{tabular} \\ \hline
\multicolumn{1}{|l|}{\cite{chen2023joint}} & \multicolumn{1}{l|}{2023} & \multicolumn{1}{l|}{\begin{tabular}[c]{@{}l@{}}Joint UAV trajectory optimization \\ and user association\end{tabular}} & \multicolumn{1}{l|}{Reinforcement Learning} & \multicolumn{1}{l|}{\begin{tabular}[c]{@{}l@{}}Multi-agent Proximal Policy \\ Optimization\end{tabular}} & \begin{tabular}[c]{@{}l@{}}Total throughput, energy \\ efficiency\end{tabular} \\ \hline
\multicolumn{1}{|l|}{\cite{liu2023drl}} & \multicolumn{1}{l|}{2023} & \multicolumn{1}{l|}{\begin{tabular}[c]{@{}l@{}}UAV trajectory planning\end{tabular}} & \multicolumn{1}{l|}{Reinforcement Learning} & \multicolumn{1}{l|}{\begin{tabular}[c]{@{}l@{}}Proximal Policy Optimization\end{tabular}} & \begin{tabular}[c]{@{}l@{}}Shortest path length, \\ energy consumption and \\ time utilization\end{tabular} \\ \hline
\multicolumn{1}{|l|}{\cite{gong2023bayesian}} & \multicolumn{1}{l|}{2023} & \multicolumn{1}{l|}{\begin{tabular}[c]{@{}l@{}}Trajectory planning and \\ network formation\end{tabular}} & \multicolumn{1}{l|}{Reinforcement Learning} & \multicolumn{1}{l|}{\begin{tabular}[c]{@{}l@{}}Multi-agent Deep Deterministic \\ Policy Gradient\end{tabular}} & \begin{tabular}[c]{@{}l@{}}Energy consumption, transmission \\ delay, data collection rate\end{tabular} \\ \hline
\multicolumn{1}{|l|}{\cite{hu2023ris}} & \multicolumn{1}{l|}{2023} & \multicolumn{1}{l|}{\begin{tabular}[c]{@{}l@{}}Path planning and jamming \\ rejection\end{tabular}} & \multicolumn{1}{l|}{Reinforcement Learning} & \multicolumn{1}{l|}{\begin{tabular}[c]{@{}l@{}}Deep Deterministic Policy \\ Gradient, Twin Delayed Deep \\ Deterministic Policy Gradient\end{tabular}} & \begin{tabular}[c]{@{}l@{}}Bit error rate, outage \\ probability\end{tabular} \\ \hline
\end{tabular}%
}\label{tab:UAVautonomous}
\vspace{-0.2cm}
\end{table*}

Precise trajectory planning is the cornerstone for autonomous UAV operations, enabling efficient navigation, agile maneuvering, and achieving mission objectives. Traditional path planning and trajectory optimization approaches, which rely mainly on classical optimization tools, categorize the available work into two categories. The first category focuses on autonomous UAV navigation, which includes conflict resolution mechanisms like collision avoidance, collective motion control through flocking algorithms, and efficient area coverage techniques. The second category targets enhancing the performance of different tasks, such as energy efficiency, beamforming, and throughput, via UAV trajectory optimization. Unfortunately, conventional optimization solutions used in designing autonomous navigation algorithms and optimizing UAV trajectories encounter significant drawbacks. Firstly, formulating an optimization problem necessitates a radio propagation model that is both accurate and manageable. Recent studies~\cite{Cai2020Joint,You20193D,HuAoI2021} have commonly employed statistical models such as simplified LoS-dominated models, probabilistic LoS models, and angle-dependent Rician fading models. Yet, these models typically offer predictions only in an average sense and cannot ensure performance in the specific local environment where UAVs are deployed. Secondly, trajectory designs based on offline optimization assume the availability of precise CSI derived explicitly from a particular radio propagation model. However, acquiring perfect CSI in practical scenarios becomes challenging due to uncertainties related to the UAV's position and the dynamic nature of the communication environment~\cite{Wang2022Trajectory}. Lastly, many optimization problems related to UAV-assisted communication systems are highly non-convex and present difficulties in efficient resolution. Recently, AI techniques and, more specifically, RL algorithms have been serving as efficient solutions to optimize the trajectory of UAVs in SAGINs, thanks to their ability to design optimal strategies based on collected samples in real-time when no prior information about the environment exists. Table~\ref{tab:UAVautonomous} lists some of the main works in the literature that leverage AI to optimize the trajectory and enhance the autonomous navigation of UAVs. 

In~\cite{huang2020}, the authors propose a deep Q-learning-based scheme for UAV navigation through massive MIMO. They treat individual UAV-ground connections as agents and compute the optimal locations for the UAVs based on signal power at reception. The authors of~\cite{liu2018efficient} propose a DRL-based method for UAV navigation, optimizing the energy efficiency defined as a function of communication coverage, fairness, energy consumption, and connectivity. The proposed method utilizes two robust DNNs to learn about the environment and its dynamics to make informed decisions. Simulation results demonstrate that this approach consistently outperforms commonly used baseline methods regarding coverage and fairness. Similar work is proposed in~\cite{chen2023joint}, where the authors present a decentralized approach combining the coalition formation game and multi-agent DRL PPO algorithm to jointly optimize the trajectories of UAVs and the ground user associations to maximize the total throughput and energy efficiency. Minimizing energy consumption while performing trajectory optimization using a PPO-based approach is also studied in~\cite{liu2023drl}. 

In~\cite{khamidehi2019}, the authors investigate the trajectory optimization problem in UAV-assisted networks to maximize the sum rate of users each UAV serves. Two sub-problems are identified: UAV trajectory optimization and joint power and sub-channel assignment, for which the authors develop a distributed algorithm based on Q-learning. The results of simulations show that Q-learning efficiently optimizes trajectories using reward signals from the network's topology. Maximizing the communication rate while maintaining energy efficiency is also studied in~\cite{gong2023bayesian}. This paper proposes a two-step iterative approach that employs a heuristic algorithm for adaptive network formation and multi-agent DDPG for trajectory optimization to optimize the multi-hop UAV network topology for data collection from ground users. 

A decision tree model-based RL approach for autonomous UAV navigation is presented in~\cite{imanberdiyev2016}. The algorithm can efficiently learn a trajectory in just a few iterations and outperform Q-learning methods in terms of action reward. Another study investigates how UAVs can track ground targets while avoiding obstacles using DDPG and LSTM algorithms~\cite{li2020}. Additionally, extensive research is conducted on cooperative approaches to learning the environment that optimizes the number of steps needed for the Q-learning-based algorithm to converge, which is detailed in~\cite{pham2018cooperative}. A multi-agent Q-learning-based approach is employed in similar works to perform joint trajectory planning and power control. It utilizes the echo state network's predictions about users' mobility information. Other relevant studies using Q-learning are documented in~\cite{liu2019opttraj, nguyen20223d}, while those employing a Q-network are referenced in~\cite{munoz_2019, zeng2021simultaneous, guo2021, mei20223d}. Additionally, variants of policy gradient algorithms are examined in~\cite{wang2019, bouhamed2020aut, zhu2019deep, saxena2019optimal, li2020trajectory, mei20223d, zhang2022, zhang2022game, li2020uav, hu2023ris} encompassing diverse applications related to UAV path design and optimization strategies. 

All works summarized in Table~\ref{tab:UAVautonomous} rely on RL methods that offer various advantages to optimizing UAV navigation for effective communications. DQN is a simple and efficient algorithm suitable for discrete spaces and adept at handling partial observability, while PPO and DDPG are more effective in handling continuous action spaces. Multi-agent DDPG suits scenarios with multiple cooperating UAVs, enabling decentralized decision-making. Model-based RL algorithms are more sample-efficient, allowing proactive trajectory optimization strategies. Choosing the right method depends on environment complexity, action space type, communication complexity, and computational resources. For instance, DQN might suffice for simple environments, while PPO or DDPG may be required for complex scenarios. 

Hybrid approaches, combining RL with evolutionary algorithms, planning algorithms, or other ML techniques, can further enhance performance. Successful implementation necessitates designing effective reward functions, utilizing realistic simulation environments, and rigorous testing and evaluation of navigation efficiency and communication quality. As there are hundreds of papers that tackle the trajectory optimization of UAVs using AI, this subsection presents a representative subset capturing recent papers covering different AI methods and objectives. However, a more detailed survey on the use of RL can be found in~\cite{Bai2023Towards} and on the use of ML in~\cite{Kurunathan2023Machine}.
\vspace{-0.2cm}
\subsection{Scheduling and Resource Management}\label{sec:uav_scheduling}
\begin{table*}[]
\caption{Summary of AI-Aided UAV Communications Solutions for the Scheduling and Resource Management Challenge}
\resizebox{\textwidth}{!}{%
\begin{tabular}{|llllll|}
\hline
{\textbf{Publication}} & \multicolumn{1}{l|}{\textbf{Year}} & \multicolumn{1}{l|}{\textbf{Objective}} & \multicolumn{1}{l|}{\textbf{AI type}} & \multicolumn{1}{l|}{\textbf{AI algorithm}} & \textbf{Performance metrics} \\ \hline
\multicolumn{6}{|c|}{\textbf{Scheduling and Resource Management}} \\ \hline
\multicolumn{1}{|l|}{\cite{cui2019}} & \multicolumn{1}{l|}{2019} & \multicolumn{1}{l|}{\begin{tabular}[c]{@{}l@{}}Dynamic resource allocation \\ for multiple UAVs\end{tabular}} & \multicolumn{1}{l|}{Reinforcement Learning} & \multicolumn{1}{l|}{Q-learning} & Average reward \\ \hline
\multicolumn{1}{|l|}{\cite{yang2019}} & \multicolumn{1}{l|}{2019} & \multicolumn{1}{l|}{Task scheduling} & \multicolumn{1}{l|}{Reinforcement Learning} & \multicolumn{1}{l|}{Deep Q-Network} & \begin{tabular}[c]{@{}l@{}}Delay efficiency, transmission \\ power, collision probability\end{tabular} \\ \hline
\multicolumn{1}{|l|}{\cite{munaye2019deep}} & \multicolumn{1}{l|}{2019} & \multicolumn{1}{l|}{Throughput estimation} & \multicolumn{1}{l|}{Supervised Learning} & \multicolumn{1}{l|}{Long Short-Term Memory} & Prediction accuracy \\ \hline
\multicolumn{1}{|l|}{\cite{bouhamed2020}} & \multicolumn{1}{l|}{2020} & \multicolumn{1}{l|}{\begin{tabular}[c]{@{}l@{}}Spatiotemporal scheduling \\ framework\end{tabular}} & \multicolumn{1}{l|}{Reinforcement Learning} & \multicolumn{1}{l|}{Deep Q-Network} & \begin{tabular}[c]{@{}l@{}}Energy consumption, event \\ time, delays\end{tabular} \\ \hline
\multicolumn{1}{|l|}{\cite{lin2020}} & \multicolumn{1}{l|}{2020} & \multicolumn{1}{l|}{\begin{tabular}[c]{@{}l@{}}Spectrum interaction technology \\ of the flight formations of UAVs\end{tabular}} & \multicolumn{1}{l|}{\begin{tabular}[c]{@{}l@{}}Supervised Learning and \\ Reinforcement Learning\end{tabular}} & \multicolumn{1}{l|}{\begin{tabular}[c]{@{}l@{}}Long Short-Term Memory \\ and Deep Q-Network\end{tabular}} & \begin{tabular}[c]{@{}l@{}}Throughput, average collision \\ rates\end{tabular} \\ \hline
\multicolumn{1}{|l|}{\cite{faraci2020}} & \multicolumn{1}{l|}{2020} & \multicolumn{1}{l|}{Network slicing with UAVs} & \multicolumn{1}{l|}{Reinforcement Learning} & \multicolumn{1}{l|}{\begin{tabular}[c]{@{}l@{}}Optimal policy with Bellman   \\ Optimality Equation\end{tabular}} & \begin{tabular}[c]{@{}l@{}}Power consumption, job loss, \\ delay\end{tabular} \\ \hline
\multicolumn{1}{|l|}{\cite{zhu2020}} & \multicolumn{1}{l|}{2020} & \multicolumn{1}{l|}{Computation offloading} & \multicolumn{1}{l|}{Reinforcement Learning} & \multicolumn{1}{l|}{Actor-Critic network} & \begin{tabular}[c]{@{}l@{}}Average response time, queuing \\ time, communication time, \\ processing time\end{tabular} \\ \hline
\multicolumn{1}{|l|}{\cite{kim2020}} & \multicolumn{1}{l|}{2020} & \multicolumn{1}{l|}{Computation offloading} & \multicolumn{1}{l|}{Reinforcement Learning} & \multicolumn{1}{l|}{Q-learning} & \begin{tabular}[c]{@{}l@{}}Total processing time, energy \\ consumption\end{tabular} \\ \hline
\multicolumn{1}{|l|}{\cite{faraci2020green}} & \multicolumn{1}{l|}{2020} & \multicolumn{1}{l|}{\begin{tabular}[c]{@{}l@{}}Wireless power transfer and UAV \\ fleet management\end{tabular}} & \multicolumn{1}{l|}{Reinforcement Learning} & \multicolumn{1}{l|}{\begin{tabular}[c]{@{}l@{}}Optimal policy with Bellman \\ Optimality Equation\end{tabular}} & \begin{tabular}[c]{@{}l@{}}Bandwidth, charging times, \\ wasted energy\end{tabular} \\ \hline
\multicolumn{1}{|l|}{\cite{liu2020location}} & \multicolumn{1}{l|}{2020} & \multicolumn{1}{l|}{\begin{tabular}[c]{@{}l@{}}Location-aware predictive \\ beamforming\end{tabular}} & \multicolumn{1}{l|}{Supervised Learning} & \multicolumn{1}{l|}{\begin{tabular}[c]{@{}l@{}}Long Short-Term Memory-based\\ Recurrent Neural Network\end{tabular}} & \begin{tabular}[c]{@{}l@{}}Location and angle prediction \\ accuracy, communication rate\end{tabular} \\ \hline
\multicolumn{1}{|l|}{\cite{jiang2020learning}} & \multicolumn{1}{l|}{2020} & \multicolumn{1}{l|}{Throughput prediction} & \multicolumn{1}{l|}{Supervised Learning} & \multicolumn{1}{l|}{Recurrent Neural Network} & Average prediction accuracy \\ \hline
\multicolumn{1}{|l|}{\cite{yuan2020learning}} & \multicolumn{1}{l|}{2020} & \multicolumn{1}{l|}{\begin{tabular}[c]{@{}l@{}}Predictive beamforming with \\ jittery\end{tabular}} & \multicolumn{1}{l|}{Supervised Learning} & \multicolumn{1}{l|}{\begin{tabular}[c]{@{}l@{}}Long Short-Term Memory-based \\ Recurrent Neural Network\end{tabular}} & \begin{tabular}[c]{@{}l@{}}Angle estimation error, \\ communication rate\end{tabular} \\ \hline
\multicolumn{1}{|l|}{\cite{ding20203d}} & \multicolumn{1}{l|}{2020} & \multicolumn{1}{l|}{\begin{tabular}[c]{@{}l@{}}Trajectory design and frequency \\ band allocation\end{tabular}} & \multicolumn{1}{l|}{Reinforcement Learning} & \multicolumn{1}{l|}{\begin{tabular}[c]{@{}l@{}}Deep Deterministic Policy \\ Gradient\end{tabular}} & Fairness and throughput \\ \hline
\multicolumn{1}{|l|}{\cite{nguyen2021}} & \multicolumn{1}{l|}{2021} & \multicolumn{1}{l|}{Energy harvesting scheduling} & \multicolumn{1}{l|}{Reinforcement Learning} & \multicolumn{1}{l|}{Deep Q-Network} & \begin{tabular}[c]{@{}l@{}}Energy harvested, quality \\ of service\end{tabular} \\ \hline
\multicolumn{1}{|l|}{\cite{mao2021optimizing}} & \multicolumn{1}{l|}{2021} & \multicolumn{1}{l|}{Computation offloading} & \multicolumn{1}{l|}{Supervised Learning} & \multicolumn{1}{l|}{Long Short-Term Memory} & \begin{tabular}[c]{@{}l@{}}Task completion rate and \\ system computation rate\end{tabular} \\ \hline
\multicolumn{1}{|l|}{\cite{chen2022deep}} & \multicolumn{1}{l|}{2022} & \multicolumn{1}{l|}{Energy optimization} & \multicolumn{1}{l|}{Supervised Learning} & \multicolumn{1}{l|}{Long Short-Term Memory} & \begin{tabular}[c]{@{}l@{}}Trajectory and received \\ power prediction accuracy\end{tabular} \\ \hline
\multicolumn{1}{|l|}{\cite{wang2023efficient}} & \multicolumn{1}{l|}{2023} & \multicolumn{1}{l|}{\begin{tabular}[c]{@{}l@{}}Efficient resource allocation\end{tabular}} & \multicolumn{1}{l|}{Reinforcement Learning} & \multicolumn{1}{l|}{\begin{tabular}[c]{@{}l@{}}Multi-agent constrained \\ attention Soft Actor-Critic\end{tabular}} & \begin{tabular}[c]{@{}l@{}}Bandwidth utilization, delay\end{tabular} \\ \hline
\multicolumn{1}{|l|}{\cite{ding2023online}} & \multicolumn{1}{l|}{2023} & \multicolumn{1}{l|}{\begin{tabular}[c]{@{}l@{}}Offloading and resource management\end{tabular}} & \multicolumn{1}{l|}{Reinforcement Learning} & \multicolumn{1}{l|}{\begin{tabular}[c]{@{}l@{}}Actor-Critic network\end{tabular}} & \begin{tabular}[c]{@{}l@{}}Computing performance, security\end{tabular} \\ \hline
\multicolumn{1}{|l|}{\cite{seid2023multi}} & \multicolumn{1}{l|}{2023} & \multicolumn{1}{l|}{\begin{tabular}[c]{@{}l@{}}Resource allocation\end{tabular}} & \multicolumn{1}{l|}{Reinforcement Learning} & \multicolumn{1}{l|}{\begin{tabular}[c]{@{}l@{}}Multi-agent Deep Deterministic \\ Policy Gradient\end{tabular}} & \begin{tabular}[c]{@{}l@{}}Latency, energy, bandwidth\end{tabular} \\ \hline
\end{tabular}%
}\label{tab:UAVscheduling}
\vspace{-0.3cm}
\end{table*}
The constrained resources and hardware limitations inherent in UAVs, combined with their constant movement and exposure to unpredictable environmental factors, impose several constraints on the connectivity and QoS of UAV networks~\cite{zeng2016}. Specifically, the success of UAV networks depends heavily on the effectiveness of resource allocation and optimization strategies. In particular, efficient management of power, spectrum, and storage resources plays a critical role in maximizing network performance and achieving mission objectives~\cite{pasandideh2023}. Traditional optimization and game theory techniques are used extensively to maximize the system throughput and fairness and address the limited availability of resources while respecting the UAV hardware constraints. For instance, the authors in~\cite{liu2023} formulate a joint resource optimization problem to maximize the energy efficiency of UAVs by optimizing communication scheduling, transmit power, and motion parameters. Another work addresses energy-efficient joint scheduling and resource management in UAV-enabled multi-cell networks~\cite{yang2020}, proposing both coordinated and uncoordinated convex optimization approaches to optimize UAV locations and resource management to maximize network energy efficiency and evaluate their performance in dynamic and stationary scenarios. For SAGINs operating in dynamic and uncertain contexts, traditional optimization methods such as game theory and convex optimization often prove to be cumbersome and suboptimal. AI, however, emerges as a more efficient and effective approach. By enabling SAGINs to learn and adapt in real-time, AI empowers them to reach optimal solutions even in unpredictable situations. Moreover, AI's capability for multi-objective optimization allows for simultaneous resource allocation across diverse system aspects, maximizing overall performance and efficiency. Additionally, AI enables proactive management while effectively addressing uncertainty to make UAV communication systems more efficient and reliable. Table~\ref{tab:UAVscheduling} presents the main works considering optimizing scheduling and resource allocation in UAV networks using AI.  

In~\cite{bouhamed2020}, the authors explore the capacity of UAVs to compute their routes using real-time learned data from the surrounding environment. They propose a spatio-temporal scheduling framework for autonomous UAVs based on Q-learning, enabling them to autonomously organize their schedules to cover a maximum number of pre-scheduled events within a specific geographical area and during a set time period. In unforeseen emergencies, the framework can adapt to the planned schedules. Additionally, they introduce a reward function that considers battery capacity constraints, event time frames, and navigation delays between events caused by UAVs. The spectrum interaction technology of the flight formation of UAVs is utilized to address the issue of spectrum sharing in~\cite{lin2020}. Priority allocation determines the importance of UAV tasks, and the current state is assessed before each time slot, followed by action selection, information transmission, and policy updating (formed by combining DRL and LSTM models) at the end of each round. The suggested method converges rapidly and demonstrates strong performance in dynamic channel allocation and time slot allocation models.

In device-to-device (D2D) communications, UAVs may be deployed to enhance user experience and network performance. However, the UAVs' constant movement, limited energy, and flight duration cause challenges for their real-time use. To solve this issue, the work in~\cite{nguyen2021} offers a DRL model for determining the most efficient energy harvesting schedule in UAV-assisted D2D communications. The system considers random user movement, randomly altering channel status information in each slot, and the UAV flying around a central point. The results reveal that the proposed scheme outperforms existing schemes in processing speed using an off-the-shelf processor employing trained NNs, indicating its capabilities in handling real-time resource allocation problems in UAV networks. In another work~\cite{faraci2020green}, RL is used to optimize the wireless charging scheme of a drone fleet to reduce wasted energy and increase the bandwidth provided to the users. High user mobility is also studied in~\cite{wang2023efficient}, in which the authors use a multi-agent constrained SAC RL algorithm to efficiently allocate resources in multi-UAV-assisted vehicular networks while accounting for spectrum efficiency and computing power with security constraints and attention mechanisms. 

Another RL method based on the optimal policy is proposed in~\cite{faraci2020}. It aims to minimize power consumption, job loss, and delay while offering an architecture for extending 5G network slices with UAVs. This is particularly crucial in network slicing within 5G-enabled systems with stringent latency constraints that require computing resources at the edge. As data increases and distances to the network's edge increase, this task becomes increasingly challenging. The joint optimization of UAV communication latency, energy consumption, and age of information in the context of RL-aided resource allocation in UAV-enabled Internet of Medical Things networks is also studied in~\cite{seid2023multi}, in which authors use a multi-agent DDPG algorithm. Another study~\cite{ding2023online} investigates the performance of MEC using UAVs, focusing on security. The approach utilizes the Dinkelbach method and DRL to optimize terminal users' binary offloading decisions and resource management while ensuring dynamic task data queue stability and minimum secure computing requirements. 

Beyond traditional static approaches, AI algorithms are revolutionizing resource management in UAV networks. Computation offloading, powered by algorithms like RL actor-critic network and LSTM in~\cite{zhu2020, mao2021optimizing}, dynamically shifts processing tasks from resource-constrained UAVs to the ground, dramatically reducing average response times and overall energy consumption. RL, in general, is widely employed for scheduling and resource management. For instance, onboard Q-learning can empower UAVs to directly manage resource allocation, maximizing user QoS metrics as in~\cite{cui2019}. Furthermore, DDPG-driven trajectory planning and frequency band allocation, as in~\cite{ding20203d}, prioritize energy-efficient operation, extending flight time and network coverage. Task-scheduling algorithms, exemplified by a DQN~\cite{yang2019}, enable real-time adjustments to UAV task strategies based on live data, optimizing resource utilization and mission fulfillment. AI algorithms like LSTM and RNN can dynamically estimate UAV-user throughput to ensure consistent QoS, as seen in~\cite{munaye2019deep, jiang2020learning}. Predictive beamforming algorithms, based on the same LSTM and RNN algorithms and presented in~\cite{liu2020location, yuan2020learning}, can precisely calculate optimal communication location and angle, focusing transmissions for efficient data delivery.  Finally, LSTM can also enable flexible control of transmission energy, further optimizing resource usage and network performance~\cite{chen2022deep}.
\vspace{-0.2cm}
\subsection{Summary and Lessons Learnt}\label{sec:uav_summary}
Integrating AI, particularly DRL and DL techniques, proves to be a game-changer for UAV communications. It overcomes challenges associated with dynamic environments and network constraints while optimizing positioning, deployment, channel estimation, interference management, autonomous navigation, trajectory, scheduling, and resource management for UAVs. The discussion on AI in UAV communications highlights its role in real-time adjustments based on live data for significant performance improvements.

DRL techniques are the main focus of this section, mostly due to their ability to improve dynamic control algorithms. DRL and DL algorithms contribute to predictive positioning, path planning, and trajectory optimization for UAV deployment. This minimizes the need for manual intervention while enhancing performance. Additionally, AI methods such as DNNs and RNNs improve accuracy in channel estimation and enable real-time processing, which is essential for adapting to dynamic UAV channels. Furthermore, AI-driven approaches like deep Q-learning and CNNs can optimize interference management, adapt to changing environments, and enhance network reliability. Autonomous navigation benefits from both DRL and DL by allowing UAVs to learn from data sources to optimize paths autonomously in response to dynamic situations. In scheduling and resource management domains, AI technologies like DRL empower UAV systems with improved schedules within dynamic environments while bolstering network efficiency.

AI has limitations when solving UAV communications-related challenges, especially in dynamic and unpredictable environments. While DRL shows promise in optimizing network performance, it still faces challenges in rapidly changing conditions. Due to unique obstacles, privacy, safety, and air traffic management also require specialized AI solutions. Beyond technical challenges, the lack of transparency inherent to many ML models raises concerns about interpretability and accountability in UAV communication systems. Ensuring clear and explainable decision-making processes is crucial for guaranteeing the safety and reliability of communication protocols in these critical applications. 

The future of AI in UAV communications holds exciting possibilities, but certain research directions and challenges need attention. One potential avenue for advancement is the exploration of SI algorithms in the context of UAV communications. Inspired by the collective behavior of biological systems, SI algorithms could offer innovative solutions for optimizing the coordination and communication patterns among multiple UAVs in dynamic environments. This approach may enhance UAV networks' adaptability, fault tolerance, and scalability. Moreover, developing hybrid AI methods and combining different AI techniques can also be a key focus for future research. These hybrid approaches leverage the strengths of different AI techniques and other emerging technologies, such as Blockchain and 6G, to enhance UAV communications capabilities further. FL is a promising approach that enables decentralized intelligence gathering and decision-making among multiple UAVs in SAGINs. Explainable AI and trustworthy systems are essential for ensuring transparency, interpretability, and ethical considerations in AI-driven UAV operations. Integration with edge computing, Blockchain, multi-modal perception, and human-AI collaboration can be identified as key research directions to enhance the capabilities of UAV communication systems. Open challenges include data security and privacy concerns associated with the increasing reliance on data collection in AI-powered UAV networks. Energy efficiency and resource management remain critical challenges, urging the development of energy-aware AI algorithms. Safety and reliability in dynamic environments, scalability, interoperability, and ethical considerations in regulation are key challenges that must be addressed to deploy AI-powered UAV networks responsibly.
\vspace{-0.2cm}
\section{AI for Optimizing Space-Air-Ground Integrated Networks}
\label{sec:optimizing}
\begin{table*}[htbp!]
\caption{Summary of AI-Aided SAGIN integration optimization solutions} 
\resizebox{\textwidth}{!}{%
\begin{tabular}{|llllll|}
\hline
\multicolumn{1}{|c|}{\textbf{Publication}} & \multicolumn{1}{c|}{\textbf{Year}} & \multicolumn{1}{c|}{\textbf{Objective}} & \multicolumn{1}{c|}{\textbf{AI type}} & \multicolumn{1}{c|}{\textbf{AI algorithm}} & \multicolumn{1}{c|}{\textbf{Performance metrics}} \\ \hline
\multicolumn{6}{|c|}{\textbf{Orchestration and Topology Management}} \\ \hline
\multicolumn{1}{|l|}{\cite{qiu2019}} & \multicolumn{1}{l|}{2019} & \multicolumn{1}{l|}{\begin{tabular}[c]{@{}l@{}}Dynamic management of networking,\\ caching, and computing resources\end{tabular}} & \multicolumn{1}{l|}{Reinforcement Learning} & \multicolumn{1}{l|}{Deep Q-learning} & Expected utility per resource \\ \hline
\multicolumn{1}{|l|}{\cite{lee2020integrating}} & \multicolumn{1}{l|}{2020} & \multicolumn{1}{l|}{\begin{tabular}[c]{@{}l@{}}Optimizing Satellite-HAPS network \\ topology\end{tabular}} & \multicolumn{1}{l|}{Reinforcement Learning} & \multicolumn{1}{l|}{Deep Q-Network} & \begin{tabular}[c]{@{}l@{}}End-to-end data rate, \\ convergence curve\end{tabular} \\ \hline
\multicolumn{1}{|l|}{\cite{han2020uav}} & \multicolumn{1}{l|}{2020} & \multicolumn{1}{l|}{\begin{tabular}[c]{@{}l@{}}Anti-jamming trajectory control for \\ UAVs in satellite-UAV coordination \\ networks\end{tabular}} & \multicolumn{1}{l|}{Reinforcement Learning} & \multicolumn{1}{l|}{Q-learning} & Utility and convergence \\ \hline
\multicolumn{1}{|l|}{\cite{han2022satellite}} & \multicolumn{1}{l|}{2022} & \multicolumn{1}{l|}{UAV trajectory control in SAGIN} & \multicolumn{1}{l|}{Reinforcement Learning} & \multicolumn{1}{l|}{Q-learning} & Simulation results \\ \hline
\multicolumn{1}{|l|}{\cite{wei2022satellite}} & \multicolumn{1}{l|}{2022} & \multicolumn{1}{l|}{\begin{tabular}[c]{@{}l@{}}UAV trajectory optimization for \\ IoT Information Collection\end{tabular}} & \multicolumn{1}{l|}{Reinforcement Learning} & \multicolumn{1}{l|}{Soft Actor-Critic} & Age of Information \\ \hline
\multicolumn{1}{|l|}{\cite{chen2022trajectory}} & \multicolumn{1}{l|}{2022} & \multicolumn{1}{l|}{\begin{tabular}[c]{@{}l@{}}UAV trajectory design and \\ UAV/Satellite link selection\end{tabular}} & \multicolumn{1}{l|}{Reinforcement Learning} & \multicolumn{1}{l|}{\begin{tabular}[c]{@{}l@{}}Graph Neural Network-\\ enhanced Q-learning\end{tabular}} & \begin{tabular}[c]{@{}l@{}}Number of served users and \\ minimum downlink rate constraint\end{tabular} \\ \hline
\multicolumn{1}{|l|}{\cite{saafi2022ai}} & \multicolumn{1}{l|}{2022} & \multicolumn{1}{l|}{\begin{tabular}[c]{@{}l@{}}Energy-centric topology management \\ and energy-efficient scheduling\end{tabular}} & \multicolumn{1}{l|}{Supervised Learning} & \multicolumn{1}{l|}{Long Short-Term Memory} & Packet delay, energy efficiency \\ \hline
\multicolumn{1}{|l|}{\cite{guo2023drl}} & \multicolumn{1}{l|}{2023} & \multicolumn{1}{l|}{\begin{tabular}[c]{@{}l@{}}Co-optimized performance of \\ RIS-assisted SAGINs\end{tabular}} & \multicolumn{1}{l|}{Reinforcement Learning} & \multicolumn{1}{l|}{\begin{tabular}[c]{@{}l@{}}Multi-Objective Deep \\ Deterministic Policy Gradient\end{tabular}} & \begin{tabular}[c]{@{}l@{}}System achievable rate and \\ UAV energy consumption\end{tabular} \\ \hline
\multicolumn{1}{|l|}{\cite{Cao2023Cooperative}} & \multicolumn{1}{l|}{2023} & \multicolumn{1}{l|}{\begin{tabular}[c]{@{}l@{}}Cooperative task offloading and \\ dispatching optimization\end{tabular}} & \multicolumn{1}{l|}{Reinforcement Learning} & \multicolumn{1}{l|}{Deep Q-learning} & \begin{tabular}[c]{@{}l@{}}Service capacity and UAV energy \\ consumption\end{tabular} \\ \hline
\multicolumn{1}{|l|}{\cite{Khoshkbari2023User}} & \multicolumn{1}{l|}{2023} & \multicolumn{1}{l|}{User association in SAGINs} & \multicolumn{1}{l|}{Reinforcement Learning} & \multicolumn{1}{l|}{Deep Q-learning} & User sum-rate \\ \hline
\multicolumn{1}{|l|}{\cite{Lee2023Integrating}} & \multicolumn{1}{l|}{2023} & \multicolumn{1}{l|}{\begin{tabular}[c]{@{}l@{}}UAV trajectories and UAV-Satellite \\ association optimization\end{tabular}} & \multicolumn{1}{l|}{Reinforcement Learning} & \multicolumn{1}{l|}{Multi-agent Actor-Critic} & \begin{tabular}[c]{@{}l@{}}End-to-end throughput, energy \\ consumption\end{tabular} \\ \hline
\multicolumn{1}{|l|}{\cite{arani2023hapsuav}} & \multicolumn{1}{l|}{2023} & \multicolumn{1}{l|}{\begin{tabular}[c]{@{}l@{}}UAV trajectory optimization and \\ UAV/HAPs-user channel allocation\end{tabular}} & \multicolumn{1}{l|}{Reinforcement Learning} & \multicolumn{1}{l|}{Deep Q-Network} & \begin{tabular}[c]{@{}l@{}}Uplink sum rate, number of served \\ users, uplink data rate\end{tabular} \\ \hline
\multicolumn{1}{|l|}{\cite{xie2023online}} & \multicolumn{1}{l|}{2023} & \multicolumn{1}{l|}{Access control optimization} & \multicolumn{1}{l|}{Reinforcement Learning} & \multicolumn{1}{l|}{\begin{tabular}[c]{@{}l@{}}Graph Neural \\ Network-enhanced \\ Policy-based agent\end{tabular}} & \begin{tabular}[c]{@{}l@{}}Total network revenue, \\ computational complexity\end{tabular} \\ \hline
\multicolumn{1}{|l|}{\cite{khoshkbari2023deep}} & \multicolumn{1}{l|}{2023} & \multicolumn{1}{l|}{\begin{tabular}[c]{@{}l@{}}User association in a Satellite-\\ HAPS-terrestrial network\end{tabular}} & \multicolumn{1}{l|}{Reinforcement Learning} & \multicolumn{1}{l|}{\begin{tabular}[c]{@{}l@{}}Long Short-Term Memory \\ Deep Q-Network\end{tabular}} & \begin{tabular}[c]{@{}l@{}}Sum-rate, convergence speed, \\ and fairness index\end{tabular} \\ \hline
\multicolumn{6}{|c|}{\textbf{Scheduling and Collaborative Resource Management}} \\ \hline
\multicolumn{1}{|l|}{\cite{liu2019}} & \multicolumn{1}{l|}{2019} & \multicolumn{1}{l|}{\begin{tabular}[c]{@{}l@{}}Antenna pointing and mobile \\ tracking\end{tabular}} & \multicolumn{1}{l|}{Reinforcement Learning} & \multicolumn{1}{l|}{Deep Q-Network} & Peaking signal strength \\ \hline
\multicolumn{1}{|l|}{\cite{jia2020}} & \multicolumn{1}{l|}{2020} & \multicolumn{1}{l|}{Intelligent spectrum sharing} & \multicolumn{1}{l|}{Supervised Learning} & \multicolumn{1}{l|}{\begin{tabular}[c]{@{}l@{}}Support Vector Machine and \\ Convolutional Neural Network\end{tabular}} & \begin{tabular}[c]{@{}l@{}}Interference, spectrum \\ efficiency\end{tabular} \\ \hline
\multicolumn{1}{|l|}{\cite{zhang2021space}} & \multicolumn{1}{l|}{2021} & \multicolumn{1}{l|}{Resource scheduling} & \multicolumn{1}{l|}{Reinforcement Learning} & \multicolumn{1}{l|}{\begin{tabular}[c]{@{}l@{}}Virtual network embedding \\ and Deep Q-Network\end{tabular}} & \begin{tabular}[c]{@{}l@{}}Long-term average revenue, \\ long-term revenue-cost ratio, \\ and virtual network request \\ acceptance rate\end{tabular} \\ \hline
\multicolumn{1}{|l|}{\cite{li2021novel}} & \multicolumn{1}{l|}{2021} & \multicolumn{1}{l|}{Radio resources scheduling} & \multicolumn{1}{l|}{Unsupervised Learning} & \multicolumn{1}{l|}{Genetic Algorithm} & \begin{tabular}[c]{@{}l@{}}Resource efficiency, successfully \\ scheduled tasks\end{tabular} \\ \hline
\multicolumn{1}{|l|}{\cite{liu2022machine}} & \multicolumn{1}{l|}{2022} & \multicolumn{1}{l|}{User scheduling} & \multicolumn{1}{l|}{Supervised Learning} & \multicolumn{1}{l|}{Deep Neural Network} & \begin{tabular}[c]{@{}l@{}}Sum-rate subject to \\ user-connectivity, backhaul, \\ and power constraints\end{tabular} \\ \hline
\multicolumn{1}{|l|}{\cite{zhang2022network}} & \multicolumn{1}{l|}{2022} & \multicolumn{1}{l|}{Network resource allocation} & \multicolumn{1}{l|}{Reinforcement Learning} & \multicolumn{1}{l|}{Deep Q-Network} & \begin{tabular}[c]{@{}l@{}}Long-term average reward, \\ acceptance ratio, long-term \\ reward/cost\end{tabular} \\ \hline
\multicolumn{1}{|l|}{\cite{guven2022}} & \multicolumn{1}{l|}{2022} & \multicolumn{1}{l|}{\begin{tabular}[c]{@{}l@{}}Channel estimation and \\ synchronization\end{tabular}} & \multicolumn{1}{l|}{Supervised Learning} & \multicolumn{1}{l|}{Convolutional Neural Network} & Data rate throughput, service quality \\ \hline
\multicolumn{1}{|l|}{\cite{Wu2023Joint}} & \multicolumn{1}{l|}{2023} & \multicolumn{1}{l|}{\begin{tabular}[c]{@{}l@{}}Joint beamforming and RIS phase \\ shift optimization design\end{tabular}} & \multicolumn{1}{l|}{Reinforcement Learning} & \multicolumn{1}{l|}{\begin{tabular}[c]{@{}l@{}}Assymetric Long Short-Term \\ Memory-Deep Deterministic \\ Policy Gradient\end{tabular}} & \begin{tabular}[c]{@{}l@{}}System sum rate, energy \\ consumption, age of information\end{tabular} \\ \hline
\multicolumn{1}{|l|}{\cite{dahrouj2023machine}} & \multicolumn{1}{l|}{2023} & \multicolumn{1}{l|}{User scheduling optimization} & \multicolumn{1}{l|}{Unsupervised Learning} & \multicolumn{1}{l|}{Ensembling Deep Neural Network} & \begin{tabular}[c]{@{}l@{}}Sum-rate performance, time \\ complexity, training time\end{tabular} \\ \hline
\multicolumn{1}{|l|}{\cite{seid2023hdfrl}} & \multicolumn{1}{l|}{2023} & \multicolumn{1}{l|}{\begin{tabular}[c]{@{}l@{}}Energy-efficient resource \\ allocation\end{tabular}} & \multicolumn{1}{l|}{Reinforcement Learning} & \multicolumn{1}{l|}{\begin{tabular}[c]{@{}l@{}}Hierarchical Deep Federated \\ Reinforcement Learning\end{tabular}} & \begin{tabular}[c]{@{}l@{}}Overall energy consumption, \\ system reward, convergence speed\end{tabular} \\ \hline
\multicolumn{1}{|l|}{\cite{alamgir2023fixed}} & \multicolumn{1}{l|}{2023} & \multicolumn{1}{l|}{UAV-based network analysis} & \multicolumn{1}{l|}{Reinforcement Learning} & \multicolumn{1}{l|}{Multi-armed bandit} & \begin{tabular}[c]{@{}l@{}}Received data, coverage and \\ data rate\end{tabular} \\ \hline
\multicolumn{6}{|c|}{\textbf{Routing and Flexible Mobility Management}} \\ \hline
\multicolumn{1}{|l|}{\cite{kato2019optimizing}} & \multicolumn{1}{l|}{2019} & \multicolumn{1}{l|}{Routing strategy} & \multicolumn{1}{l|}{Supervised Learning} & \multicolumn{1}{l|}{Convolutional Neural Network} & \begin{tabular}[c]{@{}l@{}}Network throughput, packet \\ loss rate\end{tabular} \\ \hline
\multicolumn{1}{|l|}{\cite{zhang2020}} & \multicolumn{1}{l|}{2020} & \multicolumn{1}{l|}{Handover management} & \multicolumn{1}{l|}{Supervised Learning} & \multicolumn{1}{l|}{Convolutional Neural Network} & Number of required handovers \\ \hline
\end{tabular}%
}
\label{tab:SAGIN1}
\vspace{-0.3cm}
\end{table*}

\begin{table*}[htbp!]
\caption{Summary of AI-Aided SAGIN integration optimization solutions (Continued)} 
\resizebox{\textwidth}{!}{%
\begin{tabular}{|llllll|}
\hline
\multicolumn{1}{|c|}{\textbf{Publication}} & \multicolumn{1}{c|}{\textbf{Year}} & \multicolumn{1}{c|}{\textbf{Objective}} & \multicolumn{1}{c|}{\textbf{AI type}} & \multicolumn{1}{c|}{\textbf{AI algorithm}} & \multicolumn{1}{c|}{\textbf{Performance metrics}} \\ \hline
\multicolumn{6}{|c|}{\textbf{Caching and Computation Offloading}} \\ \hline
\multicolumn{1}{|l|}{\cite{masood2021content}} & \multicolumn{1}{l|}{2021} & \multicolumn{1}{l|}{Content caching prediction} & \multicolumn{1}{l|}{Supervised Learning} & \multicolumn{1}{l|}{Deep Neural Network} & Prediction accuracy \\ \hline
\multicolumn{1}{|l|}{\cite{mao2021optimizing}} & \multicolumn{1}{l|}{2021} & \multicolumn{1}{l|}{\begin{tabular}[c]{@{}l@{}}Computation offloading in \\ Satellite-UAV networks\end{tabular}} & \multicolumn{1}{l|}{Supervised Learning} & \multicolumn{1}{l|}{Long Short-Term Memory} & \begin{tabular}[c]{@{}l@{}}Task success ratio and \\ system computation rate\end{tabular} \\ \hline
\multicolumn{1}{|l|}{\cite{jaiswal2022novel}} & \multicolumn{1}{l|}{2022} & \multicolumn{1}{l|}{\begin{tabular}[c]{@{}l@{}}Computation offloading in\\ Satellite-UAV networks\end{tabular}} & \multicolumn{1}{l|}{Supervised Learning} & \multicolumn{1}{l|}{Long Short-Term Memory} & \begin{tabular}[c]{@{}l@{}}Task success ratio and \\ system computation rate\end{tabular} \\ \hline
\multicolumn{1}{|l|}{\cite{Nguyen2022Multi}} & \multicolumn{1}{l|}{2022} & \multicolumn{1}{l|}{Task offloading} & \multicolumn{1}{l|}{Reinforcement Learning} & \multicolumn{1}{l|}{\begin{tabular}[c]{@{}l@{}}Multi-agent Deep Deterministic \\ Policy Gradient\end{tabular}} & \begin{tabular}[c]{@{}l@{}}Energy consumption and \\ task execution delay\end{tabular} \\ \hline
\multicolumn{1}{|l|}{\cite{Zhang2023Multi}} & \multicolumn{1}{l|}{2023} & \multicolumn{1}{l|}{Multiagent Task offloading} & \multicolumn{1}{l|}{Reinforcement Learning} & \multicolumn{1}{l|}{Multiagent Actor-Critic} & \begin{tabular}[c]{@{}l@{}}Task completion within delay \\ constraints, energy utilization, \\ and robustness\end{tabular} \\ \hline
\multicolumn{1}{|l|}{\cite{Liu2023Energy}} & \multicolumn{1}{l|}{2023} & \multicolumn{1}{l|}{\begin{tabular}[c]{@{}l@{}}Energy-efficient IoT task \\ offloading\end{tabular}} & \multicolumn{1}{l|}{Reinforcement Learning} & \multicolumn{1}{l|}{\begin{tabular}[c]{@{}l@{}}Deep Reinforcement Learning \\ Federated Deep Q-Network\end{tabular}} & \begin{tabular}[c]{@{}l@{}}Energy consumption, task \\ processing delay\end{tabular} \\ \hline
\multicolumn{1}{|l|}{\cite{lakew2023intelligent}} & \multicolumn{1}{l|}{2023} & \multicolumn{1}{l|}{Offloading and resource allocation} & \multicolumn{1}{l|}{Reinforcement Learning} & \multicolumn{1}{l|}{\begin{tabular}[c]{@{}l@{}}Multiagent Deep Deterministic \\ Policy-Gradient\end{tabular}} & \begin{tabular}[c]{@{}l@{}}Service satisfaction, energy \\ consumption\end{tabular} \\ \hline
\multicolumn{1}{|l|}{\cite{gong2023computation}} & \multicolumn{1}{l|}{2023} & \multicolumn{1}{l|}{\begin{tabular}[c]{@{}l@{}}Computation offloading in \\ Satellite-HAPS networks\end{tabular}} & \multicolumn{1}{l|}{Reinforcement Learning} & \multicolumn{1}{l|}{\begin{tabular}[c]{@{}l@{}}Lyapunov-guided Multi-Agent \\ Proximal Policy Optimization\end{tabular}} & \begin{tabular}[c]{@{}l@{}}Convergence speed, average \\ sum-rate, battery backup level\end{tabular} \\ \hline
\multicolumn{1}{|l|}{\cite{zhang2023learning}} & \multicolumn{1}{l|}{2023} & \multicolumn{1}{l|}{\begin{tabular}[c]{@{}l@{}}Hybrid offloading policy for IoT \\ devices\end{tabular}} & \multicolumn{1}{l|}{Reinforcement Learning} & \multicolumn{1}{l|}{Deep Q-learning} & \begin{tabular}[c]{@{}l@{}}Throughput performance and \\ outage performance\end{tabular} \\ \hline
\end{tabular}%
}
\label{tab:SAGIN2}
\vspace{-0.3cm}
\end{table*}
SAGINs, which consist of satellites, HAPS, UAVs, and terrestrial networks, represent one of the most complex network architectures globally. These components collaborate to deliver enhanced and adaptable end-to-end user services within the framework of the new 6G paradigm. Even though SAGINs are designed to attain uninterrupted coverage across all areas at all times while facilitating high-rate and dependable transmission over a larger area compared to traditional terrestrial networks~\cite{ye2020outage}, some performance degradation is inevitable unless optimization takes place. While each component of the SAGIN architecture has the potential to optimize its performance individually, it is not sufficient to optimize them separately. Joint optimization of these layers is essential to fully exploit the potential of SAGINs~\cite{kato2019optimizing}. Integrating these components necessitates carefully considering factors such as orchestration and topology management, scheduling and collaborative resource management, routing and flexible mobility management, and caching and computation offloading. Tables~\ref{tab:SAGIN1} and~\ref{tab:SAGIN2} provide a comprehensive summary of studies that have leveraged AI methodologies to tackle the challenges associated with optimizing SAGINs.

\subsection{Orchestration and Topology Management}\label{sec:integrated_topology}
Due to the distributed nature of SAGINs, topology control is a critical aspect that significantly impacts the overall network performance. Distributed topology management schemes can minimize bottlenecks and response time but increase network complexity due to cooperative operation across different SAGIN components. On the other hand, centralized control can simplify the network layout but may lead to substantial response delays, further degrading the network performance. Additionally, SAGIN's extremely dynamic network topology results in increased link changes, making data transmission via fixed traffic impossible. Thus, topology management for SAGINs requires careful consideration of user association and aerial node trajectory optimization levels~\cite{giordani_2021}. User association is crucial in determining the connections between users and network segments, influencing overall network topology and resource allocation. This has a substantial impact on network performance and efficiency. Simultaneously, optimizing the flight paths of aerial platforms, such as UAVs, enhances coverage and connectivity, reduces congestion, and minimizes energy consumption. Moreover, this approach ensures comprehensive networking adaptability to dynamic environmental conditions. Realizing a truly responsive and versatile network in SAGINs requires a reconfigurable architecture capable of intelligent resource allocation based on situational awareness. This architectural design should enable SAGINs to rapidly activate satellite services, optimize routing paths for long-distance communication needs, and dynamically dispatch UAVs according to the demands of emergency situations. Traditional optimization tools may be the first choice to formulate architectural design and topology problems. For example, the work in~\cite{li2020risassisted} proposes a reconfigurable intelligent surface (RIS)-assisted UAV communication system to maximize the average achievable rate by jointly designing UAV trajectory and passive beamforming using successive convex approximation. Another paper~\cite{alsharoa2020improvement} addresses optimizing HAPS locations and user association to maximize users' throughput by employing binary linear optimization.

However, the inherent trade-off between data complexity and optimization efficiency in SAGINs presents a major obstacle for conventional topology and orchestration management tools. Thankfully, advancements in AI, particularly ML and DNNs, are paving the way for a paradigm shift. These data-driven methods offer robust and scalable network and topology optimization solutions, revolutionizing how we manage complex SAGIN infrastructures. For instance, the study conducted in~\cite{qiu2019} demonstrates how deep Q-learning can establish a framework for dynamically managing satellite-terrestrial networks while controlling caching and computing resources. Deep Q-learning is also applied in~\cite{Khoshkbari2023User}, which addresses user association in SAGINs with delayed CSI to maximize the sum rate. The authors utilize deep Q-learning to enable the satellite controller to decide whether a user should be associated with a HAPS or a terrestrial BS based on the network's state and achieve higher performance than traditional methods. Additionally, another paper~\cite{Cao2023Cooperative} proposes task-dynamic processing through a multi-UAV cooperative strategy for optimizing UAV deployment and allocating computing resources by offloading tasks to other UAVs and HAPS. The challenge addressed is UAVs' limited computing resources and energy, which are resolved using the deep Q-learning algorithm for optimized task offloading while enhancing service capacity and reducing energy consumption. In~\cite{han2020uav}, the authors present a trajectory control approach based on Q-learning to tackle the difficulties posed by the unpredictable and uncertain environment and the presence of malicious jamming in satellite-UAV coordination networks. Focusing on optimizing the operation of hybrid satellite-terrestrial networks augmented by UAVs,~\cite{chen2022trajectory} explores efficient trajectory design and link selection strategies. It addresses this challenge through Q-learning to adjust UAV locations and determine optimal link selection using a graph NN. Similarly, another publication~\cite{han2022satellite} deals with optimizing multi-satellite association and multi-UAV trajectories in time-varying non-terrestrial network topologies. It employs multi-agent Q-learning to integrate satellite networks efficiently with UAV relaying, aiming to maximize system throughput while minimizing energy consumption.

DQNs are widespread in addressing challenges such as trajectory optimization, link association, and topology management of SAGINs. For example, the authors in~\cite{khoshkbari2023deep} propose a DQN-based method to tackle the partially observable user association problem in a satellite-HAPS-terrestrial network to maximize the sum rate while reducing CSI overhead. The proposed algorithm outperforms existing approaches and underscores the importance of incorporating action vectors in the policy network. Additionally, another publication~\cite{arani2023hapsuav} addresses SAGIN optimization considering non-terrestrial network complexity, mobility, and heterogeneity by utilizing the DQN algorithm for UAV trajectories and resource management optimization to maximize HAPS-UAV-terrestrial network performance as well as user satisfaction.

Similarly, an approach based on policy-based DRL is employed in~\cite{xie2023online}. This study focuses on optimizing access control in aerial networks involving satellites and HAPS to maximize total network revenue while minimizing computational complexity. Another publication~\cite{guo2023drl} tackles optimizing the performance of RIS-assisted satellite-UAV-terrestrial networks. The method proposed utilizes multi-objective DDPG for joint optimization of UAV trajectory, RIS configuration, and downlink beamforming to maximize the system's achievable rate and minimize UAV energy consumption. The work done in~\cite{Lee2023Integrating} discusses optimizing association and trajectories in a time-varying topology for satellite-UAV networks. It employs a multi-agent actor-critic network to enhance communication efficiency while reducing energy consumption, resulting in notable performance enhancements. Another study by~\cite{wei2022satellite} focuses on minimizing the age of information in IoT data collection using SAC to optimize UAV trajectories and address the challenge of information freshness in wide-area IoT networks. Finally, LSTM can be supplied with different user profile information (position, power availability, and energy usage) as well as configuration data for aerial-terrestrial-maritime networks to enhance the network and decrease packet delay while enhancing energy effectiveness~\cite{saafi2022ai}. 

\subsection{Scheduling and Collaborative Resource Management}\label{sec:integrated_scheduling}
SAGINs present unique challenges compared to terrestrial networks, primarily due to their constantly shifting topology, diverse network elements, and intricate relational patterns within the network architecture. Additionally, the dynamic resource constraints imposed by the high mobility of SAGIN components necessitate flexible and adaptive resource management strategies. In addition, SAGINs allow for different services requiring different requirements and needs. Thus, intelligent and collaborative resource scheduling is the key technology for maximizing network performance and application responsiveness, considering the dynamic network topology, varying communication ranges, and heterogeneous resource capabilities of SAGIN systems~\cite{zhou2023}. Sub-optimal scheduling by focusing on only parts of the SAGIN architecture may cause data bottlenecks and congested links, affecting critical applications such as remote surgery or emergency response. Additionally, inefficient use of resources can lead to wasting valuable bandwidth and processing power while struggling to meet demand in certain areas. Lastly, frequent switching between resources and unnecessary data relays can deplete limited onboard power, especially for UAV operations.

Classical SAGIN scheduling and resource management approaches rely on rule-based algorithms and static optimization techniques. These methods often struggle with the dynamic nature of SAGINs, leading to sub-optimal solutions. For instance, in~\cite{cao2022edge}, the angle-based diversity strategy is utilized in priority-based and load-balancing scheduling algorithms that use software-defined networking and network function virtualization as a foundation. Another work~\cite{li2019energy} utilizes a progressive convex approximation to enhance the system's energy efficiency through combined optimization of sub-channel selection, uplink transmission power control, and UAV relay deployment in SAGINs. These classical approaches often face computational complexity, especially for large networks or scenarios, and limited scalability; they rely on accurate channel models. On the other hand, AI models can learn from network behavior to dynamically predict traffic patterns, enabling proactive scheduling. Also, DRL algorithms can interact with the environment in real-time without any models or historical data. Thus, leveraging AI algorithms, service-driven resource management paradigms emerge as a powerful tool for SAGINs. These intelligent and collaborative strategies allow real-time adaptation to dynamically evolving service demands, ensuring flexible and efficient management of SAGIN's diverse network offerings.

Many proposed AI-based methods heavily rely on supervised learning techniques. For instance, the work in~\cite{liu2022machine} utilizes a DNN to carry out user scheduling in multiple layers of SAGIN with a specific emphasis on HAPS. Furthermore, intelligent spectrum sharing in satellite-terrestrial networks can be achieved using supervised learning algorithms such as a combination of SVM and CNN algorithms~\cite{jia2020}. These methods are utilized for making spectrum predictions based on wideband and narrow-band measurements, demonstrating higher spectrum management efficiency when used together. Yet another supervised learning approach is suggested to tackle the channel estimation and synchronization issues within a HAPS-LEO network~\cite{guven2022}. Here, CNN-based estimators are employed to minimize channel equalization and carrier frequency offset, resulting in improved data rate throughput per second and superior service quality due to an agile signal reconstruction process. In~\cite{dahrouj2023machine}, the authors optimize user scheduling in satellite-HAPS-terrestrial networks, focusing on the complexity and stochastic nature of these networks. The adopted approach employs unsupervised learning, particularly ensembling DNNs, to achieve higher sum-rate performance with reduced computational complexity, tackling the real-time processing challenges in these networks. Another study~\cite{li2021novel} introduces an innovative approach for cross-network radio resource scheduling across all layers of SAGINs. This research contributes to the field by addressing the issues related to unified mapping and scheduling of radio resources within SAGINs. Utilizing a GA and a unified resource mapping method, this proposed scheme optimizes resource allocation and improves the performance of SAGINs, specifically in emergency logistics scenarios.

Finally, DRL is also extensively utilized to address the challenge of scheduling and collaboratively optimizing SAGINs' resources. For instance, the authors in~\cite{zhang2022network} address the SAGIN resource allocation issue using the deep Q-learning approach. Antenna pointing, mobile tracking, and resource scheduling can also be addressed with DQN, as demonstrated in~\cite{liu2019}. A similar study is conducted in~\cite{zhang2021space}, which formulates the resource scheduling problem of SAGIN as a multi-domain virtual network embedding. They leverage DQN to improve the performance of the multi-domain virtual network embedding algorithm by employing basic modules of NN to construct a five-layer policy network. Another paper~\cite{Wu2023Joint} tackles the issue of enhancing RIS-assisted hybrid free space optics/radio frequency-enabled HAPS-UAV-terrestrial networks. It uses an asymmetric LSTM-DDPG algorithm to optimize active and passive beamforming techniques simultaneously, intending to improve system performance in complex and dynamic network environments. Another study~\cite{seid2023hdfrl} focuses on energy-efficient resource allocation and privacy preservation of offloaded tasks in HAPS-UAV-enabled MEC systems. It employs hierarchical deep federated RL to optimize resource allocation and decision-making, addressing the challenge of minimizing energy consumption while meeting diverse task requirements in dynamic environments. The optimization of beam selection for UAV- and HAPS-based networks to enhance connectivity for ground vehicles in 5G systems through RL is explored in~\cite{alamgir2023fixed}. This work addresses fluctuating traffic and environmental variations using the multi-armed bandit algorithm to autonomously select optimal beams over time.

\subsection{Routing and Flexible Mobility Management}\label{sec:integrated_routing}
Routing and mobility management in SAGINs pose unique challenges due to the heterogeneous nature of network components and their mobility. Adaptive routing protocols are required to handle dynamic topology and energy constraints, while seamless mobility and handover mitigation strategies must be developed to maintain uninterrupted communication during the continuous movement of satellites and aerial nodes~\cite{AlMuallim2023Handover}. Furthermore, the expanding scale and diverse node types within SAGINs demand innovative solutions for inter-layer mobility management. While ultra-dense satellite constellations offer additional capacity, they also introduce intricate handover dynamics with increased overhead and potential delays. Navigating these complexities and maintaining consistent performance, routing and mobility protocols require significant advancements in decision-making speed, resource efficiency, and intelligent adaptation to network conditions. The traditional methods used to address the routing and mobility management challenges of SAGINs, like rule-based routing algorithms, do not have the adaptability to handle the dynamic characteristics of SAGINs and often produce less than optimal routes~\cite{qu2020lbmre}. Pre-determined handover triggers based only on signal strength may not completely capture the context of network conditions, resulting in unnecessary handovers~\cite{ren2023novel}. However, AI-enabled solutions, including ML and DRL, can optimize routing and handover decisions based on real-time network conditions and environmental factors, ensuring seamless and reliable SAGIN communication. For instance, CNN is utilized in~\cite{kato2019optimizing} to address the routing issue. The objective is for CNN to learn traffic patterns within multiple SAGIN layers and use this information to optimize routing paths. Similarly, the work in~\cite{zhang2020} also employs CNN technology to tackle handover management challenges in satellite-terrestrial networks. Reference signal received power (RSRP) values are inputted into the CNN model to learn sub-optimal handover decisions.

\subsection{Caching and Computation Offloading}\label{sec:intgrated_caching}
The heterogeneous platforms that constitute the SAGIN architecture, including satellites, HAPS, UAVs, and ground stations, have different storage capacities, processing power, and network bandwidth. This dynamic environment requires smart strategies for caching content and offloading computation to maintain application responsiveness while optimizing network efficiency. Traditional caching and offloading methods frequently do not meet the needs of SAGINs, which require collaboration and organization of caching and offloading decisions across different layers. Fixed caching strategies have difficulty adjusting to changing traffic patterns, resulting in unnecessary data transfers and higher latency~\cite{yuan2023joint,zhou2022joint}. Centralized data offloading techniques lead to congestion and bottlenecks~\cite{traspadini2022uav}. In addition, constraints in onboard storage and processing capabilities for platforms such as UAVs require resource-conscious offloading approaches. AI becomes a potent solution to provide collaborative caching and computation offloading schemes to SAGINs. AI-based caching and computation offloading methods can analyze network patterns in real-time, allowing for proactive resource allocation and adaptive adjustments. These algorithms can forecast user needs and content popularity, strategically storing data throughout the different network layers to shorten data retrieval times and decrease backhaul traffic. This may entail placing frequently accessed data nearer to users on platforms such as UAVs or edge servers, thereby reducing latency and enhancing user satisfaction. Additionally, AI models can predict computational demands and preemptively transfer tasks to the most appropriate platforms based on their available resources and prevailing network conditions.

Many studies in the literature use DRL for the optimization of caching and computation offloading in SAGINs. For instance, the authors in~\cite{Zhang2023Multi} tackle the issue of optimizing task offloading decisions made by UAVs in satellite-UAVs-assisted IoT networks. The proposed scheme applies the multi-agent actor-critic approach to empower UAVs to learn efficient offloading strategies, maximizing task completion while minimizing energy usage. Another study~\cite{Liu2023Energy} discusses the issue of enhancing energy-efficient task offloading from distant IoT settings to satellites or UAVs by using federated DQNs for immediate offloading decisions and reducing energy usage while considering task processing delay. A comparable approach based on deep Q-learning is utilized in another work~\cite{zhang2023learning} to tackle the problem of maximizing utility efficiency in hybrid offloading for IoT devices within satellite-UAV-assisted networks. In addition to DRLs, policy-based DRL methods are also extensively used in the literature. For instance, the authors in~\cite{Nguyen2022Multi} address the issue of task offloading to satellites and UAVs in hierarchical SAGINs, employing multi-agent DDPG networks to minimize energy consumption and task execution delay for IoT devices. Another study~\cite{lakew2023intelligent} addresses the challenge of task offloading in HAPS-UAV-terrestrial networks, utilizing a multi-agent DDPG approach to maximize service satisfaction while minimizing energy consumption for IoT in underprivileged areas. Another publication~\cite{gong2023computation} confronts the challenge of maximizing the sum rate of terrestrial users in dynamic satellite-HAPS network environments by using multi-agent PPO to optimize computation offloading and resource allocation, resulting in improved performance and battery backup management.

Supervised learning models can also be beneficial in optimizing the decision-making for collaborative caching and computation offloading in SAGINs. For instance, the work in~\cite{jaiswal2022novel} deals with the problem of energy-efficient computation offloading in satellite-UAV-based IoT systems, utilizing the LSTM model to enhance task success rate and energy dynamics for efficient offloading. Another study~\cite{mao2021optimizing} addresses the optimization of computation offloading in satellite-UAV-integrated IoT systems, employing LSTM to boost the completion rate of tasks delivered to edge devices while simultaneously improving the overall processing capacity of the system. The research in~\cite{masood2021content} focuses on decreasing content access delay in HAPS-assisted multi-UAV networks by predicting content popularity. In this paper, DNNs are used to predict popular content accurately for caching, thereby minimizing content access delay or reduction. 

\subsection{Summary and Lessons Learnt}\label{sec:integrated_summary}
SAGINs pose a range of complex challenges due to their diverse composition. They consist of satellites, HAPS, UAVs, and terrestrial networks, requiring dynamic optimization across multiple layers. These challenges involve dealing with a constantly changing topology influenced by the mobility of aerial platforms, handling potential link fluctuations, and managing constraints like limited storage, processing power, and network bandwidth. Quick decision-making is crucial in this scenario, as network conditions and user requirements can change rapidly. This calls for algorithms capable of learning and adapting in real-time.

AI-based solutions, including DRL and supervised learning methods, are becoming increasingly important. Traditional approaches face challenges in dealing with the complex dynamics of SAGINs, making AI essential for improving network performance and user experience. Additionally, it becomes clear that joint optimization is significant— isolated optimization of individual SAGIN components is not enough. Optimizing topology, scheduling, resource allocation, routing, and mobility collectively improves overall performance. Real-time learning and adaptation become fundamental principles, with AI algorithms constantly learning from network data to adapt strategies in response to changing conditions. Moreover, considering the limited resources on aerial platforms, efficient solutions are essential for caching, computation offloading, and scheduling. Acknowledging the diversity of network components, with optimization algorithms customized to accommodate different capabilities and constraints, is crucial.

To drive the advancement of SAGINs, there are several important areas for research and development. This includes a focus on enhancing AI algorithms by exploring more advanced DRL and supervised learning techniques tailored specifically for SAGINs. For instance, the sheer volume of data inherent in SAGIN mobility management poses a considerable obstacle to traditional techniques. Tensor-based approaches and learning algorithms represent a compelling direction, potentially enabling data compression with techniques like Tucker decomposition and accelerating convergence through learning algorithms. However, bridging the gap between intelligent decision-making and its practical application in multi-layered, large-scale SAGINs with aerial components remains an open research question. Further exploration and experimentation in this area is crucial. Furthermore, harnessing the potential of emerging technologies like edge computing and network function, virtualization offers promising avenues for improved performance and enhanced flexibility in SAGIN mobility management. Standardization efforts to establish protocols are crucial for seamless collaboration between different SAGIN components. Addressing security and privacy concerns is vital for robust protection against vulnerabilities while safeguarding user privacy within AI-powered SAGINs.

\section{SAGINs for Artificial Intelligence}
\label{sec:overlay}
\begin{figure*}
    \centering
    \includegraphics[width=0.9\textwidth]{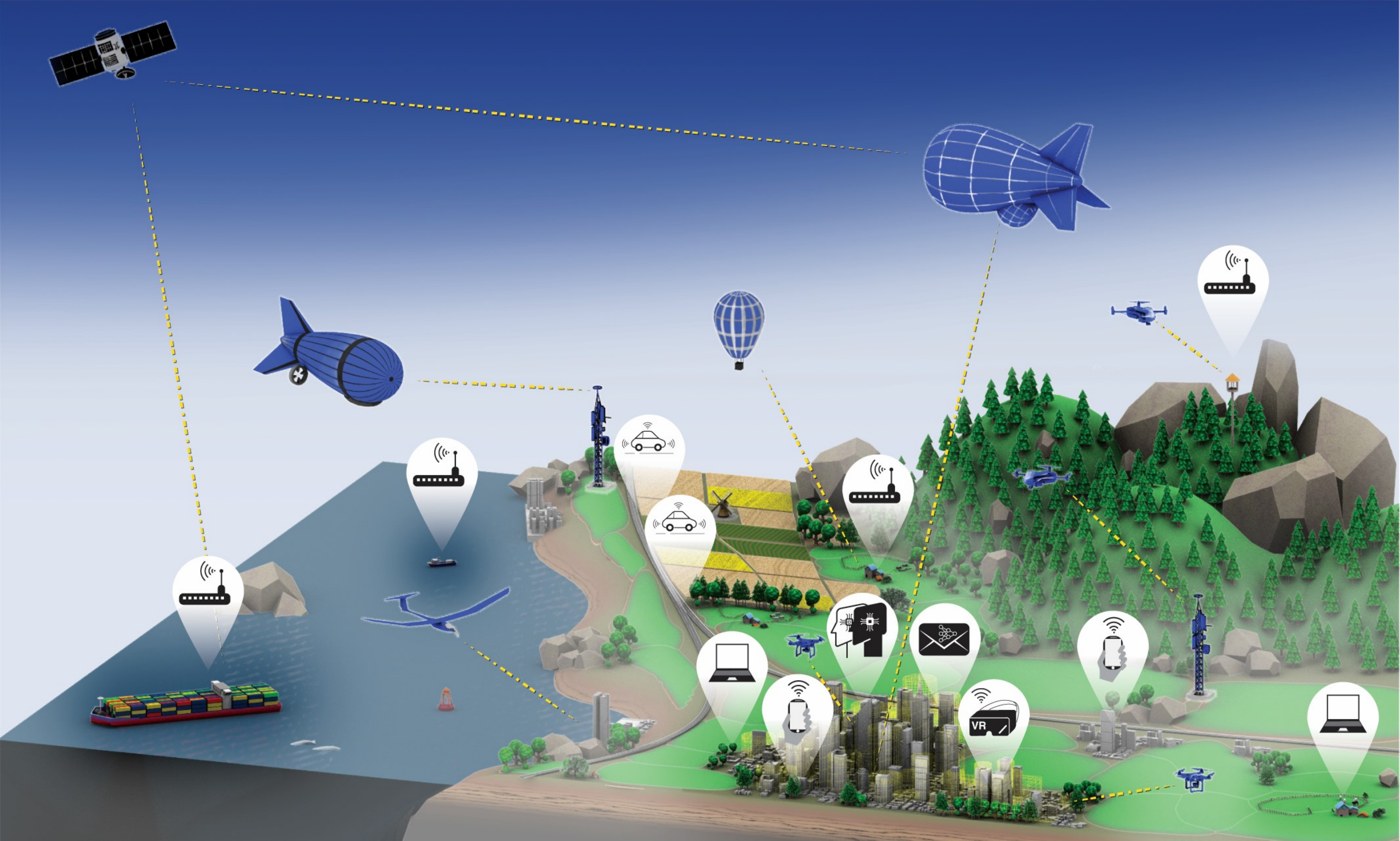}
    \caption{SAGIN-aided AI applications: Internet of Things, Autonomous vehicles, Virtual/Augmented reality, Digital twins, and Semantic communications. }
    \label{fig:large_fig}
\end{figure*}
The evolution of SAGINs and AI are intrinsically linked. While AI algorithms are crucial for tackling current limitations in SAGIN deployment and functionality, effectively managing these complex networks relies on intelligent orchestration and optimization tools. Conversely, SAGINs provide a unique platform for AI advancement. Their distributed architecture facilitates edge computing and data aggregation capabilities, fostering wider device connectivity and richer data collection. This, in turn, fuels cutting-edge data analytics and AI development, further empowering SAGINs for improved performance and adaptability. In this section, we shed light on the synergy between AI and SAGINs while focusing mainly on how SAGINs can accelerate the development of efficient AI algorithms. Specifically, we focus on FL applications and the benefits of SAGINs in allowing more devices to contribute to the learning of FL. Furthermore, we highlight some wireless techniques for optimized AI systems, i.e., analog over-the-air computation and digital RRM optimized for AI. Finally, we highlight the main SAGIN-enabled new applications relying on AI, such as autonomous cars and vehicular networks, IoT, VR/AR, digital twins, and semantic communications. Fig.~\ref{fig:large_fig} illustrates a SAGIN incorporating the previously mentioned applications to serve user demands in different environments. 

\subsection{Motivation}\label{sec:overlay_motivation}
Terrestrial networks can provide the necessary infrastructure for AI development, such as high-speed internet access, fast transfer to reliable data storage capabilities, and access to computer processing power. Additionally, they can provide secure connections to help protect sensitive data from malicious attacks. These networks are also cost-effective solutions that allow businesses to save money on expensive hardware investments while providing flexibility for scaling or expanding operations. However, with the increasing popularity of mobile devices and the exponential growth of IoT applications, terrestrial networks are heavily overloaded with the amount of traffic generated. Furthermore, the widespread of new applications such as VR/AR are burdening the capacity of terrestrial wireless networks. Such limitations add more constraints and halt the full benefit of terrestrial wireless networks for AI applications.  

SAGINs, on the other hand, hold immense promise in augmenting the capabilities of AI systems and overcoming the limitations of terrestrial networks. Here, we delve into the benefits these networks bring to the field of AI:
\begin{itemize}
    \item \textbf{Enhanced Data Accessibility:} SAGINs provide AI developers with access to extensive real-world and near-real-time data from many sources. This data diversity enables the construction of more complex and diverse datasets, enriching AI training. AI algorithms, thus exposed to various scenarios and situations, become better equipped to identify patterns and make informed decisions, even in challenging or unfamiliar environments.
    \item \textbf{Scalability without Boundaries:} Unlike traditional terrestrial networks, SAGINs offer unparalleled scalability. These networks do not depend on physical infrastructure or wiring, making adding new nodes and expanding coverage relatively easy. This scalability is essential as AI applications grow in complexity and data demands.
    \item \textbf{Advanced Problem Solving:} SAGINs enable AI systems to work collaboratively by sharing data and resources. This collaborative approach allows AI systems to tackle complex problems from different angles and leverage each other's strengths. The result is more efficient problem-solving and higher-quality outcomes.
    \item \textbf{Near Real-Time Communication:} SAGINs facilitate faster and more efficient data transmission while considering network congestion and hotspots. This capability enables near real-time communication between interconnected AI systems, a critical requirement for applications demanding timely decision-making and responsiveness.
    \item \textbf{Innovation Catalyst:} SAGINs serve as catalysts for innovation, particularly in AI-driven technologies like smart cities and autonomous vehicles. These innovations heavily rely on AI capabilities for their operation and can benefit from the collaborative power and extensive data access facilitated by SAGINs.
    \item \textbf{Privacy-Preserving AI Development:} SAGINs can provide a secure platform for testing new AI technologies without compromising privacy or overloading terrestrial networks. Developers can explore AI's potential without exposing sensitive information, making them ideal environments for developing and testing data-intensive applications like autonomous vehicles and facial recognition software.
\end{itemize}
In summary, SAGINs offer a wealth of opportunities for AI development, from enriching datasets and enabling global collaboration to fostering innovation and facilitating privacy-preserving AI research. These networks are poised to play a pivotal role in shaping the future of AI applications across various domains. However, since AI algorithms were originally conceived in centralized settings where all data is aggregated at a single location, the large-scale nature of SAGINs and the consequently distributed datasets that are produced are creating new challenges for the application of AI in SAGINs. Namely, new security and privacy concerns must be addressed in the design of the wireless protocols. Furthermore, communication and energy efficiency are yet other limitations as AI algorithms require the communication of high-dimensional models for a high number of iterations, which can introduce a bottleneck on the capacity of the network and can drain the battery of energy-limited devices.

To address these challenges, new wireless methods must be designed to carry the data needed for AI tasks. These methods aim not to deliver bits as efficiently as possible but to distill the intelligence carried within the data; thus, they are more aligned with the AI objective. Among the new methods proposed to adapt SAGINs with AI objectives, two major solutions arise: analog over-the-air aggregation and RRM optimized for AI. The details of these methods are discussed in the next sections.

\subsection{SAGINs for Edge Intelligence and Federated Learning}\label{sec:overlay_FL}
Considering the demands of the advancing 6G technology, there is a growing emphasis on edge intelligence to help create the vision of ubiquitous intelligence~\cite{letaief2022}. Edge intelligence represents a promising approach to harnessing intelligence by gathering, processing, and analyzing the vast data traffic generated at the network's edge. One of the key prerequisites for these smart services is to maintain the privacy and efficiency of data exchange, which restricts the direct transmission or collection of raw data by users~\cite{letaief2022}.
\begin{figure}
    \centering
    \includegraphics[width=\linewidth]{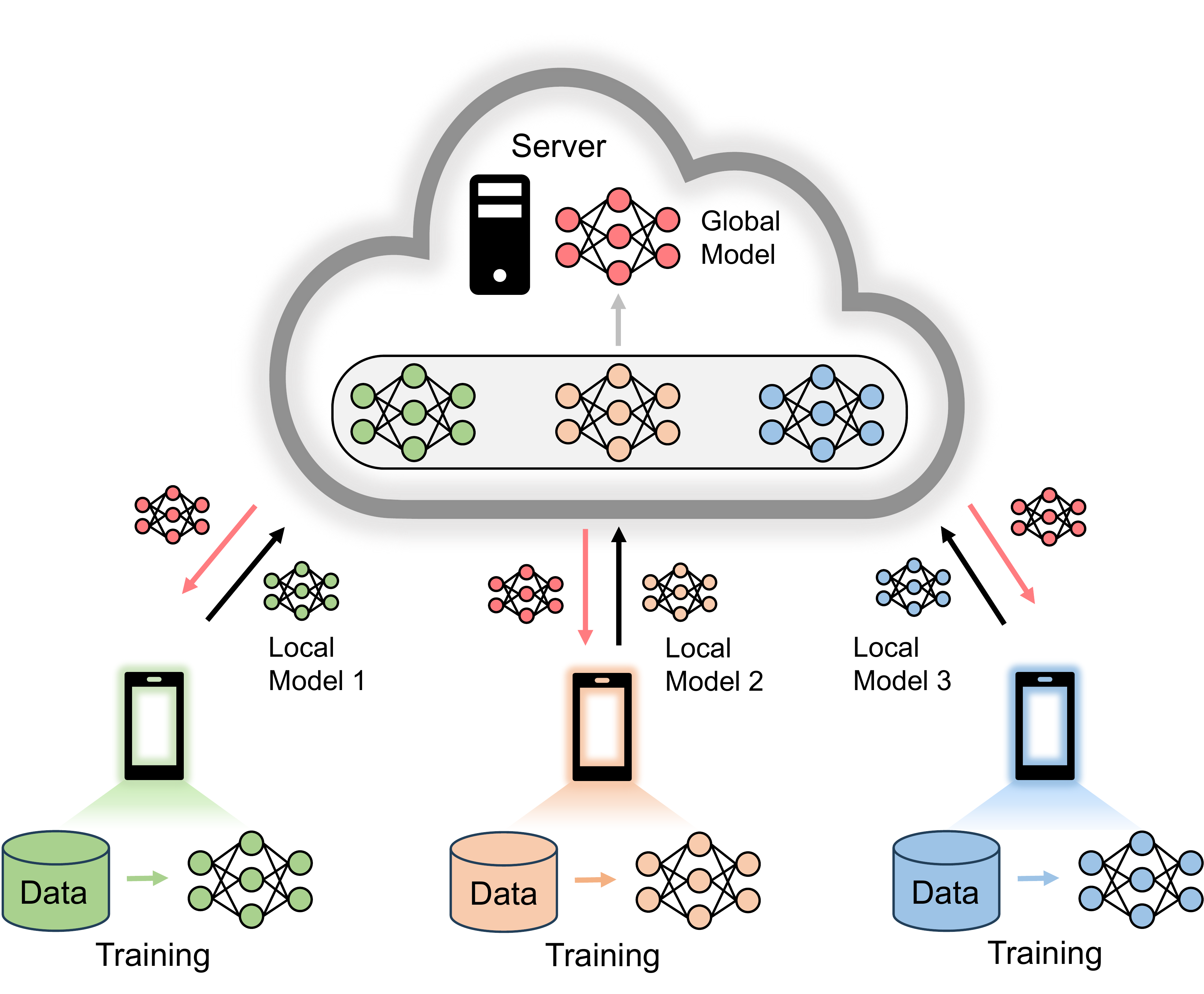}
    \caption{Sample FL scenario. Three devices collect local data, train the global model, and send the updated local models back to the server, where aggregation occurs.}
    \label{fig:fedlearn}
\end{figure}

FL has emerged as an increasingly promising ML paradigm suited to address these challenges. FL enables multiple devices to train ML models collaboratively without sharing raw data~\cite{zhao2018federated}. Each device trains a local model using its respective data in FL and transmits the updated model parameters to a central server for global model aggregation. Fig.~\ref{fig:fedlearn} illustrates the FL algorithm where a set of devices participate in the FL and send their local updates to the global server. FL stands out for its ability to provide robust security, reduce network congestion, and optimize energy consumption by integrating training capabilities throughout the network nodes~\cite{Lim2020Federated}. Consequently, when viewed from this perspective, deploying FL at the edge represents a viable paradigm to realize the much-needed pervasive edge intelligence in the context of 6G.

Terrestrial networks have been used conventionally to support edge FL scenarios. However, in remote regions distant from well-established communication hubs (such as rural or mountainous areas) or during critical situations (like large gatherings or military exercises), the limitations of terrestrial communication infrastructure can significantly impact the performance of edge FL. Fortunately, the SAGIN architecture introduces a new dimension of possibilities for FL, offering a range of benefits that can significantly enhance the efficiency and capabilities of FL in various applications. For instance, UAVs have the potential to supplant terrestrial BSs to deliver both communication and computing services to terrestrial devices, thus establishing what is referred to as air-ground integrated FL~\cite{qu2021empowering}. Several papers in the literature have considered the advantages of UAV networks in promoting efficient FL schemes. The authors in~\cite{Mestoukirdi2022UAV}, for example, optimize the trajectory of a UAV, acting as a mobile orchestrator, in an FL setting where several communities with a specific task for each exist. Graph theory and convex optimization tools were used, and the simulations highlight the benefits of UAV mobility and illustrate the out-performance of the proposed solution when compared to static UAV scenarios. In~\cite{Jing2023Exploiting}, the authors consider an air-ground integrated FL scenario and jointly optimize the UAV location and resource allocation to minimize the terrestrial users’ energy consumption and the tradeoff between energy consumption and FL training latency. Leveraging UAVs as edge servers also offers global accessibility to diverse datasets, which enables the training of more robust and generalized FL models. 

Apart from global data accessibility and remote coverage, SAGINs, with their potential for near real-time communication, can expedite the aggregation process of local model updates from edge devices, allowing for faster model convergence and decision-making. Furthermore, SAGINs often incorporate edge computing capabilities, allowing for local processing of FL tasks. This distributed approach reduces the need for transmitting extensive datasets to a central server, enhancing privacy and security while minimizing bandwidth usage. SAGINs' scalability also aligns with the distributed nature of FL. As the number of participating devices increases, SAGINs can seamlessly accommodate additional nodes, ensuring that FL can scale to meet the demands of large-scale applications without significant infrastructure changes~\cite{Matthiesen2023Federated}. The authors in~\cite{Farajzadeh2023FLSTRA} promote using HAPS to solve the issues of slow convergence and high communication delay due to limited client participation and multi-hop communications of FL implementations in terrestrial networks. Thanks to the high altitude and size of HAPS, it allows the participation of more devices with LoS links, and it has powerful computational capabilities to act as the server for local update aggregation. In~\cite{Farajzadeh2023FLSTRA}, the authors develop a joint client selection and resource allocation algorithm to minimize the FL delay and a communication and computation resource-aware algorithm to achieve the target FL accuracy.

Furthermore, SAGINs align with the security and privacy-preserving paradigm envisioned by FL. For instance, satellite and HAPS communication technologies provide an additional layer of security due to their inherent physical barriers, making it difficult for hackers or malicious actors to access sensitive information within the network itself~\cite{lu2021low}. Incorporating SAGINs into the FL ecosystem enhances the framework's adaptability, scalability, and reach. This synergy empowers FL to address a broader spectrum of applications, ranging from space exploration and environmental monitoring to smart cities and healthcare, by leveraging the unique capabilities of space, aerial, and ground networks. Several papers in the literature have considered the synergy between different components of SAGINs and FL~\cite{Matthiesen2023Federated,Wang2022Federated,Elmahallawy2022AsyncFLEO,Elmahallawy2022FedHAP}, yet remaining challenges must be addressed to achieve the full benefits of SAGINs for efficient and optimized FL. 

\subsection{Analog Over-the-air Computation}\label{sec:overlay_air}
\begin{figure}
    \centering
    \includegraphics[width=1\linewidth]{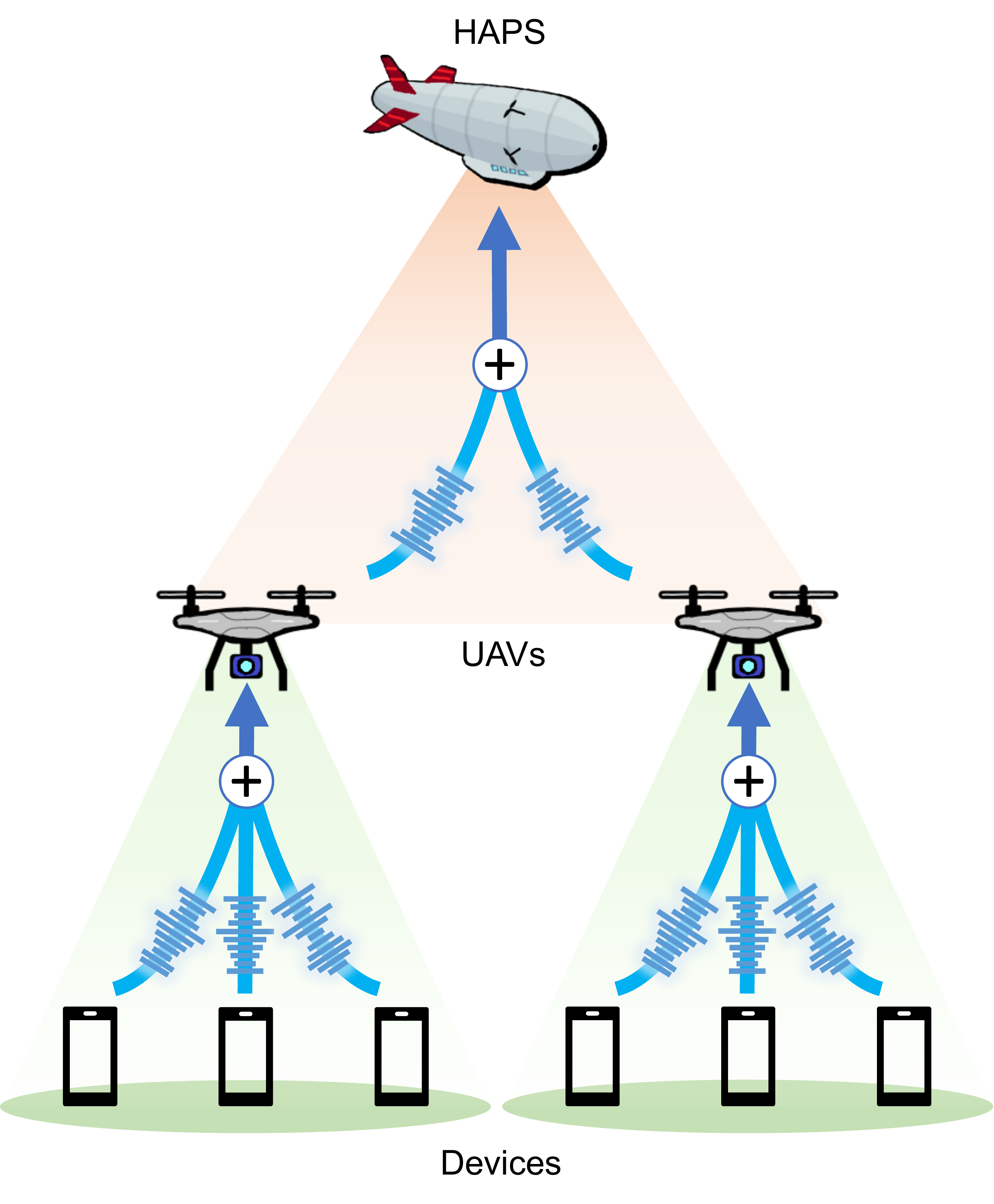}
    \caption{UAV and HAPS sensing the signal, which is the result of analog over-the-air summation of signals coming from individual devices.}
    \label{fig:analog}
\end{figure}

Several solutions are proposed in traditional wireless networks to mitigate the negative impact of interference. However, most of these solutions introduce significant reductions in the available resources for each user. Conversely, the analog over-the-air computation technique promotes interference by allocating the same time and frequency resources to multiple users. This causes the analog combining of the users' signals in the air (see Fig.~\ref{fig:analog}). The design of efficient precoding schemes at the transmitters allows for calculating useful functions (i.e., sum or average) from the superposed analog signals over the air. However, extracting the individual signals becomes nearly impossible for the receivers. While some traditional data transmissions cannot benefit from the analog over-the-air computation technique as they require the reception of individual signals, distributed ML and AI algorithms can benefit from it as they are mainly based on the computation of a weighted sum of the updates~\cite{Hellstrom2022book}. A typical example is the FL algorithm since the individual local models are not needed at any point, and the model aggregation step consists of transmitting multiple local models from users to the BS or server and then computing the weighted mean of these updates~\cite{Alphan2023Over}.

While the analog over-the-air computation technique offers several benefits to distributed AI algorithms, it also introduces distinctive challenges in terms of strong demands on CSI~\cite{Zhu2019MIMO}, stringent synchronization requirements~\cite{Goldenbaum2013Robust}, and limited peak transmission power~\cite{Cao2020Optimized}. A notable limitation of this technique in FL is the straggler problem: the need to align local gradients on the server side. In an FL setup, the server receives a linear combination of local gradients with specific coefficients, leading to a situation where devices with strong channel conditions must reduce their transmit power to accommodate devices with weaker channel conditions, commonly referred to as "stragglers." This adjustment is necessary to ensure accurate aggregation. Such biasing between strong and weak users can be solved by increasing the transmit power of weak users. However, with practical transmitters and large-scale terrestrial networks, the maximum transmit power is limited, and a large difference in the path loss between nearby and far users exists. 

In this context, SAGIN architectures, specifically UAVs, can help mitigate the straggler problem of analog over-the-air computation~\cite{Fu2023UAV,Fu2022UAV}. UAV can, therefore, act as the server and aggregate local updates from terrestrial users. Due to the LoS nature of the aerial channels, the user-to-UAV links are generally not blocked, potentially leading to good channel conditions. Furthermore, the mobility of the UAV allows it to cover a larger area and to fly to stragglers for better services. This will improve the learning accuracy of the FL algorithm and reduce the learning performance loss. In~\cite{Zhong2022UAV}, the authors design a hierarchical over-the-air computation scheme for a UAV-assisted FL system, where the UAV acts as the server to serve nearby devices and move to another location to avoid the straggler problem. A mean squared error minimization problem is formulated and solved to tune the UAV trajectory and the global aggregation coefficients. The results highlight that the proposed scheme can significantly improve complicated FL scenarios. In~\cite{Wang2022Federated}, the authors propose an over-the-air computation-based satellite FL framework in which the satellite acts as the server that collects local updates from many terrestrial users. An optimization problem is tackled to accelerate the convergence of the FL algorithm by minimizing the downlink broadcasting error and the uplink over-the-air aggregation error.

\subsection{Radio Resource Management for AI}\label{sec:overlay_RRM}
The widespread of ML and AI has brought several new advantages by relying on the potential of wireless communications. SAGINs, specifically, allow for enhanced data accessibility and promote global collaboration to further enhance the scalability of different AI and distributed ML frameworks. However, the problem with AI and distributed ML is that they differ from general data communications in different ways. These differences result in new constraints in terms of computational complexity, training time, training data, and more. In this context, existing data communication protocols perform poorly in satisfying AI needs, motivating the design of new RRM protocols for AI. SAGINs can be a great enabler, further enhancing the efficiency of the new AI-tolerated RRM protocols. Emerging RRM solutions prioritize optimized allocation of key radio resources like channels, power, and frequency within wireless systems. This strategic resource management aims to foster an environment that empowers AI algorithms to improve their performance, thus enhancing user experience. This is especially important for AI applications that require high bandwidth, minimal latency, and optimal use of network resources.

In traditional RRM, max-min fairness protocols usually sacrifice spectrum usage to ensure a minimum level of service for all users in the network. On the other side, users can be treated in a discriminatory manner in FL based on the importance of their data. Thus, such spectrum usage sacrifice is no longer needed for FL~\cite{Hellstrom2022book}. New RRM protocols are therefore required to recognize devices with valuable data and enough resources (battery, processing power, network bandwidth) for FL training participation, ensuring seamless training progress without overburdening resource-limited devices. According to network load and training needs, the RRM protocols can assign bandwidth and power dynamically to participating devices to optimize resource usage and reduce training time. Furthermore, the envisioned schemes must choose the best channels for communication between the central server and devices, which minimizes interference and ensures reliable data transmission for precise model updates. 

Another AI-driven application, autonomous vehicles, can greatly benefit from new RRM schemes that prioritize network resources for self-driving cars, ensuring critical data such as sensor readings and control signals are transmitted with minimal delay, enabling quick and accurate decisions for safe navigation. Furthermore, as vehicles move through different network cells, new RRM protocols must be developed to facilitate seamless handovers without interrupting data flow, ensuring continuous operation of AI algorithms for navigation and obstacle detection. AI-enabled IoT remote monitoring applications can also benefit from new RRM schemes to assign ample bandwidth and power to edge devices such as sensors and cameras used in remote monitoring, facilitating efficient data transfer for AI analysis and instantaneous decision-making. New RRM schemes can also be developed to allocate network resources primarily for data linked to critical events identified by AI algorithms, guaranteeing prompt intervention and response. Lastly, new RRM protocols are envisioned to adjust power levels according to network conditions and data requirements, prolonging the battery life of remote devices while lowering operational costs. These RRM schemes can also benefit from silent periods, in FL, for instance, to perform power transfer from server to devices, further enhancing their battery storage~\cite{Hellstrom2022book}.

The introduction of SAGINs opens up new opportunities for enhanced RRM protocols that maximize the effectiveness of different AI applications, particularly when these applications involve either aerial or remote devices. Such devices may encounter challenges with limited coverage, high latency, and unreliable connections in traditional networks. However, with SAGIN, the new RRM protocols can benefit from the different SAGIN components to seamlessly transfer the connection based on the location of these devices, ensuring continuous data flow~\cite{Chen2023Multi}. Recognizing the critical nature of real-time data from remote devices, new RRM schemes can allocate bandwidth and latency capabilities across various SAGIN communication nodes for faster AI analysis and quicker response times for crucial applications. Furthermore, by distributing traffic load among network segments, RRM protocols can optimize resource utilization and prevent congestion that could impede AI applications~\cite{Zhang2022Space}. At the same time, utilizing SAGINs presents distinct challenges. Each segment of SAGIN has its own features and constraints, demanding flexible adjustments of the RRM schemes. UAVs, HAPS, and satellites are in constant motion, requiring real-time resource allocation and network route modifications. Moreover, all SAGIN nodes have restricted energy and computational capacity, underscoring the importance of efficient resource utilization for sustainable operation.

\vspace{-0.2cm}
\subsection{Applications}\label{sec:overlay_application}
\begin{figure}
    \centering
    \includegraphics[width=1\linewidth]{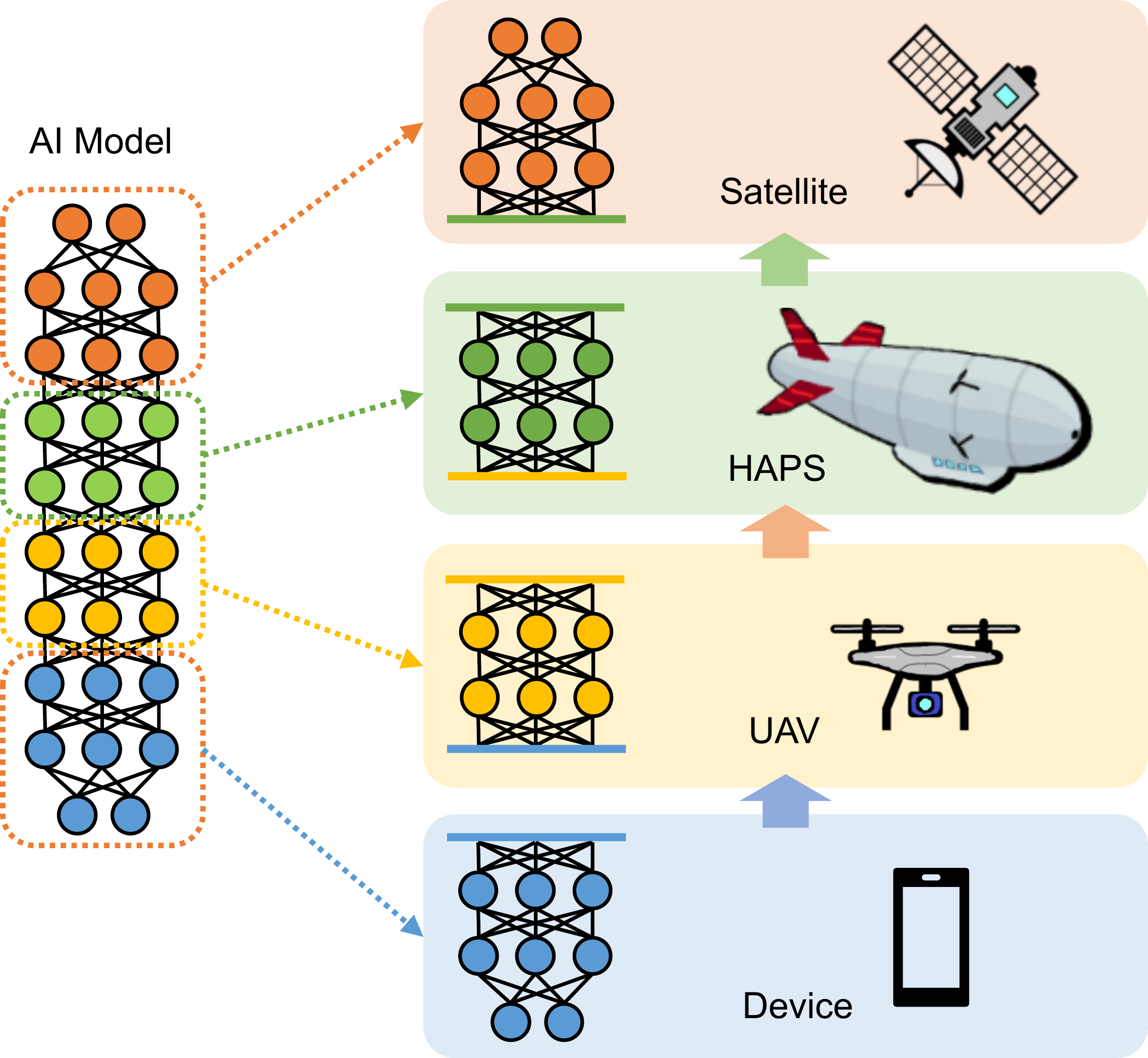}
    \caption{Example of a split learning scenario, in which different parts of the NN are processed in distinct vertical SAGIN nodes.}
    \label{fig:split}
    \vspace{-0.5cm}
\end{figure}

\subsubsection{Autonomous Cars and Vehicular Networks}\label{sec:application_car}
The past decade has seen significant progress in autonomous cars, often referred to as self-driving or driverless cars, with AI at the heart of their operation. AI empowers autonomous cars to perceive their surroundings using sensors like LiDAR, cameras, and radar, interpret this data to make informed decisions, and navigate complex environments without human intervention. ML algorithms enable these vehicles to learn from vast datasets and improve their driving abilities over time~\cite{Du2020Federated}. While AI-enabled autonomous car systems hold great promise, they also face significant challenges. Safety is one of the foremost concerns. Ensuring that these systems can reliably handle complex and unpredictable driving scenarios by efficient communication and sharing of information remains a substantial hurdle~\cite{zeng2022federated}. 

With their distinguishable features, SAGINs can provide a reliable solution to address the limitations of AI-enabled autonomous cars and vehicular networks. Non-terrestrial networks are essential for covering areas that terrestrial networks cannot reach due to geographical constraints or other obstacles~\cite{peng2019vehicular}. Satellite networks, for instance, can extend the range of vehicular communications beyond its traditional limits, allowing autonomous vehicles and other connected cars to communicate over long distances without relying on infrastructure provided by cellular companies. This enables autonomous cars, and subsequently, the AI algorithms behind them, to operate in remote or rural areas that don't have access to traditional terrestrial wireless services. Furthermore, SAGINs help handle unpredictable driving scenarios and reduce accidents by facilitating information exchange between different vehicles in real-time. Therefore, real-time traffic monitoring and management are facilitated, allowing vehicles to adapt to traffic conditions and optimize overall traffic flow.

In addition to reliable communication, SAGINs provide precise global positioning data, contributing to improved navigation and location-based services. They enable efficient data offloading, accommodating the vast amounts of data generated by autonomous vehicles and their associated sensors for advanced analytics and decision-making. For instance, the work in~\cite{ren2022} promotes using HAPS for caching and computation offloading in ITS. As HAPSs are deployed in the stratosphere and are available to provide wide coverage and strong computational capabilities, they have been regarded as efficient enablers for coordinating terrestrial resources and storing the fundamental data associated with ITS-based applications.

Another characteristic of SAGINs is their low-latency capabilities, which are crucial for real-time data exchange and autonomous vehicle decision-making. This is especially important for safety-critical applications such as automated emergency braking systems that require fast response times. Since such decision-making algorithms are mainly AI-powered, providing access to the data necessary to make a correct prediction makes SAGINs an inarguably massive driver of the accuracy and success of these AI algorithms. Furthermore, thanks to their scalability, SAGINs can accommodate the growing number of autonomous vehicles without requiring extensive terrestrial infrastructure expansion. Furthermore, SAGINs can aid in coordinating autonomous vehicles and emergency responders in emergencies. For instance, UAVs can be sent to swiftly establish communication networks when the terrestrial network is down, and autonomous cars can exchange information through UAVs acting as relays. For instance, the work presented in~\cite{naous2023} utilizes flying UAVs and terrestrial vehicles as relay nodes to improve vehicular communications' coverage, throughput, and link reliability. Multiple UAVs can assist the ground vehicular network through air-to-ground communications, offering advantages over terrestrial vehicle relays~\cite{naous2023}.

Although SAGIN architectures promise to offer several advantages to accelerate autonomous cars and vehicular networks, several challenges remain to be resolved. Specifically, the impact of the severe requirements of these technologies in terms of low latency, high reliability, and high throughput must be considered in a joint communication and learning optimization framework~\cite{Chen2021Joint}. Moreover, the impact of the quick variations in the wireless channels due to the induced mobility must be captured in the analysis of the learning convergence of the AI algorithms. Finally, resource allocation and network organization schemes must be designed to use the different SAGIN layers efficiently in optimizing the performance of autonomous cars and vehicular networks.

\subsubsection{Internet of Things}\label{sec:application_IoT}
Integrating AI in IoT systems enables IoT devices to perform advanced tasks and deliver more intelligent and autonomous services. As IoT devices generate huge amounts of data, AI can enhance data processing, allowing devices for real-time decision-making in addition to predictive and prescriptive analytics by investigating and detecting patterns, anomalies, and correlations in the IoT data~\cite{m2022}. However, integrating AI with IoT poses challenges such as data privacy, model complexity, and power consumption. Furthermore, reliable communication links must be maintained to collect the IoT-generated data from the massive number of sparsely distributed devices located in remote and hard-to-reach regions.

SAGINs can play a significant role in efficiently implementing AI-enabled IoT systems. SAGINs allow for greater control over the data transmission process from the IoT devices to the cloud processing platform; this is especially important if sensitive data needs to be securely transmitted across multiple nodes before reaching its destination point~\cite{khan2022}. Additionally, many aerial solutions offer low-power consumption options ideal for battery-powered IoT devices located at remote sites or hard-to-reach places requiring minimal maintenance costs associated with recharging or replacing batteries~\cite{khalifa2023energy,benbuk2020leveraging,saab2018uav}. Moreover, SAGINs enhance the scalability and adaptability of IoT systems by allowing a growing number of IoT devices to communicate together. This enhances further the accuracy of the AI algorithms as larger and more diverse datasets are now available. Such characteristics are mainly important for deploying smart cities that require the deployment of thousands of sensors over large regions. SAGINs also allow data fusion from various sources, such as ground sensors, aerial drones, and satellite imaging. This can enable advanced analytics and predictive modeling for sectors like agriculture, environmental monitoring, and disaster management. 

As SAGIN architectures allow for a wide range of IoT devices to communicate and share data, the heterogeneity of IoT devices and sensors is one of the main barriers to efficient AI algorithms. IoT devices are mainly different in terms of computing capability, cache size, battery power, data rate, and latency requirements. Therefore, distributed AI algorithms such as FL will be biased toward devices with better characteristics as they have a higher chance to share their updates with the server. In~\cite{zhagypar2023characterization}, the authors characterize such bias and highlight its impact on the accuracy of the FL algorithm in a UAV-enabled network. Furthermore, they propose an efficient method of unbiasing the FL algorithm to account for each IoT device's different requirements and capabilities. One possible solution to deal with the massive number of heterogeneous IoT devices is to optimally use the available resources via sparsification, allowing only a subset of devices to contribute to the learning. However, the criteria for choosing the contributing devices while maintaining high accuracy and preventing any bias is also challenging and has to be optimally designed. 

Split learning (See Fig.~\ref{fig:split}) is yet another potential scheme to mitigate the device heterogeneity issue of IoT devices as it distributes the learning vertically between different SAGIN layers~\cite{Feng2023IoTSL}. Such a distributed computing approach can also effectively address resource limitations in IoT devices. This entails performing partial model training locally on the devices, with further processing being seamlessly shifted to existing ground servers or at UAVs, HAPS, and satellites if they exist. Split learning, however, requires jointly optimizing the layer selection across the different SAGIN layers, IoT device clustering and association, and efficient resource allocation to minimize the training latency and improve the convergence of the considered ML algorithm~\cite{Wu2023Split}. Efficient resource allocation schemes are further required to mitigate the high interference that further increases the model bias, and that is mainly caused by the massive number of IoT devices contributing to the learning and communicating with the different layers of the SAGIN~\cite{Li2023Asynchronous}. 

\subsubsection{Virtual and Augmented Reality}\label{sec:application_virtual}
In VR/AR, AI tools maintain smooth virtual and real-world interactions by enabling real-time adjustments and processing. AI can also help personalize the content and create context-aware information. In addition, AI algorithms are meant to detect and minimize errors and enhance image recognition to improve performance and increase user satisfaction. In wireless networks offering VR/AR services, a sudden decline in data rate or rise in latency can impact the user experience negatively. Such interruptions in the virtual world result in breaks in presence events that can be detrimental to the users' immersive VR/AR experience~\cite{Chen2023Echo}. Moreover, most VR/AR applications require low latency and high resource allocation, creating a bottleneck at the air interface of wireless networks. Edge computing has been regarded as an efficient contributor to mitigating this bottleneck by reducing the transmission and processing latency significantly~\cite{shnaiwer2022multihop}. However, the fixed locations of the edge servers limit the resource management for VR-based applications~\cite{Nguyen2023Contract}. Moreover, while 5G networks operate at high frequencies and utilize adaptable MIMO configurations and frame structures to improve throughput and minimize latency, the susceptibility of high-frequency communication links to blockages can diminish the quality of experience for VR and AR users. Offering connectivity from the sky is an efficient solution to mitigate the impact of blockages by providing alternative LoS links.  
 
SAGINs can assist in meeting the heavy communication demands imposed by VR/AR applications. One of the many improvements that SAGINs promise, in addition to enhancing link quality and extending coverage, is to reduce the latency and increase the data processing capabilities of terrestrial networks. Undoubtedly, the salient features of automation, flexibility, and better SNR make SAGINs a promising enabler for AI-based VR/AR applications. For instance, using UAVs is essential in reducing the latency by bringing computation and caching capabilities closer to VR/AR users~\cite{Nguyen2023Contract,Taleb2023VR,Nasir2021Latency}. One major challenge for UAV-assisted VR/AR systems is the tradeoff between maintaining high learning accuracy and minimal delay. As UAVs have limited computational resources, the resolution might suffer from large attenuation, thus resulting in low accuracy. Conversely, being close to the user improves the wireless channel and minimizes the propagation delay. Thus, joint optimization of delay and learning accuracy by optimizing the UAV deployment and trajectory is mandatory when populating terrestrial networks with UAVs for VR/AR applications~\cite{Tang2023UAV}. Furthermore, joint consideration of the SAGIN wireless environment parameters and the VR/AR user-specific metrics such as user orientation, user association, and user awareness is another critical challenge for enhancing AI-enabled VR/AR systems over SAGINs.

\subsubsection{Digital Twins}\label{sec:application_twin}
A digital twin is a virtual representation of a physical object, system, or process leveraging data and analytics to drive informed decision-making~\cite{gichane2020}. It combines data-driven insights and technological agility to create predictive models for business operations, product development, manufacturing processes, and customer experience management~\cite{gichane2020}. Digital twins are pivotal in enabling organizations to gain real-time visibility across the entire life cycle, from design to operation, ensuring optimized performance and risk mitigation through timely issue detection. Moreover, digital twins foster personalized experiences to develop deeper customer relationships tailored to individual preferences. Integrating AI with digital twins enhances their ability to learn, adapt, and provision valuable insights~\cite{li2023}, rendering them powerful tools for monitoring, analyzing, and optimizing complex physical systems in manufacturing, healthcare, transportation, and smart cities. AI and ML techniques have been used with digital twins for predictive maintenance, health monitoring, fault diagnosis, adaptive control, and operation process optimization~\cite{li2023}. Thus, real-time sensing combined with AI technologies will be essential to realize digital twins for 6G~\cite{Lin20236G}.

The advancement of digital twin technology can be further propelled by integrating SAGINs, which offer extensive data accessibility~\cite{fuller2020}. This data can be used to create accurate models and simulations of physical environments, essential for creating digital twins. Leveraging satellite imagery and remote sensing technologies could provide in-depth information on terrain and land, enabling the development of robust 3D models for various environments. Additionally, the real-time data communicated over SAGINs from the embedded sensors and IoT devices empowers the development of dynamic digital twins capable of responsive adaptations to environmental shifts. Moreover, SAGINs, characterized by accelerated connection speeds that outmatch conventional terrestrial systems, are ideal for applications necessitating high-volume or frequent updates, such as real-time monitoring and control within digital twin systems~\cite{shen2022}. Additionally, the wide coverage SAGINs provide enables more granular data capture from diverse sources, fostering the creation of intricate digital twin models.

In digital twin systems, devices will generate data and actively join the network management and optimization. While organizing terrestrial networks might be feasible with simple schemes and policies, SAGINs demand sophisticated network management and coordination strategies to optimize the performance and functionality of the various network layers. To this extent, decisions for offloading data across different layers, edge association, and resource allocation must be taken carefully based on the real-time data generated to enhance synchronization and improve the performance of such systems~\cite{Hazarika2023RADiT,Guo2023Resource}. Furthermore, the substantial computational capabilities required in digital twin systems to ensure real-time interaction and response require expanding the computing capabilities of physical devices beyond their inherent limits, leading to the adoption of split computing, an emerging technology trend in the 6G landscape. Implementing efficient split computing schemes in SAGIN architectures is yet another research gap that must be filled in the context of SAGIN-enabled digital twin frameworks. Moreover, novel physical layer security solutions, keyless transmissions, and distributed anomaly detection schemes are to be designed to maintain the trustworthiness of the SAGINs supporting the digital twins. Finally, further interest must be devoted to implementing THz communications in SAGINs so as to support the massive information exchange between the physical world and the digital world with low latency and high reliability and to benefit from THz's precise positioning capabilities for extreme precision mapping.

\subsubsection{Semantic Communications}\label{sec:application_semantic}
Semantic communication surpasses simple data transmission by incorporating context and significance into the conveyed information, thanks to advancements in AI, particularly in natural language processing~\cite{shi2023task}. This allows for efficient communication by focusing on the essence of the message rather than the raw data. In traditional relaying methods like "decode-and-forward", the relay decodes the entire message and then retransmits it, while "amplify-and-forward" boosts the signal without understanding its content~\cite{weng2021semantic}. The "process-and-forward" approach in semantic communications takes a different path, powered by AI algorithms. Instead of decoding the entire message, the relay extracts and processes only the necessary semantic meaning, resulting in a smaller, more targeted data packet for transmission. This reduces bandwidth usage and processing time, making semantic communication a powerful tool for efficient and intelligent communication.

Semantic communications in SAGINs bring numerous benefits, primarily due to the end-to-end encapsulation in terrestrial networks. However, this hinders intermediate nodes like routers from extracting and processing semantic meaning without fully decoding the message, affecting the efficiency of "process-and-forward." In contrast, SAGINs provide more flexibility and distributed intelligence as data packets can be intercepted and processed at intermediate nodes such as UAVs, HAPS, or satellites~\cite{yang2022semantic}. This enables them to extract semantic meaning before forwarding the content, supported by AI-powered semantic interpretation frameworks. This allows for efficient extraction and transmission of only relevant semantic information with accurate interpretation while maintaining efficiency and low latency. On the other hand, this method may also pose challenges in efficiently managing communication resources and limited onboard computing capabilities. AI plays a crucial role here by developing algorithms to optimize resource allocation and processing tasks, tailoring them to the specific limitations of SAGINs. Extracting and handling semantic data also gives rise to security and privacy concerns that require robust measures for user protection and prevention of unauthorized access. AI-powered security solutions can analyze and anonymize semantic data, ensuring information integrity and user privacy.

An unsynchronized background knowledge issue arises when the transmitter and receiver lack shared context, causing misinterpretations in semantic communications. It is a specific scenario where SAGINs, empowered by AI, can be especially useful. In situations where there is a lack of synchronized background knowledge, if both parties can facilitate semantic communication, airborne BSs can interpret semantics from the received signal using the transmitter's background knowledge, thanks to AI-based knowledge mapping techniques. Subsequently, it can encode the signal based on the receiver's background information, lowering the burden of synchronizing background knowledge for transmission and reception and reducing semantic noise through AI-powered filtering methods. This showcases the potential of AI in bridging the gap between different knowledge bases and promoting seamless communication even in challenging scenarios.

\section{Open Issues and Future Directions}
\label{sec:openissues}
Integrating terrestrial and non-terrestrial networks has been an area of research that has seen significant advances in recent years. With the advent of AI algorithms, this field has great potential for further development. This section explores future research directions for the interaction between AI and SAGINs.

\subsection{Integrating SAGINs and Key 6G Enablers}
\label{sec:6genablers}
It is anticipated that 6G networks will utilize terrestrial and non-terrestrial networks, thus enabling users to access services and applications from various locations~\cite{araniti_2021}. Additionally, using both networks will increase network capacity and faster data speeds. Therefore, many potential research challenges are associated with integrating the SAGIN and key 6G enablers. 

\subsubsection{THz Communications}\label{sec:future_thz}
Following the successful implementation of the mmWave technology in the 5G, THz communications are envisioned as key enablers of 6G, owing to the abundant spectrum available at the THz band~\cite{magbool2022terahertz}. THz frequencies offer aerial communications with exceptionally high data rates compared to lower frequency bands due to their ultra-wide bandwidth. Thus, implementing the THz technology in SAGINs is a natural step toward achieving service continuity, ubiquitous access, network scalability, and efficient backhauling. THz aerial links exhibit reduced spectral noise, interference, and jamming compared to terrestrial links, attributed to their high attenuation and inability to penetrate the troposphere. Moreover, THz links encounter minimal molecular absorption at higher altitudes and in free space, allowing for extended communication ranges. These benefits make using THz frequencies in SAGINs a logical and practical choice.

While using THz communications in SAGINs paves the way for multiple opportunities, several challenges and open problems need to be considered. First, the limited propagation range of THz waves due to the high absorption loss and sensitivity to blockage limits the use of THz communications for long-range air-to-ground links. Thus, air-to-ground communications can use sub-$6$~GHz and mmWave transmissions, and communications over air-to-air links can be handled over THz frequencies, allowing for higher-capacity backhaul links. Flexible spectrum-sharing techniques are therefore required to maintain suitable isolation between different network operations~\cite{Kouzayha2023Coexisting}. Second, THz propagation has a highly varying channel, i.e., its coherence time is extremely short. In the context of SAGINs, the impact of the non-stationary THz channel is even more pronounced as SAGINs components, mainly UAVs and satellites, move at high speeds, thus encountering higher propagation delays and Doppler shifts. Thus, new comprehensive channel models are required to account for such effects. These new channel models should also consider the impact of the uncertainties associated with wind, the orientation and wobbling of the antenna arrays, and the possible signal blockages by the body of the SAGIN components, calling for more multidisciplinary research efforts~\cite{Geraci2022What}. Finally, THz communications are characterized by pencil beam directionality, requiring accurate tuning despite the high Doppler shifts and increased speeds in THz-SAGINs. Other challenges in THz-SAGINs include initial access, computing limitations, and cross-layer power management.

Due to the lack of explicit models to extract the performance tradeoffs in terms of rate, reliability, and synchronization of THz-enabled SAGINs, AI-based solutions can be exploited. Thus, collected data, including channel and QoS measurements, can be used to decide on the performance of the THz-SAGIN system. Nonetheless, the non-stationary nature of the collected data due to excessive handovers and high losses complicates the prediction and generalization of the distribution patterns. Furthermore, centralized AI algorithms fail to predict the performance of THz-SAGINs due to the distributed architecture and the low latency requirements of most of the 6G applications. Thus, multi-agent AI algorithms are preferred to locally collect and learn data~\cite{Chaccour2022Seven}. Moreover, the current AI methods still require long training periods, which prevents agents from learning in real-time. Performing offline training is prohibited in THz-SAGINs architectures due to the non-stationarity of available data. Thus, a real-time multi-agent learning framework is required to predict the performance of THz-SAGIN systems. Employing such a framework instead of centralized techniques allows for a specialized understanding of specific service types, mobility patterns, or subsets of users and resources. Furthermore, incorporating multi-task and meta-learning concepts ensures a more comprehensive understanding of the distribution patterns, which allows learning agents to adjust their strategies according to the specific needs while reducing the long training periods. Challenges, however, still exist in terms of meeting the low latency demands of 6G services and dealing with the complex and non-smooth performance functions. Thus, further research is required to develop novel AI methods that can optimize and control network operations in THz-SAGIN systems.

\subsubsection{Optical Wireless Communications}\label{sec:future_optical}
Unleashing the potential of optical wireless communication (OWC), including free space optics (FSO) and visible light communications (VLC), is a transformative force in the evolution of 6G SAGINs. OWC offers terabit-per-second data rates surpassing radio frequency (RF)-based alternatives, providing ultra-high capacity suitable for bandwidth-intensive applications and lower latency critical for real-time applications such as remote surgery and autonomous vehicles~\cite{chowdhury2019role}. Additionally, its enhanced energy efficiency makes it a sustainable choice for future SAGINs, consuming significantly less power than traditional wireless technologies~\cite{wu2021hybrid}. In particular, VLC leverages light's limited propagation range and absence of electromagnetic interference to provide inherent security advantages crucial in today's increasing cyber threats~\cite{obeed2018,obeed2019optimizing}.

The integration of OWC with SAGINs extends coverage to under-served areas, enhances capacity by providing an additional data transmission channel, and mitigates congestion by offloading traffic from congested RF networks~\cite{chowdhury2020optical}. Despite these benefits, integrating OWC into SAGINs poses unique challenges. These challenges encompass the limited range of OWC compared to RF communication, susceptibility to weather effects like rain and fog, and the requirement for LoS, posing difficulties for mobile users~\cite{obeed2019optimizing}. Hence, dynamic channel conditions, atmospheric turbulence, and the mobility of SAGIN components necessitate real-time link optimization. Seamless RF-OWC coexistence demands intelligent coordination while addressing VLC security concerns requires advanced physical layer security mechanisms.

AI emerges as the key to unlocking OWC's full potential in SAGINs. Addressing mobility challenges within SAGINs and OWC-specific range, weather, and LoS limitations requires dynamic network reconfiguration, where DRL algorithms prove valuable. DRL can facilitate real-time adjustments to beam paths, transmission power, and channel allocation as SAGIN components move or undergo deployment changes, ensuring continuous connectivity. Furthermore, ML for predictive analytics can aid in achieving seamless handoffs between OWC and RF channels, predicting optimal transition times and conditions, thereby enhancing efficiency and minimizing disruptions for mobile users. AI-driven resource allocation algorithms can effectively address resource allocation and coexistence challenges. These algorithms, including GAs and PSO, can dynamically balance traffic between OWC and RF channels, optimizing network performance based on real-time considerations such as data type, user priority, and channel availability. DL models also prove beneficial in predicting interference scenarios, allowing for minimizing interference through techniques like beamforming and frequency coordination.

Integrating OWC with the SAGIN infrastructure, which faces issues like power and bandwidth limitations, can also benefit from AI solutions. ML algorithms contribute to energy-efficient communication by optimizing the energy consumption of OWC systems and adjusting transmission power levels based on predicted network demand. Additionally, AI combined with cryptography can enhance security in open environments by implementing advanced encryption algorithms and intrusion detection systems. Moreover, enhanced computation capabilities offered by various SAGIN nodes can enhance any AI-powered solution by making it possible to run in a distributed manner, focusing on optimizing each link separately. In summary, the strategic integration of DRL, ML, and DL techniques is vital for overcoming challenges in OWC-SAGIN integration to enhance adaptability, efficiency, and security in dynamic network environments.

\subsubsection{Reconfigurable Intelligent Surfaces}\label{sec:future_ris}
Reconfigurable Intelligent Surface (RIS) is recognized as an appealing technology to control electromagnetic wave propagation and tailor the propagation environment for enhanced spectral efficiency, extended coverage, and robust security, all while maintaining cost-effectiveness and energy efficiency~\cite{Abbas2023Performance,aman2023downlink}. Thanks to the new opportunities RIS technology provides, it has been envisioned as a promising solution for supporting SAGIN communications and meeting the ambitious goals they are promising. RISs, for instance, offer an energy-efficient relaying solution for low-power, small-size LEO satellite swarms deployed to maintain ubiquitous connectivity. In such scenarios, the RIS elements can be tuned to act as passive reflectors between different satellites or across the different SAGIN layers, thus maintaining low power consumption and computing overhead. Furthermore, RIS can enhance the reliability of SAGIN in-layer and cross-layer communication links by overcoming the path loss impact due to long-distance propagation by providing alternative LoS links and introducing directional beamforming. This can be achieved by efficiently tuning the passive reflective elements of the RIS to achieve the desired gains.

The path towards fully realizing the transformative potential of RIS-empowered SAGINs is paved with challenges in its deployment. This includes, for instance, the need for dynamic RIS reconfiguration due to the mobility of the satellites and UAVs, which requires sophisticated coordination schemes between the SAGIN elements in the same layer and across different layers. Furthermore, the highly dynamic nature of the SAGIN architecture results in complicated calculation processes and optimization problems. Consequently, AI techniques are expected to provide efficient tools to address the challenges of RIS-empowered SAGINs. Together with the enhanced computing and storage capabilities provided by the space, air, and ground components, along with the cooperative features facilitated by the cloudification of the radio access network, AI will empower SAGINs to optimize, organize, and heal themselves autonomously~\cite{Bariah2023RIS}. 

AI algorithms can optimize RIS placement and phase shift design across different SAGIN layers in the context of RIS-empowered SAGIN. Furthermore, AI can optimize cell/user/RIS association decisions and promote proactive resource allocation to enhance performance and reduce overall delay~\cite{Ye2022Nonterrestrial}. In this context, distributed ML algorithms, such as FL, are preferred over centralized algorithms as they prioritize privacy-preserving model training, reducing communication overhead and latency. Distributed algorithms allow for a wide contribution of clients from different SAGIN layers, thus enabling network-wide training across vertical layers, which delivers a generalized global model that fits the heterogeneous SAGIN components with diverse QoS requirements. Several works in the literature have considered the integration of RISs in different layers of the SAGIN architecture. However, further efforts are still required to design efficient AI algorithms for real-time optimization of the RIS deployment and the user access schemes to meet the time-varying QoS requirements while considering the impact of the complex and dynamic SAGIN channel environment.

\subsection{Generative Adversarial Networks for SAGIN}\label{sec:future_gan}
\begin{figure}
    \centering
    \includegraphics[width=1\linewidth]{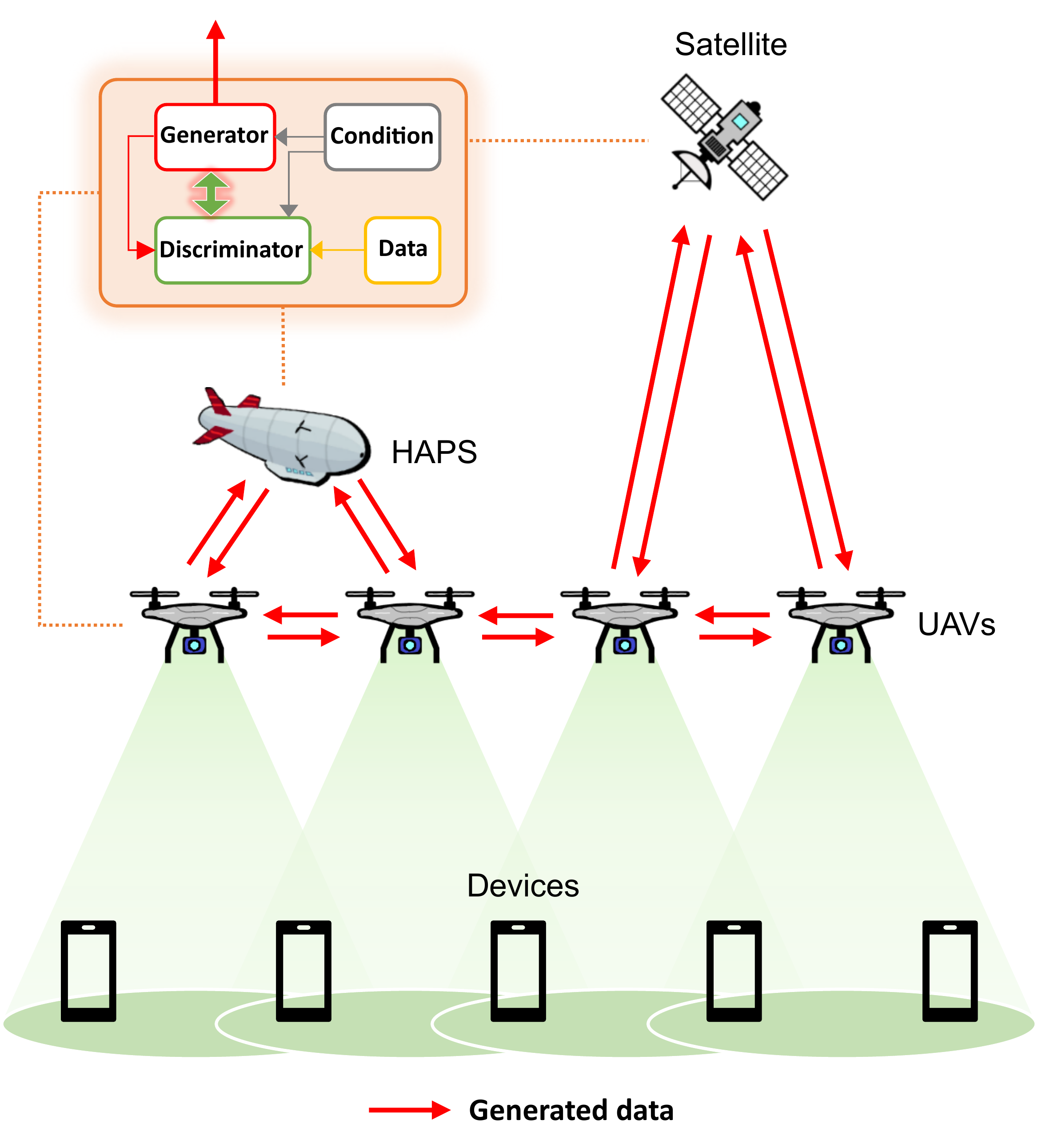}
    \caption{UAVs, HAPS, and satellites using GANs to dynamically generate the synthesized channel data and share it with neighboring communication nodes to aid with the collaborative channel learning process.}
    \label{fig:gan}
\end{figure}

Channel estimation poses a significant challenge in deploying SAGINs because of the varied movement patterns observed by different nodes, the differing propagation environments among various SAGIN layers, and the diverse types of noise and disruptions encountered within each segment. Furthermore, compared to terrestrial channels, SAGIN channels include more model parameters, such as the 3D location and dynamic orientation of the SAGIN node~\cite{zhang2022distributed}. Due to this, traditional channel estimation techniques like ray tracing and geographically estimated statistical models are not enough in SAGIN channel modeling. Data-driven methods can be employed to address the challenges of SAGIN channel estimation. ML algorithms present a viable option for achieving precise channel estimation in SAGINs in a centralized or distributed fashion. Centralized ML encounters challenges concerning intricacy, high power demands, and data privacy and is not adapted to the distributed characteristic of the SAGIN architecture. On the other side, distributed ML leverages local data at each SAGIN node for decentralized channel estimation, which might not be adequate to encompass all exceptional environmental circumstances, thus necessitating frequent channel estimation~\cite{Rasheed2023LSTM}.

GANs are a promising strategy for effective distributed channel estimation~\cite{Yang2019Generative}. Specifically, GANs are employed to produce data that closely resembles real-world scenarios, capturing the behaviors of channels and creating new samples within the dataset that represent various channel conditions~\cite{Bariah2023RIS}. By leveraging GANs, SAGIN can enhance channel data availability, improve training efficiency, and develop robust channel models adapting to diverse environmental conditions across space, air, and ground domains. Fig.~\ref{fig:gan} illustrates a SAGIN where UAVs, HAPSs, and satellites use GAN to generate data from channel estimation. Due to the ability of a generative model to understand the channel model's application scope through the dataset's temporal-spatial information during training, it offers a more effective learning framework in contrast to a discriminative method~\cite{Zou2023Generative}. However, despite its potential benefits, utilizing GANs for channel estimation in SAGIN remains an active area of research. Key aspects such as the dataset size needed for precise sample generation, the adaptability of GANs in SAGIN, and the necessary resources for integrating GANs in SAGIN have yet to be comprehensively explored.

Moreover, most of the previous research on utilizing GANs for channel estimation and modeling in SAGIN has focused mainly on local GANs models~\cite{xia2022generative}, and the literature still lacks fully distributed generative learning frameworks for addressing the challenges of data-driven SAGIN channel modeling. To bridge this gap, the development of fully distributed cooperative generative learning models is essential for accurately characterizing the SAGIN link environment. Specifically, multiple SAGIN nodes can cooperate to create the whole channel model by learning a standalone channel model by each node via GANs and then sharing synthetic channel samples generated from the local channel model with other agents. This fully decentralized approach does not require a controller and can accommodate different types of NNs while being resilient to local training errors. Furthermore, it avoids revealing the real measured data or the trained channel while allowing SAGIN nodes to learn cooperatively from each other's datasets.  

\subsection{Security Guarantees}\label{sec:future_security}
The heterogeneous and multi-dimensional nature of SAGINs brings new trust and safety considerations that must be addressed. Specifically, the wireless signal in SAGIN propagates in free space across different layers. In other words, not only legitimate users can receive the information, but also malicious users can extract secure information from the wireless signals. Furthermore, due to the distributed characteristics of SAGIN and multi-hop transmission, intrinsic trust and data reliability issues may arise in each layer or across layers. Thus, security and privacy have become essential to determine if the SAGIN can continue to evolve healthily. 

\subsubsection{Quantum-empowered FL and Quantum AI}\label{sec:future_quantum}
One of the main methods to enhance the security of SAGIN is the use of AI. AI algorithms hold the potential to proactively detect and mitigate threats before they materialize, enhancing overall network resilience. In the context of SAGINs, specific AI paradigms like federated and swarm learning offer compelling benefits. These distributed approaches enable secure and privacy-preserving data analysis directly on edge devices, minimizing communication overhead and safeguarding sensitive information while contributing to robust threat detection and efficient security solutions. Although FL improves privacy, it still faces several issues. First, due to the large size and complexity of SAGINs, an increasing number of devices with limited capabilities are required to contribute to the FL across different segments. Second, due to the long-distance transmissions that wireless signals need to propagate in SAGINs, malicious nodes have higher chances of obtaining the model gradients and inferring the user's private information.

To solve the privacy leakage issue and ensure secure transmissions, quantum computing and quantum communications are gaining more focus, especially after the advancement seen in quantum mechanics~\cite{chehimi2023foundations}. Quantum communication harnesses the distinct attributes of quantum mechanics like superposition and entanglement to enable secure transmission of information, ensuring inherent safeguards against interception and tampering. In particular, quantum teleportation, a quantum communication protocol for quantum information transfer, can be leveraged to establish secure communications in SAGIN between users and the FL server. Quantum-empowered FL can, therefore, mitigate the privacy concerns of traditional FL, facilitating its widespread adoption in SAGINs~\cite{Wang2023Quantum}. Despite Quantum-empowered FL's great advantages for SAGINs, it still faces several challenges. For instance, existing noise and decoherence can significantly affect quantum communications, limiting the communication range and resulting in unreliable FL learning outcomes. A key feature that the SAGIN architecture can provide is the abundant existence of nodes in the same layer or across layers, which can act as quantum relays due to the maneuverability and LoS transmission capabilities~\cite{Xu2022Quantum}. A potential research direction would be to study and optimize multi-hop empowered FL for secure and trustworthy SAGINs. 

In addition to quantum communications, the coupling of quantum coding with AI techniques, quantum AI, is yet another technique to enhance the security of SAGINs. For instance, quantum key distribution can be used to detect eavesdroppers trying to access the transmission more accurately and rapidly as the network dynamically adapts and learns in real-time. Quantum NNs, which also utilize quantum processing, are another efficient tool that can be exploited to increase the learning rate of different AI algorithms in SAGINs. However, the research in this area is still not mature, and several research directions can still be exploited to enable the fast utilization of quantum AI for SAGINs.

\subsubsection{AI-oriented Authentication}\label{sec:future_authentication}
Finally, AI can also contribute to enhancing the authentication of SAGINs. The authentication process in SAGINs raises questions that should be carefully considered to maintain secure and trustworthy networks. First, the long communication range significantly increases the propagation delay, reducing the QoS of latency-sensitive scenarios. Further delay is caused by the processing of conventional cryptographic authentication methods, which requires increased communication and computation overhead to cope with the new security requirements. Furthermore, these crypto-based authentication schemes verify the legitimacy only at the beginning and are therefore susceptible to different attacks. As such, new mechanisms are required for real-time and transparent security provisioning. 

Continuous authentication mechanisms have emerged recently to fulfill the security requirements and mitigate the limitations of conventional mechanisms~\cite{Yang2024AI}. Typically, continuous schemes verify the legitimacy of users based on behavior and physiological biometrics, which are user-specific. Due to the increased storage and computational resources of SAGINs, continuous authentication can be effectively implemented while relying on collecting and storing massive amounts of data. AI can be introduced here to extract unique user features from the collected data. Specifically, spatial-temporal features such as position, Doppler shift, and traffic volume can be used in the continuous authentication phase driven by the potential of SAGIN to provide vertical applications supported by heterogeneous types of devices. Therefore, AI techniques, such as DL, supervised and unsupervised learning, and RL, are key to maintaining trust while adapting to the dynamic environment. 

Although AI-oriented authentication has shown great potential in SAGINs, several research gaps remain to be investigated. For instance, dedicated work should be focused on optimizing the allocation of authentication tasks between the different layers of the SAGIN architecture while considering both communication and computation resources scarcity in addition to latency constraints. Furthermore, novel AI-based authentication algorithms must be designed to allow for simple feature extraction processes while satisfying the low computation and communication requirements of SAGINs. Furthermore, Blockchain technology can be leveraged to enhance the intelligence of the authentication algorithms further. With AI, Blockchain can create a distributed and tamper-proof ledger system for storing user credentials and audit trails, improving system transparency and accountability. This integration can also enable efficient revocation of compromised credentials and facilitate dynamic trust management between entities within the SAGIN network. Ultimately, the convergence of AI and Blockchain holds immense promise for revolutionizing authentication in SAGINs, paving the way for secure and trustworthy networks that cater to the ever-growing diverse demands of future communication systems. By addressing the critical research gaps and fostering continued innovation in this domain, we can unlock the full potential of AI-powered authentication for robust and resilient SAGIN infrastructures.

\section{Conclusion}
\label{sec:conclusion}
This paper explores the dynamic relationship between SAGINs and AI, significantly impacting the forthcoming 6G paradigm. The convergence of AI and SAGINs has great potential for communication and information accessibility. However, the complex nature of SAGINs presents challenges such as dynamic optimization across multiple layers, management of changing topologies, and resource constraints. AI-based solutions like DRL and supervised and unsupervised learning methods offer promising ways to overcome these challenges by enhancing network performance by collectively optimizing topology, scheduling, resource allocation, routing, and mobility. Real-time learning is essential as AI algorithms adapt strategies based on network data in response to evolving conditions. An efficient AI-based approach is crucial for tasks such as caching, computation offloading, and scheduling due to the limited resources on aerial platforms. Additionally, the synergy between AI and SAGINs can accelerate the development of effective AI algorithms, i.e., by leveraging FL applications and exploiting SAGIN advantages to enhance contributions to FL. Wireless techniques like analog over-the-air computation and digital RRM increase the potential, creating new opportunities for applications reliant on AI and empowered by SAGINs. In conclusion, combining AI with SAGINs offers a significant opportunity to transform communication and information access by harnessing the full range of capabilities through AI-driven optimization and adaptation across diverse applications.

\section*{Acknowledgment}
Fig.~\ref{fig:large_fig} was created by Heno Hwang, a scientific illustrator at King Abdullah University of Science and Technology (KAUST).

\ifCLASSOPTIONcaptionsoff
  \newpage
\fi

\bibliography{bibliography}
\bibliographystyle{ieeetr}
\end{document}